\documentclass[aps,pra,10pt,letterpaper,twocolumn,superscriptaddress,showpacs,preprintnumbers]{revtex4-1}

\usepackage{color}
\usepackage{graphicx}
\usepackage{bm}
\usepackage{dcolumn}
\usepackage{verbatim}
\usepackage{mathbbol}
\usepackage{dcolumn}
\usepackage{amsmath}
\usepackage{placeins}

\newcommand{\Sixj}[6]{\left\{ \!\! \begin{array}{ccc}
#1 & #2 & #3 \\
#4 & #5 & #6
\end{array} \!\! \right\}}

\definecolor{dred}{rgb}{0.75,0,0}
\definecolor{dark_purple}{rgb}{0.75,0,0.75}
\definecolor{blue}{rgb}{0,0,1}
\definecolor{dark_blue}{rgb}{0,0,0.75}
\definecolor{dark_green}{rgb}{0,0.75,0}

\newcommand{\sumint}{\sum\hspace{-12pt}\int}

\newcommand{\cvec}[1]{\mathbf{#1}}
\newcommand{\vecop}[1]{\cvec{\hat{#1}}}

\newcommand{\abs}[1]{\left|#1\right|}
\newcommand{\ket}[1]{|#1\rangle}
\newcommand{\bra}[1]{\langle#1|}
\newcommand{\braOket}[3]{\langle#1|#2|#3\rangle}

\newcommand{\Schro}{Schr\"o\-din\-ger }
\newcommand{\etal}{\emph{et~al.\@} }
\newcommand{\ie}{i.e., }
\newcommand{\eg}{e.g.\@ }

\newcommand{\au}{\,\text{a.u.}}

\newcommand{\Hep}{\ensuremath{\text{He}^{+}}}
\newcommand{\Hepp}{\ensuremath{\text{He}^{++}}}
\newcommand{\as}{\,\text{as}}
\newcommand{\fs}{\,\text{fs}}

\newcommand{\Int}{\int\limits}
\newcommand{\dd}{\mathrm{d}}
\newcommand{\dt}{\dd t}

\newcommand{\COMMENT}[1]{}

\usepackage[colorlinks,linkcolor=dred,citecolor=dark_green]{hyperref}

\begin{document}

\title{Photoionization of helium by attosecond pulses: extraction of spectra from correlated wave functions}

\newcommand{\itptuw}{Institute for Theoretical Physics, Vienna University of Technology, 1040 Vienna, Austria, EU}

\author{Luca Argenti}\email{luca.argenti@uam.es}
\affiliation{Departamento de Qu\'imica, M\'odulo 13, Universidad Aut\'onoma de Madrid, 28049 Madrid, Spain, EU}

\author{Renate Pazourek}
\affiliation{\itptuw}

\author{Johannes Feist}
\affiliation{ITAMP, Harvard-Smithsonian Center for Astrophysics, Cambridge, Massachussetts 02138, USA}
\affiliation{Departamento de F\'isica Te\'orica de la Materia Condensada, Universidad Aut\'onoma de Madrid, E-28049 Madrid, Spain}

\author{Stefan Nagele}
\affiliation{\itptuw}

\author{Matthias Liertzer}
\affiliation{\itptuw}

\author{Emil Persson}
\affiliation{\itptuw}

\author{Joachim Burgd\"orfer}
\affiliation{\itptuw}

\author{Eva Lindroth}
\affiliation{Department of Physics, Stockholm University, AlbaNova University Center, SE-106 91 Stockholm, Sweden, EU}

\date{\today}

\begin{abstract}
We investigate the photoionization spectrum of helium by attosecond XUV pulses both in the spectral region of doubly excited resonances as well as above the double ionization threshold. 
In order to probe for convergence, 
we compare three techniques to extract photoelectron spectra from the wavepacket resulting from the integration of the time-dependent Schr\"odinger equation in a finite-element discrete variable representation basis.
These techniques are: projection on products of hydrogenic bound and continuum states, projection onto multi-channel scattering states computed in a B-spline close-coupling basis, and a technique based on exterior complex scaling (ECS) 
implemented in the same basis used for the time propagation. These methods allow to monitor the population of continuum states in wavepackets created with ultrashort pulses in different regimes. 
Applications include photo cross sections and anisotropy parameters in the spectral region of doubly excited resonances, time-resolved photoexcitation of autoionizing resonances in an attosecond pump-probe setting, and the energy and angular distribution of correlated wavepackets for two-photon double ionization.
\end{abstract}


\maketitle

\section{Introduction}

During the last decade, two transformational experimental techniques, high harmonic generation~\cite{Sansone2011} and x-ray free electron lasers~\cite{McNeil2010}, have given access to femtosecond and sub-femtosecond intense light pulses in the XUV and soft x-ray energy range, thus opening the way to  time resolved studies of the correlated motion of electrons in atoms and molecules on their characteristic time-scale~\cite{Shiner2011a,Schultze2010a,Smirnova2009,Haessler2010,Feist2011a,Caillat2011,Klunder2011,Kuleff2011}. These new techniques can be used not only to monitor the electronic motion but also to steer it~\cite{Remetter2006,Mauritsson2008}. This latter capability offers the perspective of controlling dynamics at the femtosecond and sub-femtosecond timescale, such as electronic dynamics in atoms~\cite{FeiNagPaz2009,Argenti2010}, molecules~\cite{Sansone2010,Kelkensberg2011,Fischer2010}, and solids~\cite{Kruger2011}, and eventually also nuclear dynamics such as fast proton migration~\cite{Jiang2010b}.

The interpretation of experiments on attosecond dynamics, however, faces a number of difficulties and requires guidance by theory. First, sub-femtosecond pulses typically excite the target to a coherent superposition of states above the ionization threshold and across a wide range of energies. As a consequence, several different ionization regimes such as multiply excited autoionizing states, multichannel single ionization states and, possibly, multiple ionization states, are accessed at the same time. Second, in common pump-probe schemes~\cite{Sansone2010}, the strong few-cycle IR-probe pulse that follows an attosecond weak XUV-pump pulse gives rise to electronic dynamics that unfolds on a very short time-scale through non-perturbative stages, e.g.~tunneling, over-the-barrier ionization, and Rabi oscillations. 

Traditional perturbative approaches~\cite{Faisal} are clearly not well suited to describe such dynamical regimes. Moreover, since the duration of compressed IR pulses easily spans just a few~\cite{Sansone2006} or even a single~\cite{Goulielmakis2008} carrier cycle, stationary non-perturbative techniques like those based on the Floquet method~\cite{Burke1991} cannot be used either. Reliable theoretical predictions for ultrashort processes, therefore, generally require direct integration of the time-dependent Schr\"odinger equation (TDSE)~\cite{Smyth1998,LagmagoKamta2002,Laulan2004,Sanz-Vicario2006,Tong2006,Foumouo2006,Lambropoulos2008a,Lysaght2008,Feist2008,Palacios2008,Nepstad2010,Argenti2010}. Such an approach permits to reproduce faithfully the physical process under study. However, it gives rise to a multitude of problems as well. Relevant parts, if not most, of the electron dynamics triggered by sub-femtosecond pulses take place in the ionization continuum. Indeed, typical experiments are designed to 
monitor the energy and angular distribution of the photoelectrons emerging from the reaction center, e.g., with a velocity map imaging spectrometer~\cite{Eppink1997}, or even several photo-fragments in coincidence, e.g., with a reaction microscope~\cite{Ullrich2003}. Transient absorption spectroscopy~\cite{Goulielmakis2010}, which monitors quasi-bound electronic dynamics, constitutes a notable exception. One of the most prominent problems theory has to face is then how to extract from a numerical simulation, intrinsically limited in both time and space, the relevant asymptotic scattering information.

A number of alternative techniques have been used in the past to extract differential distributions of the products of a light-induced reaction from an entangled wave function: spectral analysis of autocorrelation functions~\cite{Nikolopoulos2007}, analysis of the radial flux at large distances~\cite{Kjeldsen2006},
projection of the wavepacket onto products of one-particle functions, e.g., Coulomb functions~\cite{Laulan2004,Feist2008,Madsen2007} or Volkov states~\cite{Tong2006}, asymptotic analysis of a monochromatic component of the wavepacket extracted with the resolvent technique~\cite{Catoire2012}, surface integration of the monochromatic component of the wave function extracted with a technique based on exterior complex scaling (Berkeley-ECS)~\cite{McCurdy2004b,Palacios2007a,Palacios2008,Palacios2009}, or projection on scattering states computed on the same basis used to carry out the time propagation~\cite{Kazansky2003,Foumouo2006,Madsen2007,Lambropoulos2008a,Lysaght2008,Fernandez2009a,Argenti2010}.
Very recently \cite{Tao2012,Scrinzi2012} a method that combines the flux analysis at moderate radii with the ECS technique for optical absorption at larger distances was proposed. 
Implementations for the full Coulomb two-electron problem are, however, not yet available.

In the following, we investigate three of these techniques in more detail for the prototypical three-body system, the helium atom: projection onto products of one-particle functions, projection onto exact scattering states, and the Berkeley-ECS method. For all these techniques, the extraction of asymptotic scattering information takes place in the field-free region after the laser pulse is over. As will be discussed in more detail later, each of these approaches has its strengths and drawbacks making it applicable for different photon energies and pulse durations. 
By comparing results from complementary methods we are able to assess the convergence of the simulation of dynamical observables that have become accessible by attosecond photoelectron spectroscopy.
We will present applications to three sets of benchmark data: i) photoionization cross sections and anisotropy parameters for single ionization of helium in the spectral region of doubly-excited resonances, ii) time-resolved photoionization by an attosecond pump-probe setting in the same spectral domain, and iii) the energy and angular distribution of correlated wavepackets in two-photon ionization by ultrashort pulses above the double ionization threshold.

The article is organized as follows. In section \ref{sec:FEDVR} we briefly review our method of solving the time-dependent \Schro equation (TDSE) for helium in its full dimensionality. It allows to accurately represent the electronic wavepacket generated by (a sequence of) ultrashort light pulses on a grid with a finite spatial domain and for a finite propagation time.
Three alternative methods to extract dynamical observables from such wavepackets are introduced and their applicability in different regimes is compared in section \ref{sec:methods}.
Applications to three scenarios of current interest are presented in section \ref{sec:results} followed by concluding remarks in section \ref{sec:conclusions}.
Additional technical and computational details are given in the appendix.
Atomic units are used throughout unless stated otherwise.

\section{Propagation method\label{sec:FEDVR}}

Our computational approach (see \cite{Feist2008,Fei2009,SchFeiNag2011} for a more detailed description) for solving the 
time-dependent Schr\"o\-din\-ger equation for two-electron systems,
\begin{equation}\label{eq:SchroCoord}
i\frac{\partial}{\partial t}\Psi(\cvec{r}_1,\cvec{r}_2,t) = H\Psi(\cvec{r}_1,\cvec{r}_2,t) \,,
\end{equation}
is based on a time-dependent close-coupling scheme \cite{ColPin2002,LauBac2003,HuColCol2005,PinRobLoc2007} where we expand the angular part of the six-dimensional wave function $\Psi(\cvec{r}_1, \cvec{r}_2)$ in coupled spherical harmonics $\mathcal{Y}_{l_1,l_2}^{LM}(\Omega_1,\Omega_2)$.

The interaction of a helium atom with linearly polarized light is described by the Hamiltonian
\begin{equation}\label{eq:HeHam}
H=H_a+H_{\text{em}}^{l,v}=\frac{\vecop{p}_1^2}{2}+\frac{\vecop{p}_2^2}{2}
-\frac{2}{r_1}-\frac{2}{r_2}
+\frac{1}{\abs{\cvec{r}_1-\cvec{r}_2}}
+H_{\text{em}}^{l,v} \,,
\end{equation}
where the interaction with the electromagnetic field in the dipole approximation, $H_{_{\text{em}}}^{l,v}$, is either given in \emph{length} or \emph{velocity} gauge. 
The gauge independence of the physical observables is a necessary condition for the convergence of the numerical solution.

For the discretization of the radial functions $R_{l_1,l_2}^{L}(r_1,r_2,t)$, we employ a finite-element discrete-variable representation (FEDVR)~\cite{ResMcc2000,MccHorRes2001,SchColHu2006}. We divide the radial coordinates into finite elements in each of which the functions $R_{l_1,l_2}^{L}$ are represented in a local DVR basis with a corresponding Gauss-Lobatto quadrature to ensure the continuity of the wave function at the element boundaries. This method leads to sparse matrix representations of the differential operators and to a diagonal potential matrix (within quadrature accuracy), enabling efficient parallelization. 

For the temporal propagation, we employ the short iterative Lanczos (SIL) method \cite{ParkLight86,SmyParTay1998,Lefo90} with adaptive time-step control. 
The initial He ground state $1${$^1$S}$(1s^2)$ is obtained by relaxing an arbitrary test function in imaginary time. 
For an initial $2${$^1$S}$(1s2s)$ metastable state we directly solve the eigenvalue problem of the field-free Hamiltonian (\ref{eq:HeHam}) in a small box using the SLEPc library \cite{HerRomVid2005}. 
The radial grid covers a range $[0,r_\mathrm{max}]$, with typical values of $r_\mathrm{max} \approx 150\au$, although much larger values are possible.
The temporal integration extends to a maximum time $t_\mathrm{max}$ which exceeds at least the pulse length $t_\mathrm{max} > \tau_p$ ($\tau_p \approx 1600\as \simeq 65 \au$ for $T_{\scriptstyle\mathrm{FWHM}}$=$200\as$ of a Gaussian intensity envelope) but may be increased  much further (up to $t_\mathrm{max} \approx 8\fs$), as discussed below.
At the conclusion of the time propagation, the wavepacket $\Psi(\cvec r_1, \cvec r_2, t_\mathrm{max})$ contains all accessible scattering information.
The extraction of this information on the asymptotic scattering state, \ie the $t \to \infty$ limit, is a non-trivial numerical task for which we discuss in the following three, partially complementary, techniques.

\section{Alternative extraction methods \label{sec:methods}}

In order to extract the information on the energy and angular distribution of emergent photofragments, it is crucial to
establish a correspondence between the experimental observables and the simulated finite domain wavepacket dynamics. In principle, this is straightforward: 
Upon conclusion of the electromagnetic pulse, the numerical solution of the TDSE yields the wave function $\Psi(t)$ of the wavepacket. The dynamics of $\Psi(t)$ is then governed  by the total field-free atomic Hamiltonian, $H_a$. The state $\Psi(t)$ is composed of a bound and an unbound part. Let us indicate by $\Lambda$ the 
projector on the bound states of $H_a$, and with $\Psi'(t)=(\mathbb{1}-\Lambda)\Psi(t)$ the unbound part of $\Psi(t)$. The fragments detected are associated with the long-time limit of the unbound component $\Psi'(t)$. It consists of a superposition of unbound eigenstates $\varphi_{\alpha E}$ of the sum of the Hamiltonians of the separated fragments, $H_0$. 
The projection amplitudes 
\begin{equation}\label{eq:c_alpha}
c_{\alpha E}(t)\,=\,\langle\varphi_{\alpha E}\,|\,\Psi'(t)\rangle\,,
\end{equation}
where $\alpha$ designates the collective set of quantum numbers beyond the total energy $E$, uniquely characterize the final state of the system. These include the fragmentation channel of the target, the asymptotic angular distribution of the photofragments, and their internal quantum numbers (e.g.~spin and angular momentum).

The expansion coefficients $c_{\alpha E}(t)$ in (\ref{eq:c_alpha}) may not necessarily converge in the infinite-time limit. In particular, when two or more fragments are charged, the phase of $c_{\alpha E}(t)$ diverges due to the long-range character of the Coulomb field. Notwithstanding this difficulty the probability density $|c_{\alpha E}(t)|^2$ will still converge in the sense of distributions (i.e., when convoluted with a test function) yielding a well-defined asymptotic distribution of the fragments. 
Hence, the probability distribution $P_{\alpha}(E)$ for the detection of fragments in channel $\alpha$ and with energy $E$ at the end of the propagation is given by 
\begin{equation}
P_{\alpha}(E)=\,\lim_{t\to\infty} |c_{\alpha E}(t)|^2\label{eq:coefficient_convergence3}.
\end{equation}
\COMMENT{More details on this result can be found in Sec.~XI.9 of~\cite{ReedSimonIII}. Moreover, a substantial body of numerical evidence~\cite{Feist2009,Nepstad2010,Madsen2007,Lambropoulos2008a,Foumouo2008} indicates that the more lenient result~(\ref{eq:coefficient_convergence3}) holds in the case of atomic double photoionization as well.
For the time being we will thus assume that, with suitable precautions, \autoref{eq:coefficient_convergence3} is applicable.}
The probability distribution $P_{\alpha}(E)$ can then be computed as
\begin{equation}
P_{\alpha}(E)=\lim_{t\to\infty}\big|\langle\varphi_{\alpha E}|\mathbb{1}-\Lambda|\Psi(t)\rangle\big|^2.\label{eq:P_alpha}
\end{equation}
At first sight, the prescription (\ref{eq:P_alpha}) to extract experimental observables from a wavepacket has the appeal of simplicity since the uncoupled states $\varphi_{\alpha E}$ are usually more easily obtainable than the continuum eigenstates of the full Hamiltonian. This simplicity, though, is misleading since the projector $\Lambda$ of the total Hamiltonian requires at least a certain number of bound states of the fully interacting system to be known. Since the bound states of $H_a$ are not orthogonal to $\varphi_{\alpha E}$, their elimination is essential. Otherwise they would give rise to spurious contributions to the ionization channels which do not vanish for large times.

A more serious drawback of \autoref{eq:P_alpha}, however, stems from the imposed asymptotic time limit where the propagation algorithm is limited to $t_\mathrm{max}$. Even by the inclusion of part of the long-range interactions between photofragments into the channel Hamiltonian $H_0$ this problem can only be marginally alleviated rather than solved. More generally, all the methods that require the wavepacket to reach the asymptotic region, i.e. the region where the dynamics governed by $H_0$ and $H_a$ become equivalent, face the same problem, namely the propagation of the fully correlated wave function for long times and at large distances which may be computationally prohibitively expensive. This limitation becomes particularly severe in a number of circumstances frequently encountered in atomic photoionization. For example, when the wavepacket spectrum is distributed across an ionization threshold, the wavepacket comprises components with vanishingly small kinetic energy that take exceedingly long times 
to reach the 
asymptotic region. Even more severe, when several channels with different thresholds are simultaneously open, both slow and fast photoelectrons are present at the same time. Hence, in order for the slowest part of the wave function to reach the asymptotic region, the propagation box must be large enough to accommodate for the fastest components as well. A further difficulty with \autoref{eq:P_alpha}, perhaps the most relevant in the present context, is provided by resonances in general and by Rydberg series of doubly excited states in particular. \COMMENT{Such doubly excited states are a general feature in  atomic  photoionization spectra.} First, the convergence of the resonant profiles in those channels where the excited resonances decay requires a propagation time proportional to the lifetime of the longest lived resonance which is excited in the simulation. Second, doubly excited states have, in general, non-vanishing scalar products with all the eigenfunctions of $H_0$, including 
those 
belonging to closed ionization channels.

\subsection{Projection onto asymptotic channel eigenstates} \label{sec:proj_channel}

The simplest and most straightforward implementation of \autoref{eq:P_alpha} is to extend the temporal propagation time $t_\mathrm{max}$ to the largest value computationally feasible and subsequently project onto (approximate) asymptotic channel eigenstates $\varphi_{\alpha E}$,
\begin{equation}
H_0 \ket{\varphi_{\alpha E}} = E \ket{\varphi_{\alpha E}} \, .
\end{equation}
In the present case of excitation-ionization and double ionization of helium there are two alternative choices for $H_0$ and $\ket{\varphi_{\alpha E}}$.
One choice consists of taking $H_0$ to be a hydrogenic Hamiltonian with $Z=1$ for the continuum electron and $Z=2$ for the bound electron in the case of single ionization, and $Z=2$ for both electrons in the case of double ionization.
Alternatively, asymptotic channel functions are 
obtained from the diagonalization of the total Hamiltonian in the configuration space where one of the two electrons is frozen in an hydrogenic bound state of the parent ion. These channel functions differ from the products of Coulomb functions in the radial region where the bound electron density is not zero. The effective potential felt by the free electron deviates from that of a nuclear charge with Z=1. Yet, at larger distances, these channel functions converge to phase-shifted Coulomb functions. Therefore, for the purpose of evaluating the absolute value of the projection of an outgoing wavepacket (\autoref{eq:P_alpha}), these channel functions and pure Coulomb functions are practically equivalent.
As mentioned above, projections onto such asymptotic channel functions at $t_\mathrm{max}$ may fail in the presence of long-lived resonances.

\subsection{Projection onto exact scattering states} \label{sec:PSS}

One avenue to circumvent some of the limitations associated with \autoref{eq:P_alpha} is to use, instead of the asymptotic states $\varphi_{\alpha E}$, the exact scattering states $\psi_{\alpha E}^{-}$ of the total atomic Hamiltonian $H_a$~\cite{Newton}. The $\psi_{\alpha E}^-$ states fulfill so-called incoming boundary conditions~\cite{Breit1954,Altshuler1956}, which are appropriate to the context of photoionization, since photoelectrons are observed in the positive time limit.

Such scattering states satisfy the Lippmann-Schwinger equation with advanced Green's functions~\cite{Newton}
\begin{eqnarray}
\psi^-_{\alpha E}
&=&\varphi_{\alpha E}+G_0^-(E)H'\psi^-_{\alpha E} \label{eq:lsG0}\\
&=&\varphi_{\alpha E}+G^-(E)H'\varphi_{\alpha E} \label{eq:lsG}
\end{eqnarray}
where the operators $G_0^-(E)=(E-H_0-i0^+)^{-1}$ and $G^-(E)=(E-H_a-i0^+)^{-1}$ are the resolvents of the channel ($H_0$) and full Hamiltonian ($H_a$), while $H'=H_a-H_0$ is the corresponding perturbation (the interactions not accounted for by the channel Hamiltonian $H_0$).

The $\psi^-_{\alpha E}$ states form a complete orthonormal basis for the unbound states of $H_a$,
\begin{equation}
H_a\ket{\psi^-_{\alpha E}}=E\ket{\psi^-_{\alpha E}},\quad\langle\psi^-_{\alpha E}|\psi^-_{\beta E'}\rangle=\delta_{\alpha\beta}\delta(E-E') \, ,
\end{equation}
\begin{equation}
\mathbb{1}-\Lambda=\sum_\alpha\int d\epsilon\,|\,\psi_{\alpha\epsilon}^-\rangle\,\langle\,\psi_{\alpha\epsilon}^-\,| \, .
\end{equation}
We can thus write the scattering component $\Psi'(t)$ of the wavepacket at any time $t$ after the external field is over as
\begin{eqnarray}
|\Psi'(t)\rangle&=&e^{-iH_a(t-t_\mathrm{max})}(\mathbb{1}-\Lambda)|\Psi(t_\mathrm{max})\rangle=\nonumber\\
&=&\sum_\alpha\int d\epsilon\,|\psi_{\alpha\epsilon}^-\rangle\,e^{-i\epsilon (t-t_\mathrm{max})}\,c^-_{\alpha \epsilon}(t_\mathrm{max})\label{eq:scatexp}
\end{eqnarray}
where
\begin{equation}
c^-_{\alpha E}(t_\mathrm{max})=\langle\psi^-_{\alpha E}|\Psi'(t_\mathrm{max})\rangle.
\label{eq:scatcoef}
\end{equation}
A crucial aspect of \autoref{eq:scatexp} is that, in the large time limit, the states $\psi^-_{\alpha \epsilon}$ can be replaced by their asymptotes $\varphi_{\alpha\epsilon}$~\cite{Breit1954,Altshuler1956}. This is one defining feature of the $\psi^-_{\alpha \epsilon}$ states, sometimes referred to as control of $\psi^-_{\alpha \epsilon}$ by the future. In particular,
\begin{equation}\label{eq:control_future}
\lim_{t\to\infty}|c_{\alpha E}(t)|=|c_{\alpha E}^-(t_\mathrm{max})|
\end{equation}
To compute the distribution $P_\alpha(E)$ we now combine~\autoref{eq:coefficient_convergence3}, \autoref{eq:control_future}, and \autoref{eq:scatcoef} and obtain the \emph{exact} result
\begin{equation}
P_\alpha(E)=\big|\langle\psi^-_{\alpha E}|\Psi(t_\mathrm{max})\rangle\big|^2.
\label{eq:Pascat}
\end{equation}
In contrast to \autoref{eq:P_alpha}, \autoref{eq:Pascat} requires neither a projection onto the bound states of the system, to which the scattering states are orthogonal, nor the evaluation of a long-time limit. The convenience of using scattering states $\psi^-_{\alpha E}$, instead of the asymptotic limits $\varphi_{\alpha E}$, is thus apparent: \autoref{eq:Pascat} can be evaluated as soon as the external time-dependent field is over, without having to wait until the ionizing wavepacket reaches the asymptotic region (\autoref{eq:P_alpha}).

\COMMENT{In~\cite{Madsen2007}, the use of scattering states has been deemed impractical but in the simplest cases, due to the additional workload required to solve \autoref{eq:lsG0} or \autoref{eq:lsG}. Indeed, no systematic procedure to generate accurate scattering states for the double ionization of atoms has been reported to date.} For single-ionization processes the calculation of scattering states is straightforward. This task has been tackled successfully in the course of the last four decades~\cite{pra.14.2159,pra.43.3474,pra.43.1301,Moccia1991,pra.44.R13}. Today several efficient methods capable of computing multi-electron single ionization scattering states are available. They include the R-matrix~\cite{Burke_book}, J-matrix~\cite{Vanroose2002}, K-matrix~\cite{Cacelli1991,Argenti2006}, Feshbach projection~\cite{Martin1993}, and inverse iteration~\cite{Venuti1996}. Furthermore, the computational overhead for generating scattering states is easily compensated whenever a large 
number of different simulations must be carried out, as is the case, e.g., of time-delay scans in pump-probe schemes. 
Some of the details of our implementation for helium are given in appendix \ref{sec:appendix_PSS}.
The biggest drawback of these methods is that they only work for total energies below the double ionization threshold
as no systematic procedure to generate accurate scattering states for double ionization is known to date.

\subsection{Berkeley-ECS method} \label{sec:ECS}

A third elegant strategy for computing $P_\alpha(E)$, based on exterior complex scaling, was put forward by Palacios, McCurdy and Rescigno~\cite{McCurdy2004b,Palacios2007a,Palacios2008,Palacios2009}. 
In this approach, referred to in this work as the Berkeley-ECS method, the monochromatic component $\Psi_\textrm{sc}(E)$ at energy $E$ of the wavepacket $\Psi(t_\mathrm{max})$ is extracted by applying to $\Psi(t_\mathrm{max})$ the retarded resolvent $G^+(E)$ of the total Hamiltonian
\begin{equation}
\label{eq:ECS_driven_eq}
  (E-H_a)\ket{\Psi_\textrm{sc}(E)} = \ket{\Psi(t_\mathrm{max})}\, ,
\end{equation}
the realization of which in a finite radial domain is provided by the resolvent of an exterior-complex-scaled Hamiltonian $H_\theta$. From the function $\Psi_\textrm{sc}(E)$ the distribution of the photofragments is extracted in the asymptotic region. 

Our implementation of the Berkeley-ECS method closely follows that of Palacios \emph{et al.}~\cite{Palacios2007a,Palacios2008,Palacios2009}, which builds on earlier work reviewed in~\cite{McCurdy2004b}. We give a brief overview here and refer the reader to the original papers for details. The central idea is to solve the inhomogeneous linear system \autoref{eq:ECS_driven_eq}
in the basis used for temporal propagation (in our implementation using the PETSc package~\cite{petsc-web-page}). The scattered wave $\Psi_\textrm{sc}(E)$ is equivalent to the Fourier transform of the time-dependent wavepacket from $t=t_\mathrm{max}$ to $t=\infty$ and corresponds to the application of the retarded Green's function of the total Hamiltonian on the wavepacket,
\begin{equation}
\ket{\Psi_\textrm{sc}(E)}=G^+(E)\ket{\Psi(t_\mathrm{max})}.
\end{equation}
Purely outgoing boundary conditions are enforced by an exterior complex scaling (ECS) transformation for each of the radial coordinates. As the wavepacket at the end of the pulse is a square-integrable function, the asymptotic behavior of the scattered wave can be deduced from the asymptotic form of the Green's function. For single ionization, the amplitude $c_{\alpha E}(t_\mathrm{max})$ can be expressed as \cite{McCurdy2004b},
\begin{equation}
c_{\alpha E}(t_\mathrm{max}) = \bra{\varphi_{\alpha E}} E - H_0 \ket{\Psi_\textrm{sc}(E)},
\end{equation}
where $\varphi_{\alpha E}$ is an asymptote of the channel Hamiltonian $H_0$. 
The latter should contain the monopolar long-range interaction between the fragments to suppress spurious contributions. Using Green's theorem allows one to express the single ionization amplitude as a surface integral in the non-scaled region of space,
\begin{equation}
\label{eq:ECS_surf_int}
c_{\alpha E}(t_\mathrm{max}) = \frac{1}{2} \int_S (\varphi_{\alpha E} \boldsymbol\nabla \Psi_\textrm{sc}(E) - \Psi_\textrm{sc}(E) \boldsymbol\nabla \varphi_{\alpha E}) \cdot \mathrm{d}\cvec{S},
\end{equation}
where $\boldsymbol\nabla = (\nabla_1,\nabla_2)$ is the six-dimensional gradient operator (in the present case of helium). Since the integral~(\autoref{eq:ECS_surf_int}) is evaluated in a radial region far from the atom and $\Psi_{\textrm{sc}}(E)$ is an outgoing wavepacket by construction, it is sufficient that $\varphi_{\alpha E}$ satisfies the same outgoing boundary conditions as the eigenstates of $H_0$. Thus, in single ionization, $\varphi_{\alpha E}$ can be taken as the symmetrized product of an ionic bound state and a Coulomb wave with $Z=1$. For double ionization, a similar expression can be found~\cite{Palacios2007a,McCurdy2004b}.

\subsection{Range of applicability}

\begin{figure}[tb]
\begin{center}
\includegraphics[width=\linewidth]{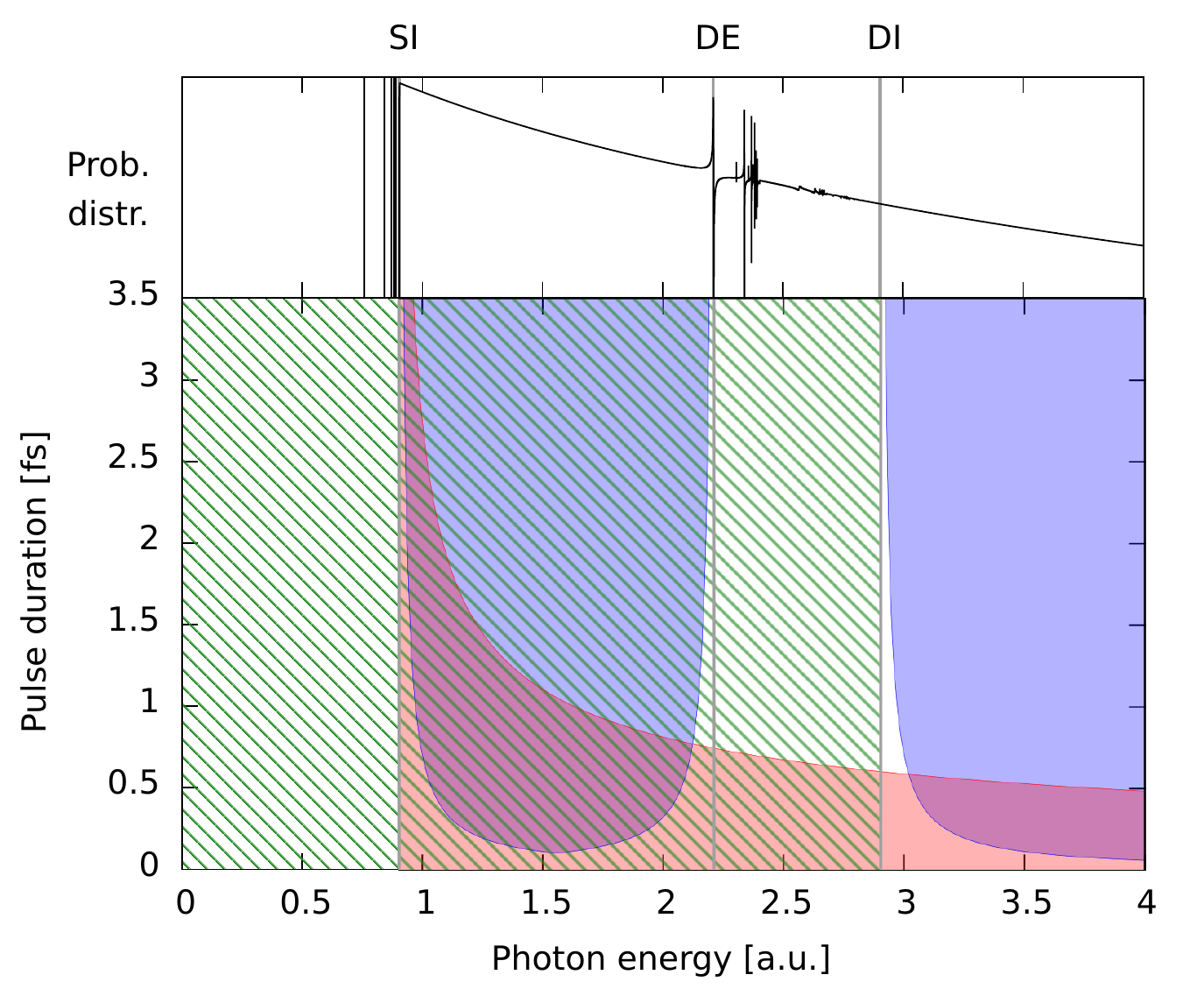}
\caption{\label{fig:si_regimes} Approximate regions of applicability of different extraction methods for the single ionization continuum of helium as a function of photon energy and pulse duration for a single-photon transition. SI: single ionization, DE: double excitation, DI: double ionization. Green shaded: projection on scattering states (PSS), blue solid: projection onto asymptotic channel functions, red solid: Berkeley-ECS method. See text for details. The upper part shows the (logarithmic) ionization probability for single ionization and positions of excited bound states.}
\end{center}
\end{figure}

The accuracy of the three extraction methods strongly varies in different spectral regimes.
As will be illustrated in more detail below in connection with applications of current interest, the range of applicability can be summarized as follows (\autoref{fig:si_regimes}).

The projection onto asymptotic channel functions discussed in \autoref{sec:proj_channel} works well
when the wavepacket only contains electrons that are already well-separated in coordinate space. While this is relatively straightforward to achieve at high energies (such as above the double ionization continuum), it becomes prohibitive close to the doubly excited resonances that are located for helium in the energy interval from around $-0.7\,$a.u.\ to $0\,$a.u., and which decay only after many femtoseconds. The main advantages of this approach are its analytical and numerical simplicity.
For double ionization, the asymptotic channel functions are Coulomb functions with $Z=2$ and the method can be applied even for long driving pulses when large computational boxes are employed.
In the case of single ionization, either hydrogenic channel functions with $Z=1$ (continuum) and $Z=2$ (bound state) or the closely related eigenstates obtained by freezing the inner electron in the ionic state can be used.
Projection on scattering states (PSS, \autoref{sec:PSS}) works for any separation of the 
ionized electron from the core,
and thus can be used also when doubly excited resonances are excited and before they have decayed.
Fully differential photoelectron spectra can be 
extracted from the wavepacket arising in photoionization by means of a simple projection. However, scattering states are not available above the double ionization threshold as the boundary conditions for double ionization~\cite{Brauner1989} cannot be easily enforced due to the infinite set of constraints they entail (as opposed to the finite number of constraints of single ionization problems). In addition, the scattering states become exceedingly expensive to calculate as the double ionization threshold is approached from below due to the presence of many double Rydberg series.
Finally, the Berkeley-ECS method is applicable for both single and double ionization, and also works when resonances and other long-lived states are involved. Its main drawback is the computational complexity: for each desired final energy, a large linear system representing the exterior-complex-scaled Hamiltonian acting on the final wavepacket has to be iteratively solved (\autoref{eq:ECS_driven_eq}). This approach becomes computationally expensive when the system has large spatial extent and when the wavepacket to be analyzed covers a wide range of energies. With currently available supercomputers, this becomes impractical for linear systems with a dimension of more than a few million. For typical simulations in helium this limits the box sizes to the range of $\sim\!r_{max}=250\,$a.u. and thus to simulations where only relatively short pulses are used.

\section{Applications}\label{sec:results}

We present in this section three benchmark applications for the extraction of scattering information from the wavepacket formation and propagation generated by the interaction of helium with an attosecond XUV pulse absorbing either one photon
\begin{equation}\label{eq:SPSI}
 \mathrm{He} + \gamma \longrightarrow \Hep + \mathrm{e}^-
\end{equation}
or two photons
\begin{equation}\label{eq:TPDI}
 \mathrm{He} + \gamma + \gamma \longrightarrow \Hepp + \mathrm{e}^- + \mathrm{e}^- \, .
\end{equation}
Helium is the simplest system that features autoionizing states, excited-threshold openings, and a double ionization continuum. Hence the differences between the various methods to extract asymptotic observables from an ionization wavepacket are particularly transparent in this case.
Moreover, a comparison with experimental data and other calculations is possible in some cases.
We have carried out two separate simulations, one starting from the ground state $1${$^1$S}~$(1s^2)$ and one starting from the $2${$^1$S}~$(1s2s)$ metastable state (lifetime 19.7~ms~\cite{Drake1969}) of the atom. A short ($T_{\scriptstyle\mathrm{FWHM}}$=$200$~as), moderately intense ($I_{peak}=10^{12}$~W/cm$^2$) Gaussian XUV pulse with carrier frequency $\omega=2.4$\,a.u.~and $\omega=1.65$~a.u., respectively, was employed.
This Fourier-limited broadband excitation pulse gives access to a large number of doubly excited {$^1$P$^o$} resonances.
Moreover, for two-photon absorption, the {$^1$S} and {$^1$D} double ionization continuum is accessed.
It also allows for monitoring of time-resolved photoexcitation and ionization near Fano resonances in a pump-probe setting.

\subsection{Photoionization spectrum below the double ionization threshold}

The broad spectral width of the attosecond pulse 
\begin{equation}\label{eq:Fourier}
 F_\mathrm{XUV}(t) = \Int_0^\infty \dd \omega \left( \tilde F_\mathrm{XUV}(\omega) e^{-i\omega t} + \mathrm{c.c.} \right)
\end{equation}
covers many resonances in the single ionization continuum. For low pulse intensities,
where depletion and multi-photon processes can be neglected, it is therefore 
possible to extract the partial photoionization cross sections $\sigma_\alpha$ 
and the dipole anisotropy parameters $\beta_\alpha$ directly from the partial 
differential emission probabilities $P_{\alpha E}$ and amplitudes $c_{\alpha E{\ell_E}}$,
\begin{equation}\label{eq:cs}
  \sigma_\alpha(\omega) = \frac{\omega P_{\alpha E} }{ j(\omega) },
\end{equation}
with
\begin{equation}\label{eq:def_prob}
P_{\alpha E}=\sum_{\ell_E}|c_{\alpha E{\ell_E}}|^2\, ,
\end{equation}
and $E=E_i+\omega$. The index $\alpha$ here characterizes only the remaining quantum numbers of the ionic state while the angular momentum of the continuum electron is now explicitly denoted by $\ell_E$. 
In \autoref{eq:cs} $j(\omega)$ is the current density of photons in the pulse with energy 
$\omega$, 
$j(\omega) = |\tilde{F}_{XUV}(\omega)|^2\,c$.
The emission probability $P_{\alpha E}$ is related to the transition matrix element usually employed in the calculation of $\sigma_\alpha$ through first-order perturbation theory,
\begin{multline}\label{eq:P}
 P_{\alpha E}(\omega) = \abs{\Int_{-\infty}^{\infty} \dt F_\mathrm{XUV}(t) e^{i(E-\epsilon_i)t} \braOket{\varphi_{E \alpha}}{z}{\varphi_i}}^2 \\
 = 4\pi^2 \abs{ \tilde F_\mathrm{XUV}(\omega = E-\epsilon_i)}^2 \abs{\braOket{\varphi_{E \alpha}}{z}{\varphi_i}}^2\, .
\end{multline}
The anisotropy parameters can be expressed by
\begin{eqnarray}\label{eq:betaS}
\beta_{\alpha}(\omega)&=&\sqrt{6}\sum_{\ell_E\ell'_E}\Pi_{\ell_E\ell'_E} C_{\ell_E 0,\ell'_E 0}^{2\,0}(-1)^{\ell_E}
\Sixj{\ell_E}{\ell'_E}{2}{1}{1}{\ell}\,\times\nonumber\\
&\times&\frac{c_{\alpha E\ell_E}^{\phantom{*}} c_{\alpha E\ell'_E}^*}{P_{\alpha E}}\, ,
\end{eqnarray}
where $\Pi_{ab}=\sqrt{(2a+1)(2b+1)}$, $C_{\ell_1m_1,\ell_2m_2}^{LM}$ is a Clebsch-Gordan coefficient, and the curly brackets denote a Wigner 6j-symbol.
Equation \ref{eq:betaS} applies to initial states with zero angular momentum
and absorption of a single linearly polarized photon.
In the present context, the attosecond duration of the XUV pulse permits 
to compute $\sigma_\alpha(\omega)$ and $\beta_\alpha(\omega)$ in an energy 
range spanning tens of electronvolts.
The three methods described in Sec.~\ref{sec:methods} differ in the extraction methods for the amplitudes $c_{\alpha E \ell_E}$ (see  \autoref{eq:c_alpha} for projection onto Coulomb states, \autoref{eq:scatcoef} for the PSS method, and \autoref{eq:ECS_surf_int} for the Berkely-ECS method). 

\begin{figure}[tb]
\begin{center}
\includegraphics[width=\linewidth]{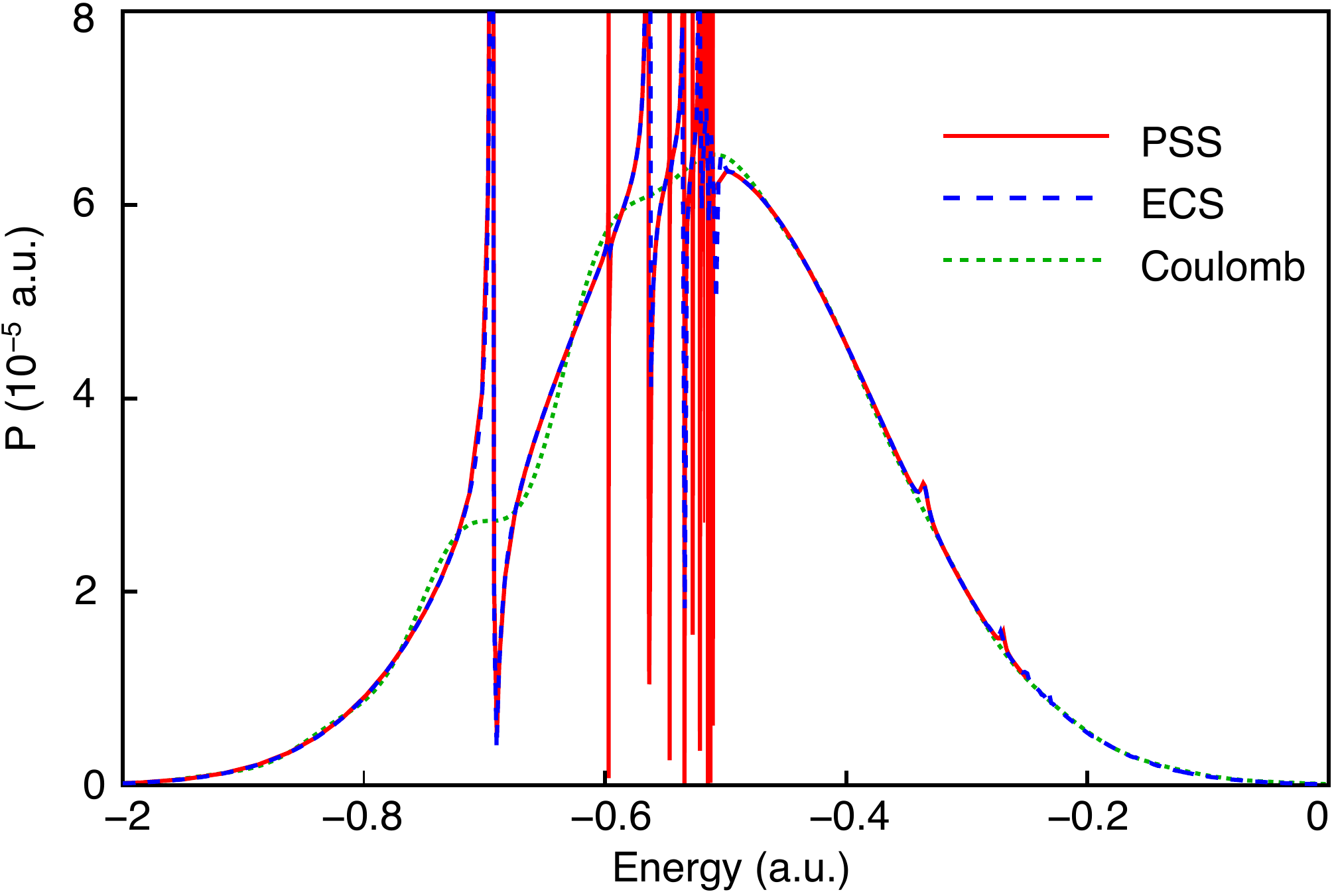}
\caption{\label{fig:1s2ECS-SCAT-COU_N1} Photoelectron distribution in the $1s$ {$^1$P$^o$} channel resulting from the photoionization of helium from the ground state with an XUV pulse with $\omega=2.4$~a.u., $I=10^{12}$W/cm$^2$, and $T_{\scriptstyle\mathrm{FWHM}}$=200~as. Projection onto scattering states (PSS), red solid line; Berkeley-ECS, blue dashed line; projection on the product of Coulomb functions at $t=1590$~as after the center of the XUV pulse, green dotted line.}
\end{center}
\end{figure}

\begin{figure}[tb]
\begin{center}
\includegraphics[width=\linewidth]{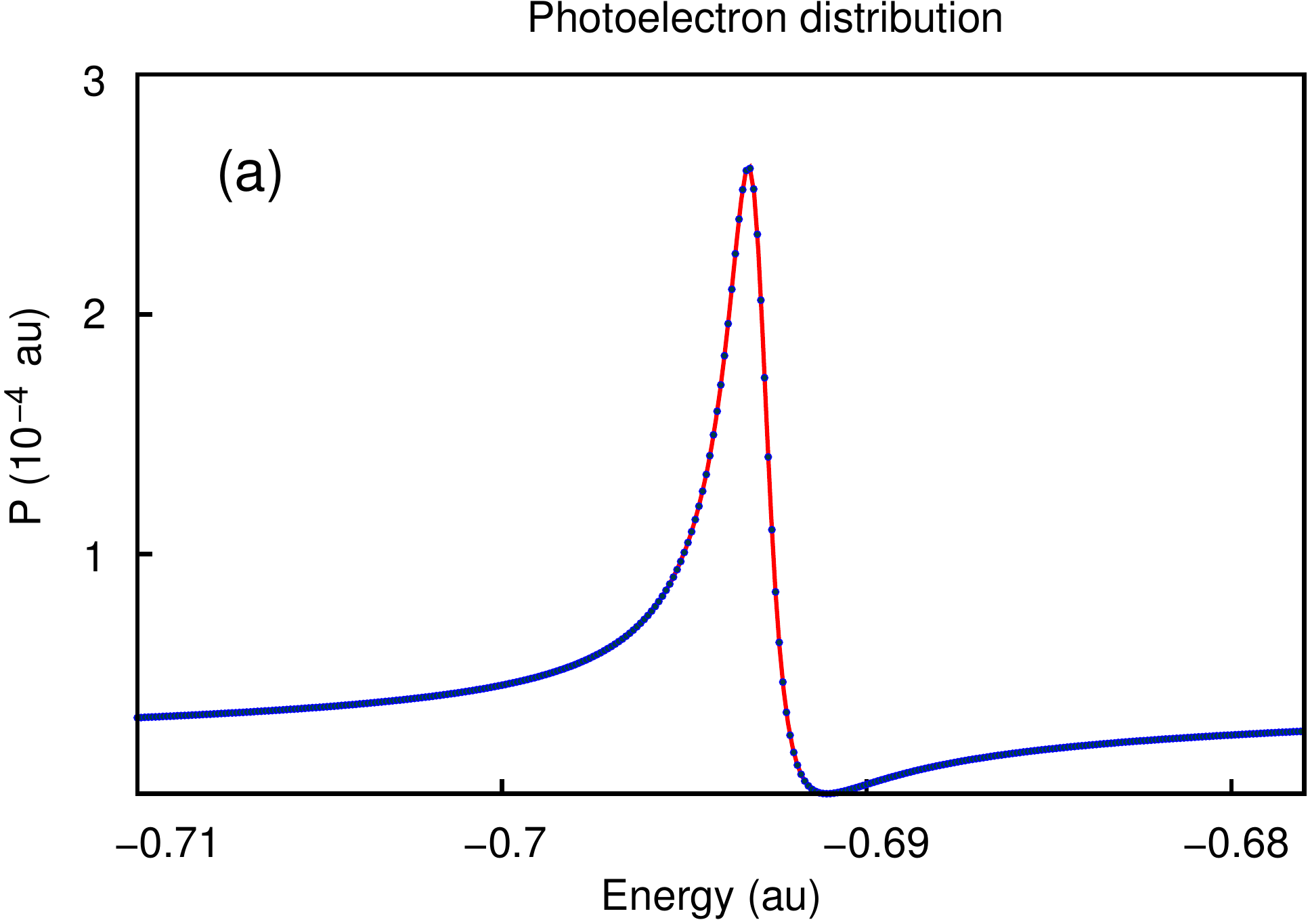} \\
\includegraphics[width=\linewidth]{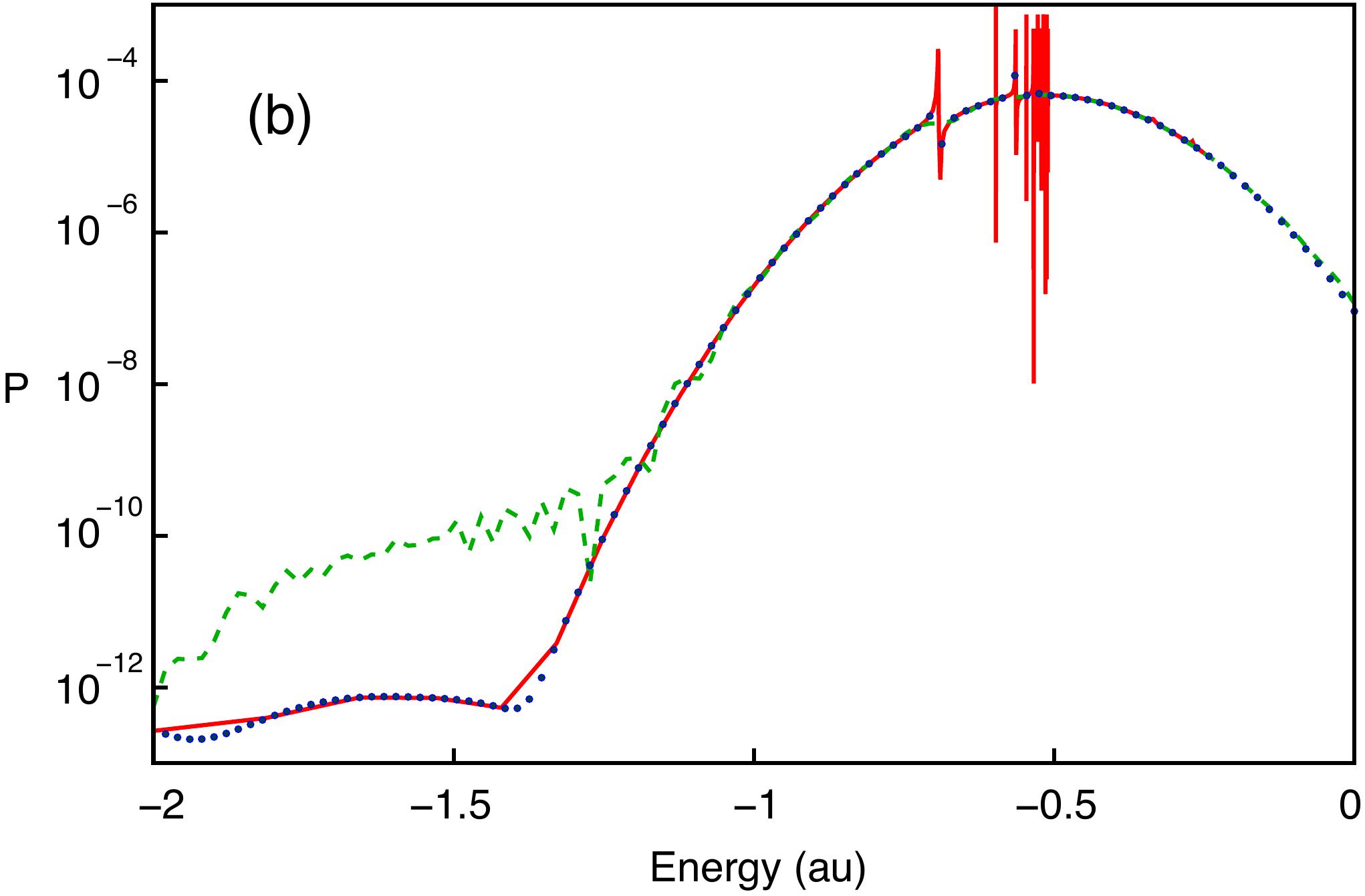}
\caption{\label{fig:1s2ECS-SCAT-COU_N1_details} Photoelectron spectrum as in \autoref{fig:1s2ECS-SCAT-COU_N1} with (a) a close-up near the $sp_2^+$ resonance, and (b) the spectrum on a logarithmic scale to highlight small deviations in the tails of the direct ionization peak.}
\end{center}
\end{figure}

\begin{figure}[tb]
\begin{center}
\includegraphics[width=\linewidth]{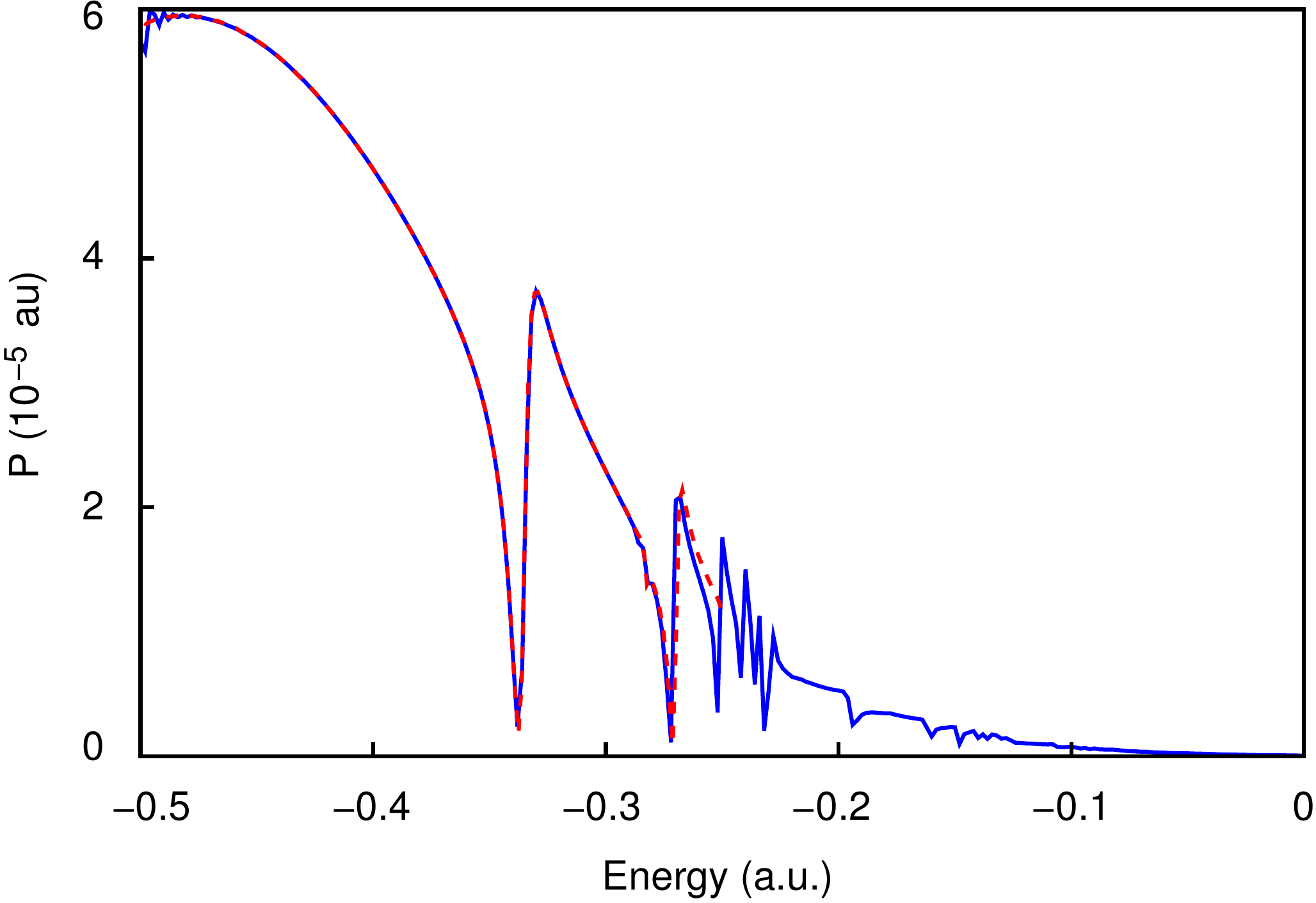}
\caption{\label{fig:1s2ECS-SCAT_N2} Photoelectron distribution in the $N=2$ {$^1$P$^o$} channel resulting from the photoionization of helium from the ground state with an XUV pulse with $\omega=2.4$\,a.u., $I=10^{12}$W/cm$^2$, and $T_{\scriptstyle\mathrm{FWHM}}$=200~as. PSS: red dashed line; ECS: blue solid line.}
\end{center}
\end{figure}

\begin{figure}[tb]
\begin{center}
\includegraphics[width=\linewidth]{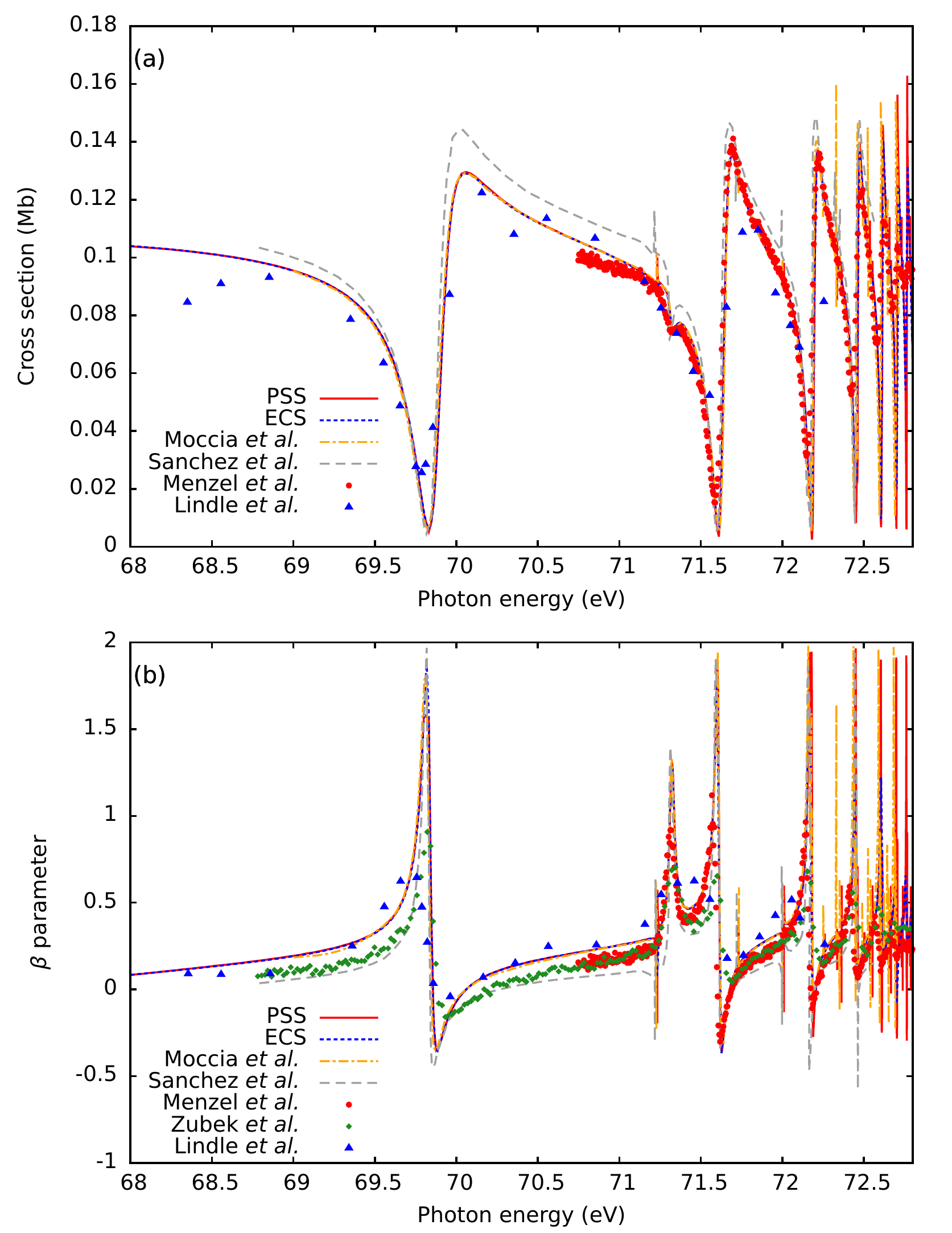}
\caption{\label{fig:1s1s_cross_beta_comp} 
Photoionization cross section $\sigma_2(\omega)$ (\autoref{eq:cs}) and anisotropy parameter $\beta_2$ (\autoref{eq:betaS}) for ionization of helium from the ground state to N=2 excited ionic states, in the region of autoionizing resonances between the N=2 and the N=3 thresholds. 
We compare our results obtained by the Berkeley-ECS and PSS methods with the theoretical values in velocity gauge by Moccia and Spizzo \cite{Moccia1991}, earlier theoretical values by Sanchez \etal, \cite{SanMar1991} and \cite{SanMar1992} and experimental results by Menzel~\emph{et al} \cite{MenFriWhi1996}, Zubek~\emph{et al} \cite{ZubDawHal1991} and Lindle~\emph{et al} \cite{LinFerBec1985}. Our results agree perfectly with the theoretical values of Moccia \etal (and with results by Venuti~\emph{et al} \cite{VenDecLis1996}, not shown) and are in best agreement with the experimental data of Lindle~\emph{et al}.
}
\end{center}
\end{figure}

A resulting typical photoelectron spectrum as extracted by the three different methods outlined in Sec.~\ref{sec:methods} is shown in \autoref{fig:1s2ECS-SCAT-COU_N1}.
While the projection onto asymptotic channel functions can reproduce the smooth background spectrum associated with the direct ionization component very well, it fails to reproduce the sharp structures associated with the autoionizing resonances (see \autoref{fig:si_regimes}). The spectra calculated with the PSS and with the Berkeley-ECS method feature accurately a large number of resonant profiles and are to within the graphical resolution in excellent agreement with each other. 
In the following discussion of this subsection, we focus now on the latter two methods.
We emphasize that the smaller number of resonances appearing in the Berkeley-ECS method is not due to any fundamental limitation of the method but because of the use of a coarser energy grid. Additional insights can be gained from a close-up (\autoref{fig:1s2ECS-SCAT-COU_N1_details}a) and a logarithmic presentation of the photoionization probability (\autoref{fig:1s2ECS-SCAT-COU_N1_details}b).
\autoref{fig:1s2ECS-SCAT-COU_N1_details} highlights two aspects of the performance of the three methods. First, the projection on scattering states and the Berkeley-ECS method are in excellent agreement close to the resonance (\autoref{fig:1s2ECS-SCAT-COU_N1_details}a) and down to the smallest probability densities (\autoref{fig:1s2ECS-SCAT-COU_N1_details}b). The deviation of the background profile of the spectrum from a parabola on a logarithmic scale at low energies is due to the fact that the Gaussian envelope of the XUV pulse is eventually truncated. Second, the spectrum obtained through the projection on Coulomb functions clearly deviates from the background below $-1$~a.u.~, however, only when $P(E)$ is already small ($\lesssim 10^{-4}$ of the direct ionization peak).
This error due to the contamination by doubly excited states is small in the present case since they provide only a minor admixture to the wavepacket. When they have a higher relative weight, however, their spurious effect on the spectrum can be larger.
The convergence of the DES spectrum as a function of the size of the close-coupling expansion within the PSS is illustrated in \autoref{fig:1s2ECS-SCAT_N2}. Here, the PSS employs a minimal close-coupling expansion involving the open channels ($1s\epsilon_p$, $2s\epsilon_p$, $2p\epsilon_s$, and $2p\epsilon_d$) only. Clearly, since the $N=3$ channels (those corresponding to the He$^+$ parent ion in the $3s$, $3p$, and $3d$ states) are not included, the higher members of the autoionizing Rydberg series converging to the N=3 threshold are not reproduced. As a consequence, the spectrum obtained with the PSS deviates from the one obtained with the Berkeley-ECS method at energies higher than -0.3 a.u.~. On the other hand, the spectrum below the $N=3$ threshold is already well converged, i.e.~the influence of the closed channels are adequately accounted for. This observation highlights the salient feature of the close-coupling expansion with pseudostates, namely the possibility to drastically truncate the representation while still obtaining an accurate representation of the continuum states 
in a given 
single-ionization energy region.
Having established the convergence, accuracy, and excellent agreement of the PSS and the Berkeley-ECS methods for single ionization spectra below the double ionization (\autoref{fig:si_regimes}), we turn now to a comparison with available experimental data and calculations for the $N=2$ photoionization cross section $\sigma_2(\omega)$ (\autoref{eq:cs}) and anisotropy parameter $\beta_2(\omega)$ (\autoref{eq:betaS}).
For photoexcitation of the He($1s^2$) ground state, several experimental datasets for $\sigma_2(\omega)$ and $\beta_2(\omega)$ measured with synchrotron radiation are available (\autoref{fig:1s1s_cross_beta_comp}).
Moreover, this process has served over the years to benchmark theoretical descriptions of photoemission in a strongly correlated system (see, e.g., \cite{Rost1997,TanRicRos2000,DomXuePus1991,BizWuiDhe1982} and references therein). 
For clarity, we display in \autoref{fig:1s1s_cross_beta_comp} only a small selection of theoretical datasets.
While for cross sections (\autoref{fig:1s1s_cross_beta_comp}a) several experimental datasets as well as theoretical results are in close agreement with each other, there are still unresolved discrepancies for the $\beta_2$ parameters (\autoref{fig:1s1s_cross_beta_comp}b).
Increased sensitivity of $\beta_2$ to the approximations employed is not surprising as, unlike energy distributions, angular distributions depend on the relative phases between photoionization amplitudes in different channels as well as on their absolute values.
Earlier results by Sanchez and Mart{\'i}n \cite{SanMar1992} lie closer to the experimental data by Menzel \etal \cite{MenFriWhi1996}.
By contrast, the more recent calculation by Moccia and Spizzo \cite{Moccia1991} agrees significantly better with the data by Lindle \etal \cite{LinFerBec1985}.
The present two complementary methods (PSS, Berkeley-ECS) agree, within the graphical resolution, perfectly with each other and with the calculation of Moccia and Spizzo \cite{Moccia1991} and also of Venuti \etal \cite{VenDecLis1996}.
Thus, it appears that the measurements of the $\beta_2$ parameter by Lindle \etal between the $N=2$ and $N=3$ thresholds are consistent with theory.
We note that our theory curves in \autoref{fig:1s1s_cross_beta_comp} pertain to perfect spectral resolution while the experimental spectral resolution in the experiment of Lindle \etal \cite{LinFerBec1985} was $\Delta E \approx 170\mathrm{meV}$ due to monochromator broadening.
Folding our $\beta_2(\omega)$ with the experimental resolution would further improve the agreement near sharp resonances.

\begin{figure}[tb]
\begin{center}
\includegraphics[width=\linewidth]{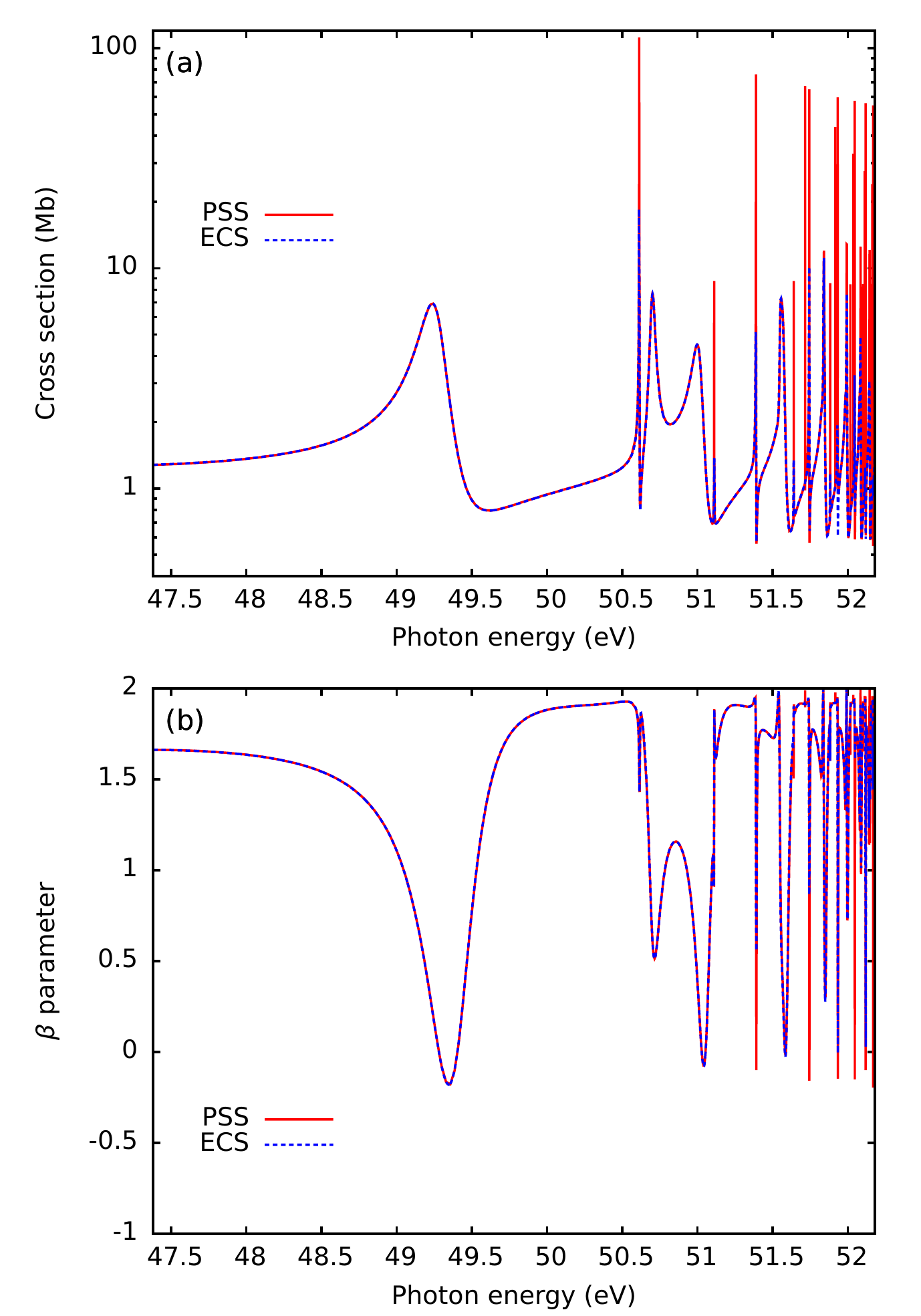}
\caption{\label{fig:1s2s_cross_beta} 
Photoionization cross section and anisotropy parameter $\beta_2$ between the N=2 and the N=3 thresholds for ionization of helium from the metastable {$^1$S}~$(1s2s)$ state. PSS: red solid line; ECS: blue dotted line.}
\end{center}
\end{figure}

\begin{figure}[tb]
\begin{center}
\includegraphics[width=\linewidth]{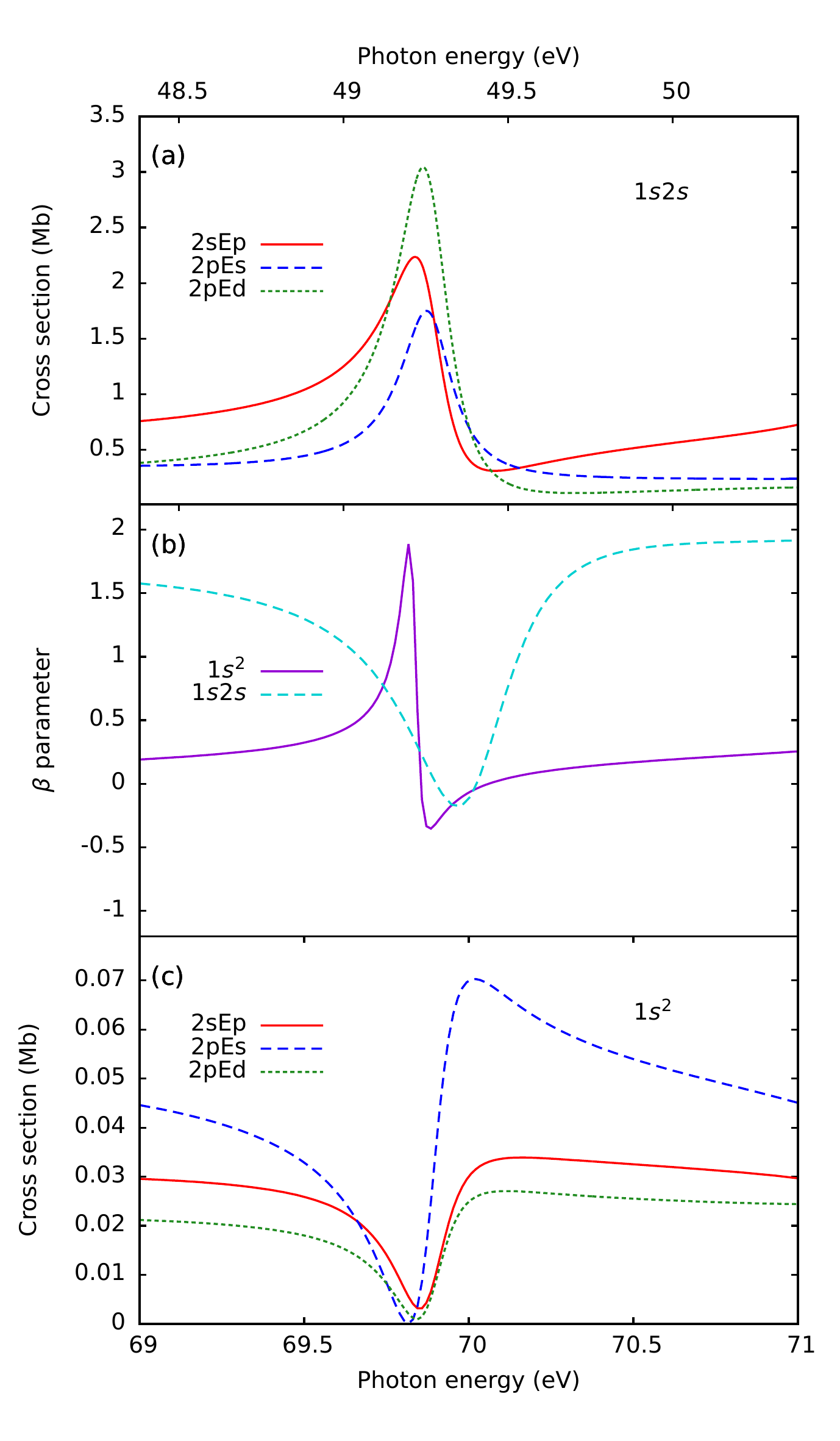}
\caption{\label{fig:sp_resonance} Partial $N\!=\!2$ photoionization cross sections for the first $^1$P$^o$ autoionizing 
resonance below the $N=3$ threshold for ionization from the metastable {$^1$S}~$(1s2s)$ state (a) and ionization from the ground state (c). In (b) the according $\beta_2$ parameters for $n\!=\!2$ are compared, where the energies are shifted such that the resonance appears at the same position. The lower energy axis corresponds to ionization from the ground state, while the upper energy axis corresponds to ionization from the metastable {$^1$S}~$(1s2s)$ state.}
\end{center}
\end{figure}

Unlike excitation from the ground state, the excitation of doubly excited 
resonances starting from the metastable {$^1$S}~$(1s2s)$ state is still an experimental challenge 
due to the difficulty of producing a sample with the
required optical thickness. Theoretical treatments of this process have also been 
scarce~\cite{Sanchez1993,FanCha2000,ZhoLin1994}. First experimental results for some 
doubly excited states below the N=2 threshold only became available recently~\cite{Alagia2009}.
Our present simulation of the excitation by an XUV pulse with central frequency $\omega=1.65$\,a.u. appears to be the first that provides information on the $\beta_2$ parameter (\autoref{fig:1s2s_cross_beta}).
The results from the PSS and Berkeley-ECS method agree to within the graphical resolution.
It is now of interest to compare the excitation spectrum and $\beta_2$ parameters for the same final energies when accessed from different initial states.
This allows to probe propensity rules for radiative and non-radiative transitions between strongly correlated excited states \cite{Rost1997}.
As a prototypical example we focus in the following on the spectrum in the proximity of the first $^1$P$^o$ autoionizing 
resonance below the $N=3$ threshold, i.e., reached for photon energies between 68~eV and 71~eV starting from the ground state, and for photon energies between 47.5~eV and 50.5~eV starting
from the metastable He($1s2s$) state (\autoref{fig:sp_resonance}).

For resonances of doubly excited states several equivalent classification schemes signifying the departure from the independent particle model are in use \cite{TanRicRos2000,Lin1986,Lin84,HerKelPol1980,FeaBri1986,HerSin1975}.
We employ in the following the so-called parabolic classification scheme within which 
each resonance of a given 
symmetry $^{2S+1}L^\pi$ is uniquely identified by the notation $[N_1N_2m]^A_n$, where $N_1$, 
$N_2$ and $m$ are the Stark quantum numbers in parabolic coordinates for a hydrogenic ion in 
an external uniform electric field, $A=\pm 1$ indicates whether the wave function has an 
anti-node $(+)$ or a node $(-)$ when the two electrons are at the same distance from the 
nucleus, $r_1=r_2$, and $n$ is the principal quantum number of the outer electron. 
The corresponding continuum scattering channels above the $N=N_1+N_2+m+1$ threshold are labeled by $[N_1N_2m]^A$ (\ie with the $n$ index dropped).
The parabolic classification scheme is particularly well suited to formulate propensity rules, \ie dominant channels in branching ratios.
The resonance investigated in \autoref{fig:sp_resonance} carries the parabolic label $[011]^+_3$. 
According to autoionization propensity rules, the $[011]^+_n$ series autoionizes to the 
$[001]^+$ continuum through the efficient configuration interaction coupling ($[011]^+_n \leftrightarrow [001]^+$) characterized by  $\Delta N_2=-1$. Indeed, the lifetime
of the $[011]^+_3$ resonance is short, $\tau=3.44$~fs, highlighting the efficient coupling.
The autoionizing branching ratios for decay of this resonance are $b_{1s\epsilon_p}=0.019$ (corresponding to the $[000]^-$ channel in the parabolic classification), and
$b_{2s\epsilon_p}=0.141$, $b_{2p\epsilon_d}=0.308$, and $b_{2p\epsilon_s}=0.532$ \cite{Moccia1991}, 
the latter corresponding to the $[001]^+$ channel.
Photoexcitation of the $[011]^+_3$ resonance from the symmetric ground state  (\ie with two equivalent electrons) follows similar propensity rules for radiative transitions \cite{Rost1997} which, in this case, predict that the transition is very weak.
Clearly, the dominant excitation channel $1s \epsilon p$ corresponding to $[000]^-$, present already at the independent particle level, does not directly couple to autoionizing resonances.
The subdominant channel $[001]^+$ couples to the resonance $[011]^+$ via configuration interaction, as discussed above.
Consequently, all partial cross sections $\sigma_\alpha$ ($\alpha\!=\!2\ell, E \ell_E$) should resemble the Fano model for 
an isolated resonance with energy $E_r$ and width $\Gamma_r$ embedded in a single-channel continuum~\cite{Fano1961}, namely,
\begin{equation}\label{eq:FanoModel}
\sigma_{\alpha}(E) = \sigma_{bg}(E) b_{\alpha}\frac{(\epsilon + q)^2}{\epsilon^2+1},\quad \epsilon \equiv 2(E-E_r)/\Gamma_r
\end{equation}
where $\sigma_{bg}(E)$ is a smooth background total cross sections and $q$ is the
Fano asymmetry parameter. The quantity $\pi q^2/2$ expresses the ratio between the probability to excite the resonance 
and that of exciting the continuum in an energy interval equal to the resonance width. Therefore, in this case, we expect
a very small asymmetry parameter $q$ giving rise to a typical window resonance shape.
In particular, \emph{all the partial cross sections should vanish, or 
almost vanish, at the same energy close to the resonance position} (\autoref{fig:sp_resonance}c).
In turn, the $\beta_2$ parameter is expected to closely follow that for the $[001]^+$ channel, $\beta_2 \approx 0$, far from the resonance as, indeed, observed (\autoref{fig:sp_resonance}b).
Close to the resonance, however, a sharp peak in $\beta_2$ is observed.
The origin of this modulation of $\beta_2$ follows directly from the variation of the partial cross section (\autoref{fig:sp_resonance}c).
Even though all $N=2$ partial cross sections approach values near zero at around $E=69.8$~eV, 
the $2s\epsilon_p$ cross section misses the zero by a small yet significantly larger
amount than the other two channels. As a result, close to the resonance, the relative proportion
of the channels changes abruptly. At the minimum of the $N=2$ cross section, the $2s\epsilon_p$ 
channel ($\beta_2=2$) dominates and the $\beta_2$ parameter has a sharp maximum close to the 
theoretical limit $2$. As soon as the Fano minimum is passed, the asymmetry parameter drops 
back to zero again.
Therefore, the sharp modulation in the $\beta_2$ parameter for $N=2$ for excitation from the ground state 
is a dramatic magnification of the \emph{slight} misalignment between the $N=2$ decay channel 
of the $[011]^+_3$ resonance and the $N=2$ dipole excitation channel
giving rise to a local break-down of the propensity rules.
A significantly different scenario applies to photoexcitation from the ($1s2s$) state.
Since the initial state features inequivalent electrons, the initial state qualifies as non-symmetric within the framework of propensity rules.
Therefore, the propensity rule
derived by the saddle-point approximation~\cite{Rost1997} no longer applies.
Yet, even simpler rules apply here: both the $1s2s\to 1s\epsilon_p$ and $1s2s\to 2s\epsilon_p$
transitions are allowed already at the level of the independent particle approximation resulting in a 50 fold increased cross section (see Figs.~\ref{fig:sp_resonance}ac).
As a consequence, the background $\beta_2$ parameter is expected to be very close to $2$ (\autoref{fig:sp_resonance}b).
Moreover, the direct transition to the $[011]^+$ channel is allowed resulting in a remarkably large $q$ parameter $\simeq -4.6$ \cite{Sanchez1993} and an almost Lorentzian resonance profile (\autoref{fig:sp_resonance}a).
Consequently, near the resonance the $\beta_2$ value drops locally to values associated with the $[001]^+$ channel, \ie $\beta_2 \approx 0$ (\autoref{fig:1s2s_cross_beta}b).
Experimental verifications of these predictions would provide sensitive tests for the applicability of propensity rules and would be of considerable interest.

\subsection{Time-resolved autoionization resonances in helium}

With the availability of attosecond XUV pulses with durations small compared to the lifetime of the resonances, the time-evolution of the excitation and decay of an autoionizing resonance can be monitored in real time.
The two-path interference between the direct ionization (\eg the $[001]^+$ channel) and the indirect ionization via quasi-bound states (the $[011]^+_3$ states in the example of the previous subsection) gives rise to non-stationary coherent dynamics in the continuum.
Quantitative features of the temporal evolution have been predicted for a generic Fano resonance model \cite{Wickenhauser2005,WicBurKra2005,Wic2006}.
With the present accurate wavepacket propagation and extraction protocol, it will be possible 
for the first time to monitor with unprecedented resolution the evolution of single-ionization 
resonant profiles for helium in a pump-probe setting employing ultrashort and intense light pulses. 
Here, the doubly excited states are populated in the initial (pumping) step and the double 
ionization continuum is accessed in the final (probing) step.

Monitoring the time evolution of the continuum portion of the wavepacket generated by the XUV attosecond pulse (see \autoref{eq:coefficient_convergence3})
\begin{equation} \label{eq:coefficient_convergence3_t}
P_{\alpha}(E,t)=|c_{\alpha E}(t)|^2
\end{equation}
necessarily requires projection at finite times, \ie in the non-asymptotic region where the asymptotic channel Hamiltonian does not yet apply.
Therefore, the observables associated to the time-resolved quantity $P_{\alpha}(E,t)$ must be obtained through an additional interrogation step akin to a pump-probe setting.
In the previous analysis of time-resolved Fano resonances, attosecond streaking employing an additional few-cycle IR pulse was proposed for the interrogation (probing) step \cite{Wickenhauser2005,WicBurKra2005,Wic2006}.
Below, we propose an alternative XUV-XUV pump-probe sequence (see \autoref{eq:TPDI}) bearing resemblance to the attosecond transient absorption (ATA) protocol \cite{Goulielmakis2010}.

\begin{figure}[hbtp!]
\centering
\includegraphics[width=\linewidth]{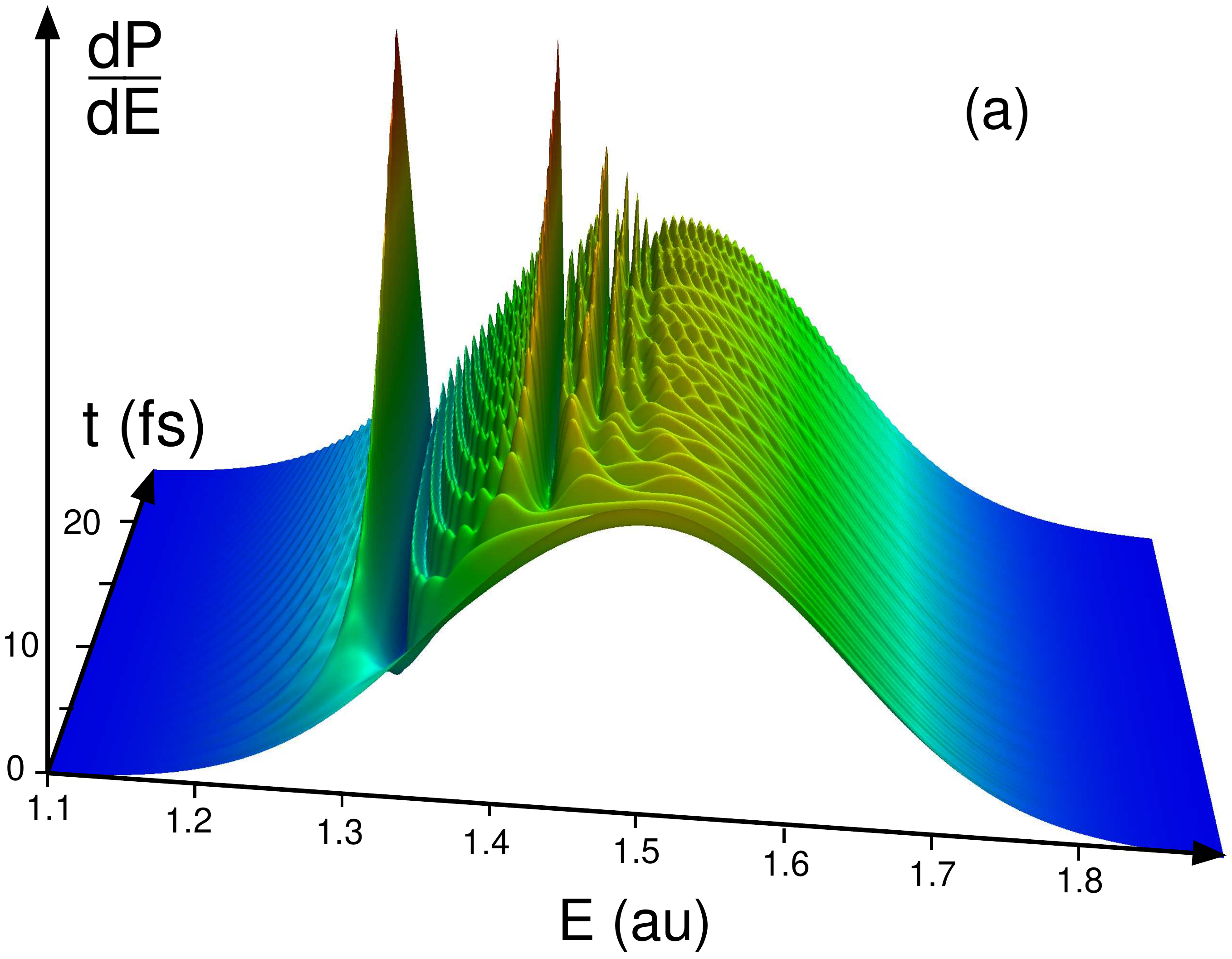} \\
\includegraphics[width=\linewidth]{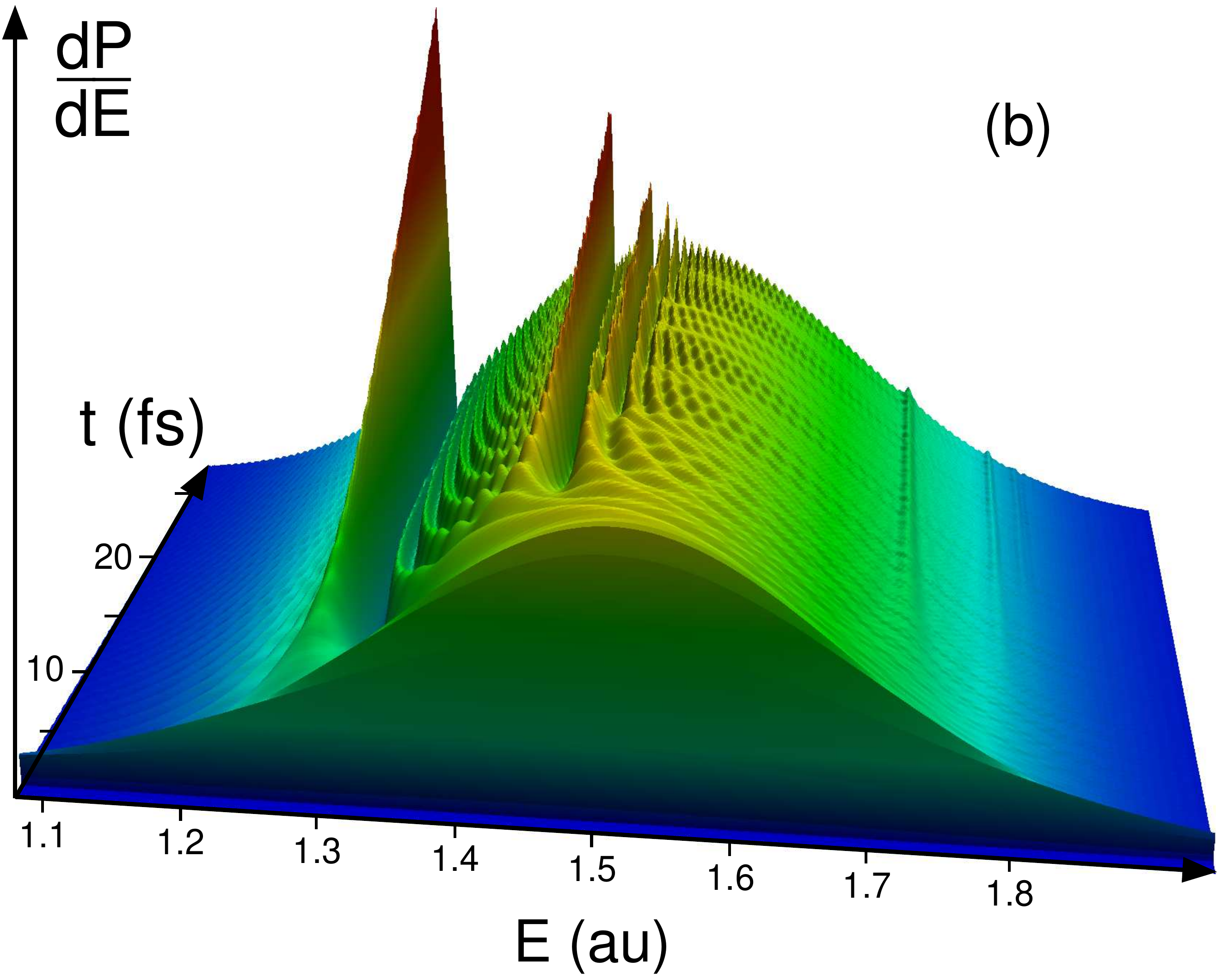}
\caption{\label{fig:td-Fano}
a) Formation of the Fano profiles for a regular series of autoionizing Rydberg states 
according to the Fano model within the impulsive approximation. The energy distribution
of the photoelectron is multiplied by a Gaussian spectral-shape function to simulate 
the effect of an attosecond pulse. This approach is justified as long as the duration 
of the pulse is much shorter than the lifetimes of the autoionizing states.
See text for more details. \\
b) Total photoelectron spectrum of the wavepacket created by the action of a sub-femtosecond pulse on the ground state of the helium atom and computed by projecting the wavepacket on products of bound (Z=2) and continuum (Z=1) Coulomb functions. Since such products are not eigenstates of the field-free Hamiltonian, the spectrum is only approximated and changes with time after the XUV pulse. As the doubly excited states populated by the pulse decay, characteristic Fano profiles build up. See text for more details.
}
\end{figure}

\begin{figure}[tb]
\begin{center}
\includegraphics[width=\linewidth]{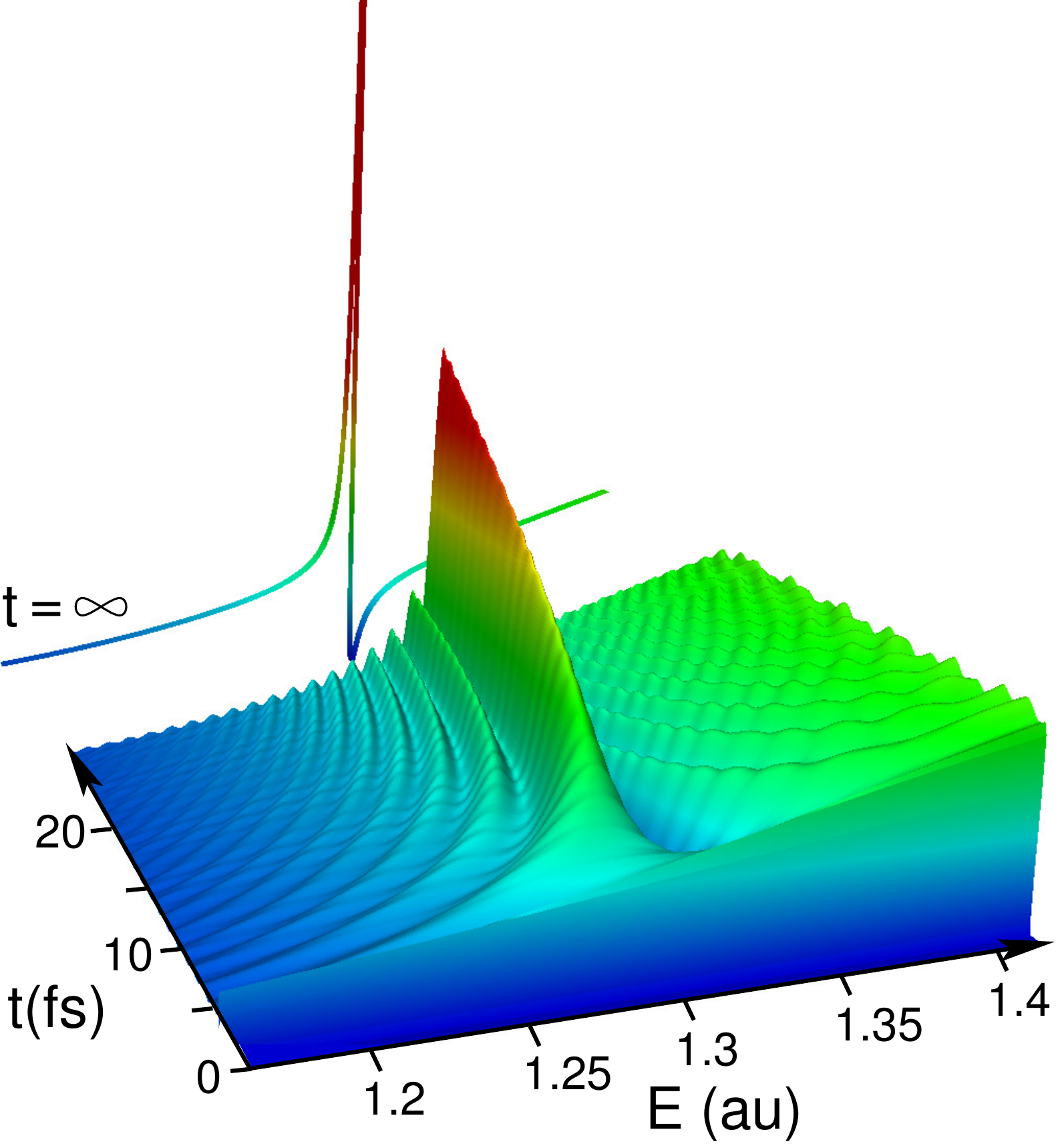}
\caption{\label{fig:CoulProj_vs_time_vs_ScatProj} Close-up of the evolution of the photoelectron spectrum determined by projection on products of Coulomb functions in comparison with the asymptotic Fano profile extracted from the same wavepacket by projection onto scattering states immediately after the end of the external pulse.}
\end{center}
\end{figure}

\begin{figure*}[tb]
\begin{center}
\includegraphics[width=\linewidth]{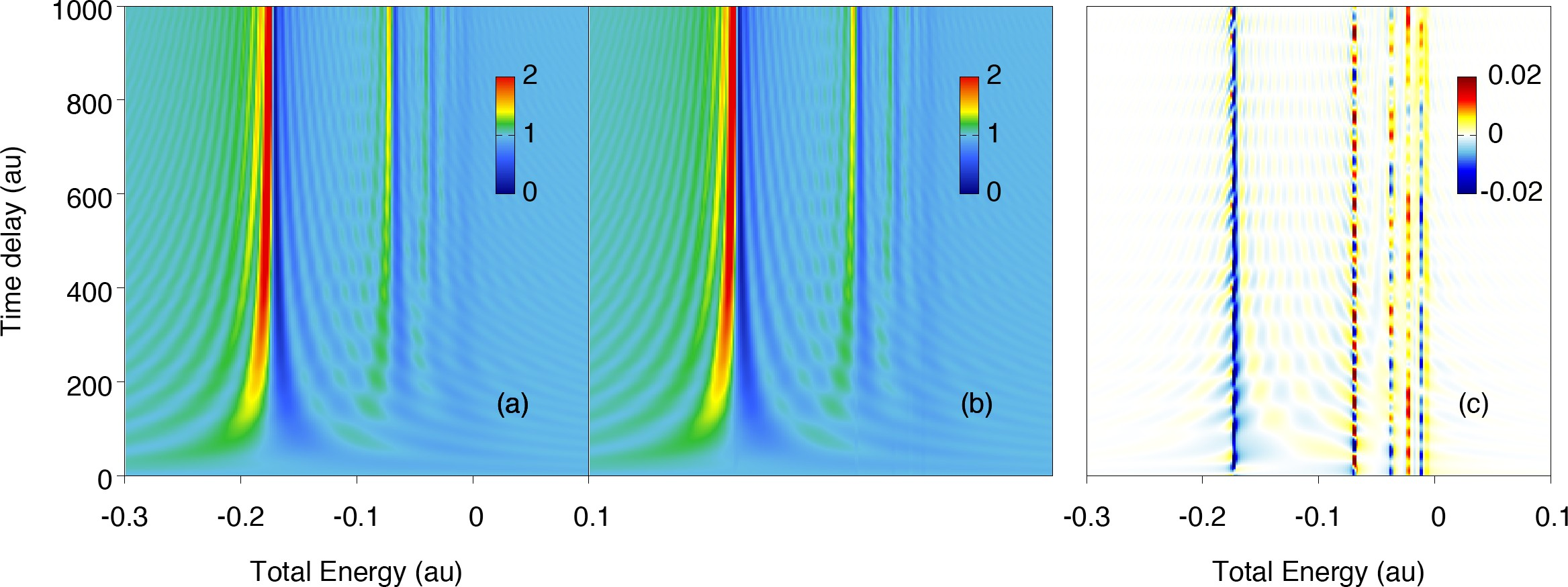}
\caption{\label{fig:ResonanceDecayModels} a) Time evolution of the photoelectron distribution $P(E,t)$
in the unperturbed continuum as a function of time after the impulsive excitation of a 
series of autoionizing resonances. b) Asymptotic photoelectron distribution $\bar P(E,\tau)$ as a function of
the time delay $\tau$ between the impulsive excitation and the impulsive depletion of the residual
localized part of the autoionizing resonances. 
c) shows the comparatively small differences that 
exist between the two spectra.}
\end{center}
\end{figure*}

It is now instructive to analyze \autoref{eq:coefficient_convergence3_t} for the wavepacket simulated by projecting at finite times onto the channel eigenstates (see \autoref{sec:proj_channel}), either onto symmetrized
products of a bound $(Z=2)$ and a continuum $(Z=1)$ Coulomb function (\autoref{fig:td-Fano}), or 
alternatively, projecting onto partial-wave channel functions  with a frozen core (see \autoref{sec:proj_channel}). 
Both lead to an almost identical quantum beat pattern lending credence to the physical significance of the ensuing interference fringes.
The build-up and decay of the dominant series $sp_n^+$ of $^1$P$^o$ autoionizing resonances corresponding to the series $[001]^+$ in the parabolic classification scheme is clearly visible (\autoref{fig:td-Fano}a).
This time-dependent spectrum $P(E,t)$ extracted from the ab-initio simulation can be directly compared with the analytic prediction for the time evolution of generic Fano resonances.
Wickenhauser \etal derived a closed expression $P(E,t)$ for an isolated resonance in the impulsive (\ie broad-band excitation) limit \cite{Wickenhauser2005,Wic2006},
\begin{equation}\label{eq:FanoImpulsiveModelSingleResonance}
P(E,t)\propto P_{E 0}
\left|1+(i-q)\frac{\Gamma}{2}\frac{e^{-i (\tilde{E}_a-E) t}-1}{E-\tilde{E}_a}\right|^2\, ,
\end{equation}
where $P_{E 0}$ is the dipole transition strength between the initial ground state and the unperturbed continuum, $q$ is the Fano asymmetry 
parameter~\cite{Fano1961}, $\Gamma$ is the resonance width and $\tilde{E}_a=\tilde{E}_a^{\Re e}-i\Gamma/2$ 
is the complex energy of the resonance.
This expression can be easily extended to the case of many isolated resonances on top of a smooth 
background
\begin{equation}\label{eq:FanoImpulsiveModelManyResonances}
P(E,t)\propto P_{E 0}
\left|
1+\sum_{j}(i-q_j)\frac{\Gamma_j}{2}\frac{e^{-i (\tilde{E}_j-E) t}-1}{E-\tilde{E}_j}
\right|^2.
\end{equation}
We apply \autoref{eq:FanoImpulsiveModelManyResonances} to the first nine terms ($n=2,3,\cdots,10$) of the $sp^+$ $^1$P$^o$ series.
The positions $E_n$, widths $\Gamma_n$, and asymmetry parameters $q_n$ of the $n$th term are
approximated by
$E_n=E_{N=2}-1/2(n-\mu)^2$, $\Gamma_n=\bar{\Gamma}/(n-\mu)^3$, and $q_n=q$, where
$\mu$, $\bar{\Gamma}$, and $q$ are the quantum defect, the reduced width and the asymmetry
parameter of the series, thereby extrapolating  the approximate values $\mu=0.3$, $\bar{\Gamma}=0.007$, 
$q=-2.6$ taken from accurate calculations for the low members of the series available in the literature~\cite{Rost1997}.
The spectrum is multiplied by a Gaussian envelope 
$\exp(-\frac{1}{2}((E-E_0)/\sigma)^2)$ with $\sigma=0.127$ to reproduce the Fourier-width of an attosecond 
pulse. This correction to the impulsive limit is justified as long as the duration of the pulse is much smaller than the
lifetime of all resonances involved.
The resulting agreement between the analytic model (\autoref{eq:FanoImpulsiveModelManyResonances}, \autoref{fig:td-Fano}a) and the ab-initio wavepacket simulation (\autoref{fig:td-Fano}b) is remarkable:
all features of the interference fringes associated with different resonances are qualitatively and, to a good degree of approximation, even quantitatively reproduced.
The details of the temporal interference fringes present in the wavepacket simulation are highlighted in the close-up of the spectrum near the $sp^+_2$ {$^1$P$^o$} (or $2s2p$ or $[001]^+_2$) resonance (\autoref{fig:CoulProj_vs_time_vs_ScatProj}).
For small times, 
the photoelectron spectrum rapidly builds-up from zero to a smooth Gaussian profile mirroring the temporal evolution of the attosecond pulse. At this stage, only the ``direct ionization'' component is visible since the duration of the pulse is much shorter
than the lifetime of the resonance.
With increasing time, hyperbolic-shaped interference fringes in the $E\!-\!t$ plane converging towards the resonance appear on both flanks.
They have been first observed for the generic time-dependent Fano-resonance model \cite{Wickenhauser2005,WicBurKra2005,Wic2006} but appear in the ab-initio simulation as well.
These quantum beats follow directly from \autoref{eq:FanoImpulsiveModelSingleResonance}.
The ridges (valleys) are given by the condition $(E_a-E)t=n\pi$.
Appearance of this quantum beat structures in the continuum is not limited to Fano resonances but is universal whenever the continuum is accessed both directly and via a (quasi) bound state.
More recently, similar structures were found in attosecond XUV-IR pump-probe electron interferometry for continuum electrons just above the first ionization threshold \cite{MauRemSwo2010}.
Finally, for $t \to \infty$ the stationary Fano-resonance profile emerges.
The point to be noted is that this asymptotic profile shown in \autoref{fig:CoulProj_vs_time_vs_ScatProj} is determined by the PSS method by projecting the wavepacket onto exact scattering states 
$\psi^-_{1s E}$ (\autoref{eq:Pascat}) right after the conclusion of the pulse, \ie during the early stages of the evolution depicted in \autoref{fig:CoulProj_vs_time_vs_ScatProj}.
The scattering states implicitly account for the time evolution of the full wavepacket to infinity and no free propagation beyond the end of the pulse is required. In turn, quantum beats observable at finite time are not visible in the PSS method as it projects into the asymptotic future. 
The experimental observation of the temporal interference pattern requires an interrogation of the system at finite times, \ie before the asymptotic  scattering regime is reached, \eg by IR streaking \cite{Wickenhauser2005,WicBurKra2005,Wic2006}. 
As an alternative to streaking 
we propose here to employ a time-delayed XUV pulse that probes the quasi-bound rather than the continuum component of the Fano resonance. 
In such an XUV-XUV pump-probe setting the population of the localized component is suddenly depleted by the second attosecond pulse.
In close analogy to the analytic model for projective interrogation of the continuum (\autoref{eq:FanoImpulsiveModelSingleResonance}), we find for the asymptotic energy distribution of the continuum electron as a function of the delay time $\tau$ at which the quasi-bound state is impulsively removed
\begin{equation}\label{eq:FanoImpulsiveModelAndDelayedDepletion}
\bar P(E,\tau)\propto
|P_{E0}|^2
\left|
+(i-q)\frac{\Gamma}{2}\frac{e^{-i(\tilde{E}_a-E)\tau}-e^{2i\varphi_q}}{E-\tilde{E}_a^*}
\right|^2,
\end{equation}
where $\varphi_q=\arctan(q)$.
This protocol resembles the attosecond transient absorption employed to time-resolve core-level dynamics \cite{Goulielmakis2010}.
In fact, Pfeifer and co-workers~\cite{Ott2012} recently implemented the attosecond transient
absorption analogue of this experimental scheme and, indeed, interference fringes similar to 
those described here were observed and theoretically confirmed by one of us with full \emph{ab-initio}
simulations~\cite{Argenti2012}.
It is now of interest to quantitatively compare the two interrogation protocols (\autoref{eq:FanoImpulsiveModelSingleResonance} and \autoref{eq:FanoImpulsiveModelAndDelayedDepletion}) displayed in \autoref{fig:ResonanceDecayModels}.
Projection onto the continuum portion (\autoref{fig:ResonanceDecayModels}a) and the complementary operation of projecting out the bound portion (\autoref{fig:ResonanceDecayModels}b) yields nearly identical quantum beat patterns for a series of autoionizing resonances. The difference between the two (\autoref{fig:ResonanceDecayModels}c) is remarkably small.
This result suggests the direct experimental observability of the temporal interference fringes by the XUV-XUV pump-probe scheme.

\subsection{Two-photon double ionization}

\begin{figure}[htb]
\begin{center}
\includegraphics[width=\linewidth]{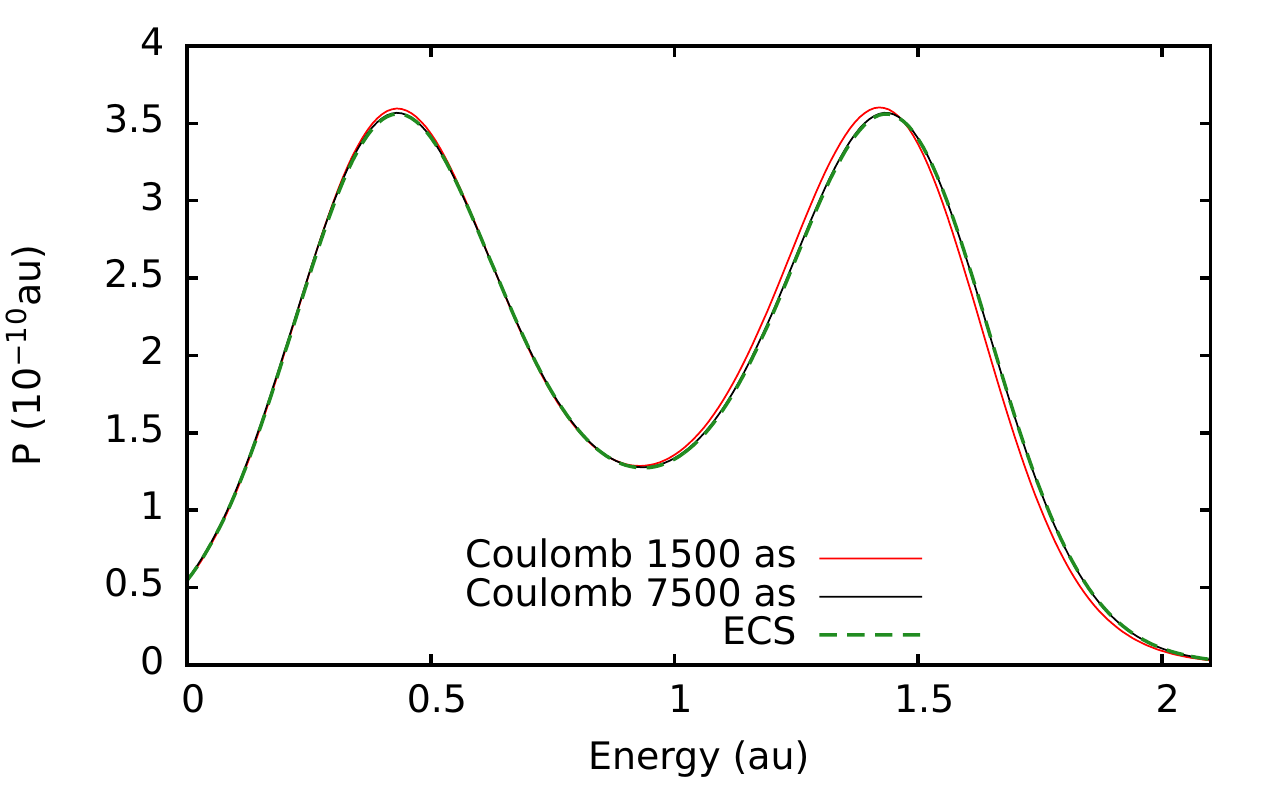}
\caption{\label{fig:DI_ECS_Coulomb_spec} Singly differential photoelectron distribution after two-photon double ionization
 of helium by a sub-femtosecond XUV pulse ($\omega=2.4$\,a.u., $I=10^{12}$W/cm$^2$, and $T_{\scriptstyle\mathrm{FWHM}}$=200\,as). 
 Two different extraction methods are compared: projection on products of Coulomb functions with $Z=2$ performed $1500$as (red solid line) and $7500$as (black solid line) after the peak of the XUV pulse and the Berkeley-ECS method (green dashed line) for a computational box with $R=244$\,a.u.\ performed $1500$as after the peak of the XUV pulse. 
}
\end{center}
\end{figure}

We turn now to the process of probing the two-electron continuum portion of the wavepacket (\autoref{eq:TPDI}). 
We analyze the
correlated wavepacket for the doubly ionized part of the spectrum caused by two-photon absorption of the previously used Gaussian XUV pulse with $T_{\scriptstyle\mathrm{FWHM}}$=$200$\,as and a carrier frequency $\omega=2.4$\,a.u.~from the helium ground state. 
In this regime, the PSS is not applicable because of the lack of accurate scattering states for the double continuum.
However, the Berkeley-ECS method (see \autoref{sec:ECS}) is able to impose the correct boundary conditions and obtain the asymptotic spectral information of a doubly ionized wavepacket directly after the completion of the laser pulse. 
The projection of the wavepacket on an uncorrelated, symmetrized product of two Coulomb functions with $Z=2$ is straightforward (see \autoref{sec:proj_channel}).
However, it requires the propagation of the wavepacket to large distances in order to control and minimize the error due to the neglect of the electron-electron interaction.
In practice this yields accurate results as long as the box and angular momentum basis are large enough to correctly represent the wave function at the time of projection. 
We first compare the singly differential photoelectron spectrum with the prominent two peaks near the expected positions for the sequential two-photon double ionization process
corresponding to the emission of the ``first'' electron (\autoref{eq:SPSI}) with a kinetic energy of $E_1=\omega - I_{p,1} \approx 1.5\au$ ($I_p$: ionization potential) and the subsequent emission of the second electron with $E_2=\omega - I_{p,2} \approx 0.4\au$, \autoref{fig:DI_ECS_Coulomb_spec}.
Note that the separation between the two maxima is smaller than predicted by the sequential limit due to the energy exchange between the two electrons enforced by strong temporal correlation between the two emission events.
In the present case, it is not the carrier frequency $\omega > 2\au$ which lies above the threshold for non-sequential ionization but the short pulse duration that controls the degree of non-sequentiality of the emission process.
The projection onto Coulomb waves $7500$\,as after the peak of the pulse shows excellent agreement with the spectrum obtained by the Berkeley-ECS method using the wavepacket immediately after the conclusion of the pulse (\autoref{fig:DI_ECS_Coulomb_spec}). 
In fact, virtually the same level of agreement is already reached for considerably smaller propagation times before the projection (not shown). Even projection directly after the conclusion of the field ($1500$\,as after the peak of the XUV pulse, red line in \autoref{fig:DI_ECS_Coulomb_spec}) gives very similar results.

\begin{figure}[htb]
\begin{center}
\includegraphics[width=\linewidth]{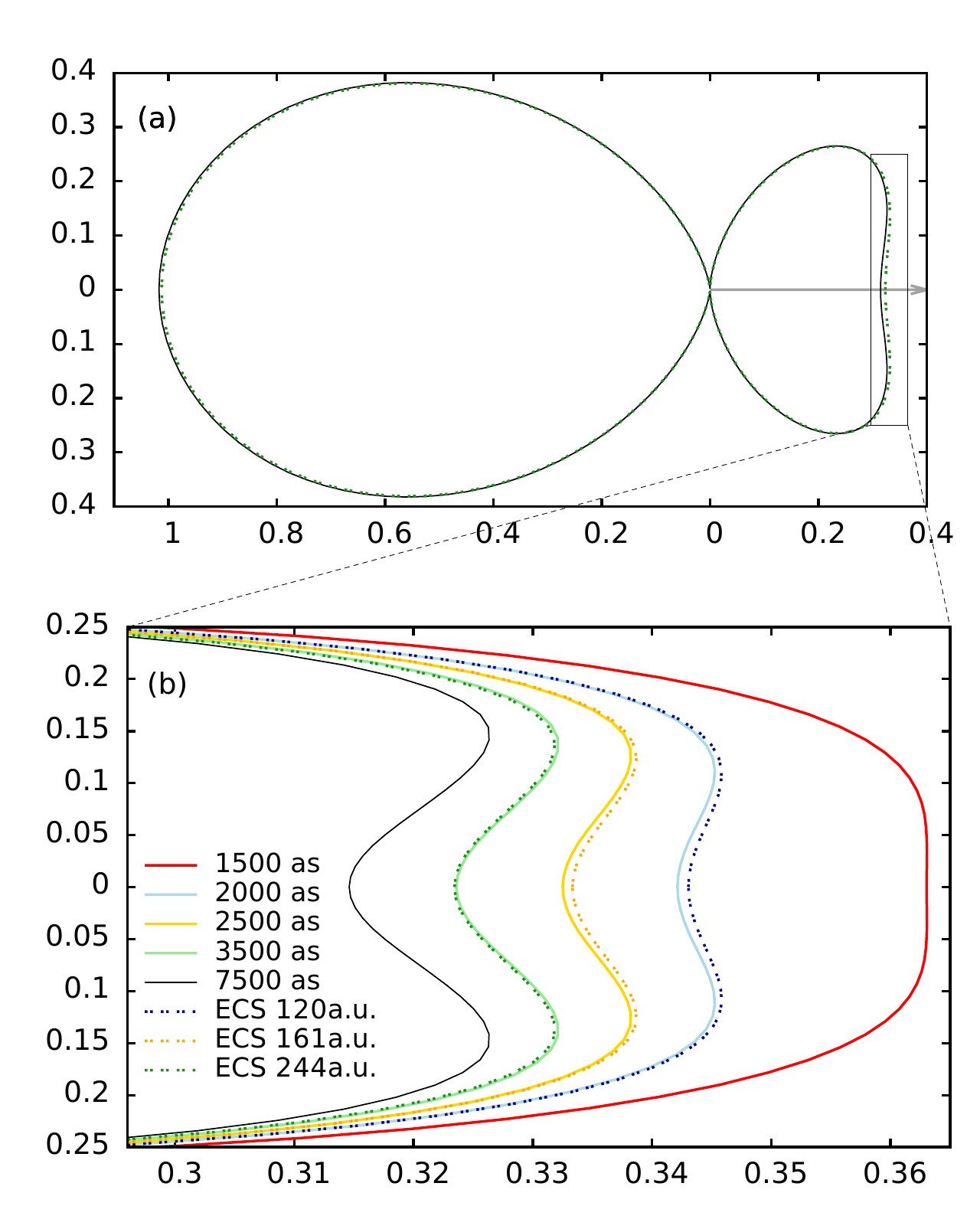}
\caption{\label{fig:DI_ECS_Coulomb_angdis} Conditional angular distribution for one electron emitted in the laser polarization axis [grey arrow in (a)] and integrated over both electron energies after two-photon double ionization
 of helium by a sub-femtosecond XUV pulse ($\omega=2.4$\,a.u., $I=10^{12}$W/cm$^2$, and $T_{\scriptstyle\mathrm{FWHM}}$=200\,as). 
 Two different extraction methods are compared (a): projection on products of Coulomb functions with $Z=2$ performed $7500$\,as (solid black line) after the peak of the XUV pulse and the Berkeley-ECS method (dashed green line) for an extraction radius of $R=244$\,a.u.\ performed $1500$\,as after the peak of the XUV pulse. In (b) a close-up of the emission of the two electrons in the same direction is shown, illustrating the convergence of the Berkeley-ECS method with extraction radius (dashed lines) and of the Coulomb projection with propagation time (solid lines). The two techniques agree when the extraction radius of the Berkeley-ECS method approximately equals the position (of the relevant parts) of the wavepacket at the time of projection on the Coulomb functions. The partial wave expansion includes angular momenta up to $l_1=l_2=15$.
}
\end{center}
\end{figure}

The fact that the pulse duration has a profound effect on the two-photon two-electron emission process that goes beyond a Fourier broadening \cite{LauBac2003,IshMid2005} becomes more apparent when one studies the angular correlation between the photofragments \cite{FeiNagPaz2009,Palacios2009,PalHorRes2010,PazFeiNag2011}.
For example the energy-integrated conditional angular emission probability for one electron when the other electron is ejected along the laser polarization axis (\autoref{fig:DI_ECS_Coulomb_angdis}a)
displays pronounced deviations from a simple dipolar pattern expected for sequential double ionization. Such structures have been previously observed both by projecting on Coulomb waves~\cite{FeiNagPaz2009} and with the Berkeley-ECS method~\cite{Palacios2009}.
We find excellent agreement between the two methods within the graphical resolution of \autoref{fig:DI_ECS_Coulomb_angdis}a.
The remaining small residual differences between the two methods can be understood by inspecting their respective convergence behavior:
the projection onto Coulomb functions becomes more accurate when the electrons have spread further apart before the spectral analysis is performed, provided the partial wave expansion of the wave function covers enough angular momenta to accurately describe the  motion of the free electrons. For the present case convergence is reached for propagation times of about $7500$\,as after the peak of the XUV pulse (see the solid lines in \autoref{fig:DI_ECS_Coulomb_angdis}b).
For the Berkeley-ECS method the extension of the computational box for the transformation to the spectral domain (i.e.~the solution of \autoref{eq:ECS_driven_eq}) and for the surface integral (cf.~\autoref{eq:ECS_surf_int}) influences the quality of the results. Thus, for accurate angular distributions comparably large radial boxes are also required for the Berkeley-ECS method (see the dashed lines in \autoref{fig:DI_ECS_Coulomb_angdis}b).
The convergence behavior of the two methods is linked: their results for the angular distributions agree when the extraction radius of the Berkeley-ECS method roughly equals the position (of the relevant parts) of the wavepacket at the time of projection on the Coulomb functions (compare solid and dashed lines in \autoref{fig:DI_ECS_Coulomb_angdis}b).
Thus, for larger ECS boxes even the small differences to the (converged) Coulomb projection in \autoref{fig:DI_ECS_Coulomb_angdis}a would vanish.
For both methods the convergence for the angular distributions is considerably slower than for the energy spectra, especially where both electrons are emitted in the same direction (\autoref{fig:DI_ECS_Coulomb_angdis}b).
The latter highlights the fact that those regions in phase space where electron-electron correlation is strongest still pose a major challenge for highly accurate ab-initio simulations.
Since this region of mutual repulsion contributes, however, little to total emission probabilities, the overall accuracy remains largely unaffected.

\section{Conclusions\label{sec:conclusions}}

We have analyzed and applied three different methods to extract the continuum component of a correlated multi-electron wave function in helium obtained from an ab-initio solution of the time-dependent Schr\"odinger equation. Each of the three methods investigated, the projection onto asymptotic channel functions, \ie Coulomb continuum states, the projection onto exact scattering states (PSS), and the Berkeley-ECS method feature both advantages and disadvantages depending on the energy range of interest, the required box size, and whether single or double ionization distributions are desired. Below the double ionization threshold, methods to extract photoelectron spectra that are based on the projection on approximated continuum channels such as Coulomb continuum functions work well when no resonances are significantly populated, but become inapplicable in spectral regions in the vicinity of narrow resonances. 
In this regime, remarkably good agreement between the PSS and the Berkeley-ECS method is observed. This finding suggests an efficient way of monitoring the composition of a complex wavepacket $\psi$ in the single-ionization channels below the double ionization threshold. Since the scattering states can be computed separately, once and for all, in an adapted basis, the extraction of the expansion coefficients of $\psi$ requires only the calculation of a simple scalar product. 
Above the double ionization threshold, the PSS becomes inapplicable while the Berkeley-ECS method agrees well with projection onto Coulomb functions provided the computational box is sufficiently large and the wavepacket is propagated into the asymptotic region.
The three applications presented explored different aspects of attosecond-pulse-driven wavepacket dynamics.
Exploiting the broad spectral width we presented converged calculations for the photoionization cross section and anisotropy parameter below the double ionization threshold including first results for spectra and angular distributions for photoionization of metastable He($1s2s$, {$^1$S}).
Comparison to the corresponding ground state process provides insights into the applicability of the propensity rules for excitation and decay of doubly excited resonances.
Exploiting the short duration of the pulse we have provided first ab-initio results for time-resolved autoionizing resonances and have suggested a new protocol for observing the ensuing quantum beats complementing attosecond streaking.
Furthermore, we have verified that two-electron wavepackets above the double ionization threshold acquire features of non-sequential emission for ultrashort pulses notwithstanding the fact that the mean photon frequency lies well above the threshold where sequential ionization should prevail.

\begin{acknowledgments}
The authors thank Dr.~Alicia Palacios for fruitful discussions. Special thanks are due to Prof. Mariusz Zubek and Prof. Alexander Menzel for providing us with original experimental data. This work was supported by the G\"oran Gustafsson Fundation, the Swedish science research council (VR), the European COST Action CM0702, and the FWF-Austria, Grant No.~P21141-N16,~P23359-N16, and the SFB ViCoM.  The computational results presented have also been achieved in part using the Vienna Scientific Cluster (VSC). LA acknowledges support from the European Research Council under the European Union's Seventh Framework Programme (FP7/2007-2013)/ERC grant agreement n$^\circ$~290853 and the MICINN project n$^\circ$~FIS2010-15127. JF acknowledges support from the NSF through a grant to ITAMP and by the European Research Council under Grant No. 290981 (PLASMONANOQUANTA). RP acknowledges support by the TU Vienna Doctoral Program Functional Matter. ML acknowledges funding by the Vienna Science and Technology Fund (WWTF) through project No.~MA09-030.
\end{acknowledgments}

\appendix

\section{Implementation of the PSS for helium}\label{sec:appendix_PSS}

Accurate scattering states (\autoref{eq:lsG0}) and corresponding projection amplitudes (\autoref{eq:scatcoef}) can be calculated for energies below the double ionization threshold.
In the present case,
the single-ionization scattering states of helium below the double ionization threshold of the atom are computed with the B-spline K-matrix method. B-splines are a convenient tool to accurately represent the radial component of continuum atomic orbitals on a finite interval~\cite{Bachau2001,Argenti2009}, while the K-matrix method~\cite{Cacelli1991} is an $L^2$ realization of configuration interaction in the continuum along the lines of Fano's pioneering paper~\cite{Fano1961}. The K-matrix method has been successfully applied for the single-photoionization spectrum of several atomic and molecular systems~\cite{Moccia1991,FanCha2000,Argenti2006,Argenti2008b,Argenti2010b}. We will provide here only a brief description of its implementation for the case of helium (details can be found elsewhere \cite{Argenti2006,Lindroth2012}).

A complete set $\psi_{\alpha E}^\mathcal{P}$ of stationary eigenfunctions of the field-free Hamiltonian $H_a$ at a given energy $E$ in the single-ionization continuum are sought in the form of a linear combination of partial-wave channel functions $\phi_{\alpha E}$ (PWC's), plus an additional component from a localized (or pseudostate) channel (LC),
\begin{equation}
\psi_{\alpha E}^{\mathcal{P}}=\phi_{\alpha E}+\sum_\gamma\sumint d\epsilon\, 
\phi_{\gamma \epsilon}\frac{\mathcal{P}}{E-\epsilon}
\mathbf{K}_{\gamma\epsilon,\alpha E} \,,
\label{eq:tm1}
\end{equation}
where the index $\alpha$ runs over the channels which are open at energy $E$, while the index $\gamma$ runs over all open and closed channels, including the localized one.

The PWC $\alpha$ is defined by coupling and antisymmetrizing a bound state of the He$^+$ parent ion with quantum numbers $N_\alpha$ and $L_\alpha$ and energy $E_\alpha$, to an electron state with angular momentum $\ell_\alpha$, the radial degree of freedom of which is otherwise unconstrained, to give a state with definite values for the total spin $S$ and angular momentum $L$.
\begin{equation}\label{eq:PWC_def}
\phi_{\alpha E}\,=\,
\hat{\mathcal{A}}\,{\Theta_{S\Sigma}}\,
\mathcal{Y}_{L_{\alpha}\ell_\alpha}^{LM}(\Omega_1,\Omega_2)\,
R_{N_\alpha L_\alpha}(r_1)\,\frac{f_{\alpha E}(r_2)}{r_2},
\end{equation}
where $\hat{\mathcal{A}}$ is the antisymmetrizer, $\Theta_{S\Sigma}$ is a two-electron spin function, $R_{N_\alpha L_\alpha}$ is the radial part of the frozen He$^+$ parent ion state, and $f_{\alpha E}$ the continuum radial function. 
Asymptotically, the $f_{\alpha E}$ are a linear combination of the regular and irregular Coulomb functions with angular momentum $\ell_\alpha$, energy $E$, and a phase shift $\delta_{\alpha E}$, determined by the short range behavior of the differential equation for $f_{\alpha E}$, which differs from that of the hydrogenic functions. This results from the deviation of the frozen-core potential from that of a pure Coulomb potential. 

The PWCs in~(\autoref{eq:tm1}) do not exhaust the state space associated with single ionization, because the set of bound states of the parent ion is not complete, and because the close-coupling expansion~(\autoref{eq:tm1}) is truncated. Nevertheless, if all single- and double-ionization closed channels were to be included in the close-coupling expansion, their contribution would decay exponentially at large radii. Therefore, instead of using a complete basis, it is sufficient to include in~(\autoref{eq:tm1}) a pseudo-state channel LC that comprises a sufficiently large number of normalized two-electron functions built from localized orbitals, to attain good accuracy. 

Equation (\ref{eq:tm1}) may be solved for the unknown coefficient matrix $\mathbf{K}$ by requiring $\psi_{\alpha E}^\mathcal{P}$ to be an eigenfunction of the complete projected Hamiltonian with eigenvalue $E$,
\begin{equation}
\langle\phi_{\beta E'}|\,E-H_a\,|
\psi_{\alpha E}^{\mathcal{P}}\rangle=0\quad\forall\,\beta,E'.\label{eq:phiHpsi}
\end{equation}
This condition leads to a system of integral equations for $\mathbf{K}$ which can be discretized and solved with standard linear algebra routines. The scattering states with definite spherical symmetry  $\psi_{\alpha E}^-$ are then computed as
\begin{equation}
\psi_{\alpha E}^-=\sum_\beta\psi_{\beta E}^{\mathcal{P}}\,\left[\frac{1}{\mathbf{1}-
    i\pi\mathbf{K}(E)}\right]_{\beta\alpha}e^{- i(\sigma_{\ell_\alpha}+\delta_\alpha-\ell_\alpha\pi/2)},
\end{equation}
where $\mathbf{K}_{\alpha\beta}(E)\equiv\mathbf{K}_{\alpha E, \beta E}$ is the on-shell reactance matrix (\S $7.2.3$ in \cite{Newton}) while $\sigma_{\ell_\alpha}$ and $\delta_\alpha$ are the Coulomb and channel phase shifts, respectively. 
Finally, the scattering states which correspond to Coulomb plane waves associated with a parent ion in a given state A, are given by
\begin{equation}
\psi_{A,E\Omega\sigma}^-=\sum_{\alpha}^{^{L_\alpha=L_A}_{N_\alpha=N_A}}
C_{L_A M_A,\ell
  m}^{LM}C_{\frac{1}{2}\Sigma_A,\frac{1}{2}\sigma}^{S\Sigma}
Y_{\ell m}^*(\Omega)\,\,\psi_{\alpha E}^{-},
\end{equation}
where $L_A$, $M_A$, and $\Sigma_A$ indicate the angular momentum and spin of the parent ion, $\Omega$ and $\sigma$ indicate the asymptotic photoelectron's direction and spin, and $C_{a\alpha,\,b\beta}^{c\gamma}$ are Clebsch-Gordan coefficients. The states $\psi_{A,E\Omega\sigma}^-$ are normalized according to
\begin{equation}
\langle \psi_{A,E\Omega\sigma}^- |
\psi_{B,E'\Omega'\sigma'}^-\rangle=\delta_{AB}\delta_{\sigma\sigma'}\delta(E-E')
\delta(\Omega-\Omega').
\end{equation}

\section{Conversion of B-spline close-coupling functions to the FEDVR basis}\label{sec:appendix_conversion}

One technical key feature of the present method of projecting onto accurate scattering states (PSS) is that the calculation of scattering states is independent of the actual simulation and can be optimized separately.
Yet, scattering states have so far been employed only in those cases where they are built in the same basis which is used to carry out the time-dependent simulations. We demonstrate here that the scattering states computed in an optimized B-spline close-coupling basis with the K-matrix method~\cite{Argenti2006} can be accurately converted to a finite-element discrete-variable (FEDVR) basis~\cite{Feist2008} optimized for time propagation.
Table~\ref{tab:boundenergy} compares the energies of $^1$S$^e$, $^1$P$^o$, and $^1$D$^e$ helium Rydberg states with principal quantum number $n$ for the outer electron up to $n\!=\!6$, obtained by diagonalizing the full configuration interaction  Hamiltonian of helium built in either the FEDVR or the B-spline basis. The FEDVR basis comprised eleven functions per finite element. The width of the first element was 2\,a.u.~and increased linearly to $4.0$\,a.u.~within the first $5$ finite elements. The grid extensions were $28$\,a.u.~for one radial coordinate
and $156$\,a.u.~for the other. The B-spline basis comprised spline functions of order $k=8$~\cite{Bachau2001} defined on a non-uniform grid of nodes, optimized at small radii to optimize the representation of the ground state of the helium atom, and with an asymptotic spacing between consecutive nodes of $0.5$\,a.u.~up to a maximum radius of 800\,a.u.\,. For both the FEDVR and the B-spline basis, the maximum orbital angular momentum $\ell_{max}=4$ (which suffices for the present comparison between methods) was used.
\begin{table}[tbh]
\caption{\label{tab:boundenergy} Comparison between the energies of the He Rydberg states with principal quantum number for the outer electron up to $n=6$, obtained by diagonalizing the Hamiltonian in either the FEDVR (upper value, $R=156$\,a.u.) or the B-spline (lower value, $R$=800\,a.u.) basis.}
\begin{ruledtabular}
\begin{tabular}{cccc}
 &\multicolumn{3}{c}{Symmetry}\\
 n &{$^1$S$^e$} & {$^1$P$^o$} & {$^1$D$^e$}\\ 
\hline\\
 1 & -2.903 5102 &             &             \\
   & -2.903 5164 &             &             \\
 2 & -2.145 9610 & -2.123 8231 &             \\
   & -2.145 9615 & -2.123 8232 &             \\
 3 & -2.061 2684 & -2.055 1399 & -2.055 6203 \\
   & -2.061 2685 & -2.055 1400 & -2.055 6203 \\
 4 & -2.033 5852 & -2.031 0669 & -2.031 2796 \\
   & -2.033 5853 & -2.031 0669 & -2.031 2796 \\
 5 & -2.021 1761 & -2.019 9046 & -2.020 0016 \\
   & -2.021 1761 & -2.019 9045 & -2.020 0016 \\
 6 & -2.014 5627 & -2.013 8331 & -2.013 8981 \\
   & -2.014 5627 & -2.013 8331 & -2.013 8981 \\
\end{tabular}
\end{ruledtabular}
\end{table}
The very good agreement between the two approaches for the Rydberg spectrum indicates that the wave functions in the two bases are represented at comparable levels of accuracy.
To assess the accuracy with which the wave functions computed in the B-spline basis are converted to the FEDVR basis we computed the norm 
of the converted Rydberg states $\langle\tilde\phi_n|\tilde\phi_n\rangle$ as well as their overlap $\langle\phi_n|\tilde\phi_n\rangle$ with the corresponding states computed directly in the FEDVR basis. Both numbers should within the numerical accuracy be close to $1$. The errors $\tilde\delta_n=1-\langle\tilde\phi_n|\tilde\phi_n\rangle$ and $\delta_n=1-\langle\phi_n|\tilde\phi_n\rangle$ for the states 
listed in table~\ref{tab:boundenergy} are between $10^{-11}$ and $10^{-7}$ (table~\ref{tab:norm}).
\begin{table}[tbh]
\caption{\label{tab:norm} Error in the norm of the bound states translated from the B-spline to the FEDVR basis $\tilde\delta_n=1-\langle\tilde\phi_n|\tilde\phi_n\rangle$, and error in the overlap $\delta_n=1-\langle\phi_n|\tilde\phi_n\rangle$ (see text for details). The notation $[n]$ is a shorthand for $10^{-n}$.}
\begin{ruledtabular}
\begin{tabular}{ccccccc}
   &\multicolumn{2}{c}{$^1$S}&\multicolumn{2}{c}{$^1$P$^o$}&\multicolumn{2}{c}{$^1$D$^e$}\\
$n$&$\tilde\delta_n$&$\delta_n$&$\tilde\delta_n$&$\delta_n$&$\tilde\delta_n$&$\delta_n$\\
\hline\\
 1 &8.0[-7]&8.2[-7]&        &        &       &       \\
 2 &7.6[-8]&7.8[-8]&7.4[-11]&3.9[-10]&       &       \\
 3 &2.4[-8]&2.5[-8]&2.5[-9] &2.7[-9] &3.3[-9]&3.3[-9]\\
 4 &1.2[-8]&2.1[-8]&3.0[-9] &1.3[-8] &3.3[-9]&9.8[-9]\\
 5 &7.6[-9]&1.7[-8]&3.1[-9] &1.1[-8] &3.3[-9]&1.3[-8]\\
 6 &6.1[-9]&1.3[-8]&4.7[-9] &1.6[-8] &4.0[-9]&1.3[-8]\\
\end{tabular}
\end{ruledtabular}
\end{table}

This confirms the accuracy of the conversion from the B-spline to the FEDVR basis. The error is larger for the ground state than for the excited states because the first finite element is still comparatively wide and could be reduced further by choosing a smaller radial span for the first few finite elements.

\FloatBarrier

\bibliography{biblio}

\begin{thebibliography}{100}%
\makeatletter
\providecommand \@ifxundefined [1]{%
 \ifx #1\undefined \expandafter \@firstoftwo
 \else \expandafter \@secondoftwo
\fi
}%
\providecommand \@ifnum [1]{%
 \ifnum #1\expandafter \@firstoftwo
 \else \expandafter \@secondoftwo
\fi
}%
\providecommand \enquote [1]{``#1''}%
\providecommand \bibnamefont  [1]{#1}%
\providecommand \bibfnamefont [1]{#1}%
\providecommand \citenamefont [1]{#1}%
\providecommand\href[0]{\@sanitize\@href}%
\providecommand\@href[1]{\endgroup\@@startlink{#1}\endgroup\@@href}%
\providecommand\@@href[1]{#1\@@endlink}%
\providecommand \@sanitize [0]{\begingroup\catcode`\&12\catcode`\#12\relax}%
\@ifxundefined \pdfoutput {\@firstoftwo}{%
 \@ifnum{\z@=\pdfoutput}{\@firstoftwo}{\@secondoftwo}%
}{%
 \providecommand\@@startlink[1]{\leavevmode\special{html:<a href="#1">}}%
 \providecommand\@@endlink[0]{\special{html:</a>}}%
}{%
 \providecommand\@@startlink[1]{%
  \leavevmode
  \pdfstartlink
   attr{/Border[0 0 1 ]/H/I/C[0 1 1]}%
   user{/Subtype/Link/A<</Type/Action/S/URI/URI(#1)>>}%
  \relax
 }%
 \providecommand\@@endlink[0]{\pdfendlink}%
}%
\providecommand \url  [0]{\begingroup\@sanitize \@url }%
\providecommand \@url [1]{\endgroup\@href {#1}{\urlprefix}}%
\providecommand \urlprefix [0]{URL }%
\providecommand \Eprint[0]{\href }%
\@ifxundefined \urlstyle {%
  \providecommand \doi [1]{doi:\discretionary{}{}{}#1}%
}{%
  \providecommand \doi [0]{doi:\discretionary{}{}{}\begingroup
  \urlstyle{rm}\Url }%
}%
\providecommand \doibase [0]{http://dx.doi.org/}%
\providecommand \Doi[1]{\href{\doibase#1}}%
\providecommand \bibAnnote [3]{%
  \BibitemShut{#1}%
  \begin{quotation}\noindent
    \textsc{Key:}\ #2\\\textsc{Annotation:}\ #3%
  \end{quotation}%
}%
\providecommand \bibAnnoteFile [2]{%
  \IfFileExists{#2}{\bibAnnote {#1} {#2} {\input{#2}}}{}%
}%
\providecommand \typeout [0]{\immediate \write \m@ne }%
\providecommand \selectlanguage [0]{\@gobble}%
\providecommand \bibinfo [0]{\@secondoftwo}%
\providecommand \bibfield [0]{\@secondoftwo}%
\providecommand \translation [1]{[#1]}%
\providecommand \BibitemOpen[0]{}%
\providecommand \bibitemStop [0]{}%
\providecommand \bibitemNoStop [0]{.\EOS\space}%
\providecommand \EOS [0]{\spacefactor3000\relax}%
\providecommand \BibitemShut [1]{\csname bibitem#1\endcsname}%
\bibitem{Sansone2011}%
  \BibitemOpen
  \bibfield{author}{%
  \bibinfo {author} {\bibfnamefont{G.}~\bibnamefont{Sansone}}, \bibinfo
  {author} {\bibfnamefont{L.}~\bibnamefont{Poletto}},\ and\ \bibinfo {author}
  {\bibfnamefont{M.}~\bibnamefont{Nisoli}},\ }%
  \bibfield{journal}{%
  \Doi{10.1038/nphoton.2011.167}{\bibinfo {journal} {Nature Photonics}}\ }%
  \textbf{\bibinfo {volume} {5}},\ \bibinfo {pages} {655} (\bibinfo {month}
  {Sep.}\ \bibinfo {year} {2011}),\ ISSN \bibinfo {issn} {1749-4885},\
  \url{http://www.nature.com/doifinder/10.1038/nphoton.2011.167}%
  \bibAnnoteFile{NoStop}{Sansone2011}%
\bibitem{McNeil2010}%
  \BibitemOpen
  \bibfield{author}{%
  \bibinfo {author} {\bibfnamefont{B.~W.~J.}\ \bibnamefont{McNeil}}\ and\
  \bibinfo {author} {\bibfnamefont{N.~R.}\ \bibnamefont{Thompson}},\ }%
  \bibfield{journal}{%
  \Doi{10.1038/nphoton.2010.239}{\bibinfo {journal} {Nature Photonics}}\ }%
  \textbf{\bibinfo {volume} {4}},\ \bibinfo {pages} {814} (\bibinfo {year}
  {2010})%
  \bibAnnoteFile{NoStop}{McNeil2010}%
\bibitem{Shiner2011a}%
  \BibitemOpen
  \bibfield{author}{%
  \bibinfo {author} {\bibfnamefont{A.~D.}\ \bibnamefont{Shiner}}, \bibinfo
  {author} {\bibfnamefont{B.~E.}\ \bibnamefont{Schmidt}}, \bibinfo {author}
  {\bibfnamefont{C.}~\bibnamefont{Trallero-Herrero}}, \bibinfo {author}
  {\bibfnamefont{H.~J.}\ \bibnamefont{W\"{o}rner}}, \bibinfo {author}
  {\bibfnamefont{S.}~\bibnamefont{Patchkovskii}}, \bibinfo {author}
  {\bibfnamefont{P.~B.}\ \bibnamefont{Corkum}}, \bibinfo {author}
  {\bibfnamefont{J.-C.}\ \bibnamefont{Kieffer}}, \bibinfo {author}
  {\bibfnamefont{F.}~\bibnamefont{L\'{e}gar\'{e}}},\ and\ \bibinfo {author}
  {\bibfnamefont{D.~M.}\ \bibnamefont{Villeneuve}},\ }%
  \bibfield{journal}{%
  \Doi{10.1038/nphys1940}{\bibinfo {journal} {Nature Physics}}\ }%
  \textbf{\bibinfo {volume} {7}},\ \bibinfo {pages} {464} (\bibinfo {year}
  {2011}),\ ISSN \bibinfo {issn} {1745-2473},\
  \url{http://dx.doi.org/10.1038/nphys1940}%
  \bibAnnoteFile{NoStop}{Shiner2011a}%
\bibitem{Schultze2010a}%
  \BibitemOpen
  \bibfield{author}{%
  \bibinfo {author} {\bibfnamefont{M.}~\bibnamefont{Schultze}}, \bibinfo
  {author} {\bibfnamefont{M.}~\bibnamefont{Fiess}}, \bibinfo {author}
  {\bibfnamefont{N.}~\bibnamefont{Karpowicz}}, \bibinfo {author}
  {\bibfnamefont{J.}~\bibnamefont{Gagnon}}, \bibinfo {author}
  {\bibfnamefont{M.}~\bibnamefont{Korbman}}, \bibinfo {author}
  {\bibfnamefont{M.}~\bibnamefont{Hofstetter}}, \bibinfo {author}
  {\bibfnamefont{S.}~\bibnamefont{Neppl}}, \bibinfo {author}
  {\bibfnamefont{A.~L.}\ \bibnamefont{Cavalieri}}, \bibinfo {author}
  {\bibfnamefont{Y.}~\bibnamefont{Komninos}}, \bibinfo {author}
  {\bibfnamefont{T.}~\bibnamefont{Mercouris}}, \bibinfo {author}
  {\bibfnamefont{C.~A.}\ \bibnamefont{Nicolaides}}, \bibinfo {author}
  {\bibfnamefont{R.}~\bibnamefont{Pazourek}}, \bibinfo {author}
  {\bibfnamefont{S.}~\bibnamefont{Nagele}}, \bibinfo {author}
  {\bibfnamefont{J.}~\bibnamefont{Feist}}, \bibinfo {author}
  {\bibfnamefont{J.}~\bibnamefont{Burgd\"{o}rfer}}, \bibinfo {author}
  {\bibfnamefont{A.~M.}\ \bibnamefont{Azzeer}}, \bibinfo {author}
  {\bibfnamefont{R.}~\bibnamefont{Ernstorfer}}, \bibinfo {author}
  {\bibfnamefont{R.}~\bibnamefont{Kienberger}}, \bibinfo {author}
  {\bibfnamefont{U.}~\bibnamefont{Kleineberg}}, \bibinfo {author}
  {\bibfnamefont{E.}~\bibnamefont{Goulielmakis}}, \bibinfo {author}
  {\bibfnamefont{F.}~\bibnamefont{Krausz}},\ and\ \bibinfo {author}
  {\bibfnamefont{V.~S.}\ \bibnamefont{Yakovlev}},\ }%
  \bibfield{journal}{%
  \bibinfo {journal} {Science}\ }%
  \textbf{\bibinfo {volume} {328}},\ \bibinfo {pages} {1658} (\bibinfo {year}
  {2010})%
  \bibAnnoteFile{NoStop}{Schultze2010a}%
\bibitem{Smirnova2009}%
  \BibitemOpen
  \bibfield{author}{%
  \bibinfo {author} {\bibfnamefont{O.}~\bibnamefont{Smirnova}}, \bibinfo
  {author} {\bibfnamefont{Y.}~\bibnamefont{Mairesse}}, \bibinfo {author}
  {\bibfnamefont{S.}~\bibnamefont{Patchkovskii}}, \bibinfo {author}
  {\bibfnamefont{N.}~\bibnamefont{Dudovich}}, \bibinfo {author}
  {\bibfnamefont{D.~M.}\ \bibnamefont{Villeneuve}}, \bibinfo {author}
  {\bibfnamefont{P.~B.}\ \bibnamefont{Corkum}},\ and\ \bibinfo {author}
  {\bibfnamefont{M.~Y.}\ \bibnamefont{Ivanov}},\ }%
  \bibfield{journal}{%
  \Doi{10.1038/nature08253}{\bibinfo {journal} {Nature}}\ }%
  \textbf{\bibinfo {volume} {460}},\ \bibinfo {pages} {972} (\bibinfo {month}
  {Aug.}\ \bibinfo {year} {2009}),\ ISSN \bibinfo {issn} {1476-4687},\
  \url{http://www.ncbi.nlm.nih.gov/pubmed/19626004}%
  \bibAnnoteFile{NoStop}{Smirnova2009}%
\bibitem{Haessler2010}%
  \BibitemOpen
  \bibfield{author}{%
  \bibinfo {author} {\bibfnamefont{S.}~\bibnamefont{Haessler}}, \bibinfo
  {author} {\bibfnamefont{J.}~\bibnamefont{Caillat}}, \bibinfo {author}
  {\bibfnamefont{W.}~\bibnamefont{Boutu}}, \bibinfo {author}
  {\bibfnamefont{C.}~\bibnamefont{Giovanetti-Teixeira}}, \bibinfo {author}
  {\bibfnamefont{T.}~\bibnamefont{Ruchon}}, \bibinfo {author}
  {\bibfnamefont{T.}~\bibnamefont{Auguste}}, \bibinfo {author}
  {\bibfnamefont{Z.}~\bibnamefont{Diveki}}, \bibinfo {author}
  {\bibfnamefont{P.}~\bibnamefont{Breger}}, \bibinfo {author}
  {\bibfnamefont{A.}~\bibnamefont{Maquet}}, \bibinfo {author}
  {\bibfnamefont{B.}~\bibnamefont{Carr\'{e}}}, \bibinfo {author}
  {\bibfnamefont{R.}~\bibnamefont{Ta\"{\i}eb}},\ and\ \bibinfo {author}
  {\bibfnamefont{P.}~\bibnamefont{Sali\`{e}res}},\ }%
  \bibfield{journal}{%
  \Doi{10.1038/nphys1511}{\bibinfo {journal} {Nature Physics}}\ }%
  \textbf{\bibinfo {volume} {6}},\ \bibinfo {pages} {200} (\bibinfo {year}
  {2010}),\ ISSN \bibinfo {issn} {1745-2473},\
  \url{http://dx.doi.org/10.1038/nphys1511}%
  \bibAnnoteFile{NoStop}{Haessler2010}%
\bibitem{Feist2011a}%
  \BibitemOpen
  \bibfield{author}{%
  \bibinfo {author} {\bibfnamefont{J.}~\bibnamefont{Feist}}, \bibinfo {author}
  {\bibfnamefont{S.}~\bibnamefont{Nagele}}, \bibinfo {author}
  {\bibfnamefont{C.}~\bibnamefont{Ticknor}}, \bibinfo {author}
  {\bibfnamefont{B.~I.}\ \bibnamefont{Schneider}}, \bibinfo {author}
  {\bibfnamefont{L.~A.}\ \bibnamefont{Collins}},\ and\ \bibinfo {author}
  {\bibfnamefont{J.}~\bibnamefont{Burgd\"{o}rfer}},\ }%
  \bibfield{journal}{%
  \Doi{10.1103/PhysRevLett.107.093005}{\bibinfo {journal} {Phys. Rev. Lett.}}\
  }%
  \textbf{\bibinfo {volume} {107}},\ \bibinfo {pages} {093005} (\bibinfo {year}
  {2011})%
  \bibAnnoteFile{NoStop}{Feist2011a}%
\bibitem{Caillat2011}%
  \BibitemOpen
  \bibfield{author}{%
  \bibinfo {author} {\bibfnamefont{J.}~\bibnamefont{Caillat}}, \bibinfo
  {author} {\bibfnamefont{A.}~\bibnamefont{Maquet}}, \bibinfo {author}
  {\bibfnamefont{S.}~\bibnamefont{Haessler}}, \bibinfo {author}
  {\bibfnamefont{B.}~\bibnamefont{Fabre}}, \bibinfo {author}
  {\bibfnamefont{T.}~\bibnamefont{Ruchon}}, \bibinfo {author}
  {\bibfnamefont{P.}~\bibnamefont{Sali\`{e}res}}, \bibinfo {author}
  {\bibfnamefont{Y.}~\bibnamefont{Mairesse}},\ and\ \bibinfo {author}
  {\bibfnamefont{R.}~\bibnamefont{Ta\"{\i}eb}},\ }%
  \bibfield{journal}{%
  \Doi{10.1103/PhysRevLett.106.093002}{\bibinfo {journal} {Phys. Rev. Lett.}}\
  }%
  \textbf{\bibinfo {volume} {106}},\ \bibinfo {pages} {093002} (\bibinfo {year}
  {2011})%
  \bibAnnoteFile{NoStop}{Caillat2011}%
\bibitem{Klunder2011}%
  \BibitemOpen
  \bibfield{author}{%
  \bibinfo {author} {\bibfnamefont{K.}~\bibnamefont{Kl\"{u}nder}}, \bibinfo
  {author} {\bibfnamefont{J.~M.}\ \bibnamefont{Dahlstr\"{o}m}}, \bibinfo
  {author} {\bibfnamefont{M.}~\bibnamefont{Gisselbrecht}}, \bibinfo {author}
  {\bibfnamefont{T.}~\bibnamefont{Fordell}}, \bibinfo {author}
  {\bibfnamefont{M.}~\bibnamefont{Swoboda}}, \bibinfo {author}
  {\bibfnamefont{D.}~\bibnamefont{Gu\'{e}not}}, \bibinfo {author}
  {\bibfnamefont{P.}~\bibnamefont{Johnsson}}, \bibinfo {author}
  {\bibfnamefont{J.}~\bibnamefont{Caillat}}, \bibinfo {author}
  {\bibfnamefont{J.}~\bibnamefont{Mauritsson}}, \bibinfo {author}
  {\bibfnamefont{A.}~\bibnamefont{Maquet}}, \bibinfo {author}
  {\bibfnamefont{R.}~\bibnamefont{Ta\"{\i}eb}},\ and\ \bibinfo {author}
  {\bibfnamefont{A.}~\bibnamefont{L'Huillier}},\ }%
  \bibfield{journal}{%
  \Doi{10.1103/PhysRevLett.106.143002}{\bibinfo {journal} {Phys. Rev. Lett.}}\
  }%
  \textbf{\bibinfo {volume} {106}},\ \bibinfo {pages} {143002} (\bibinfo {year}
  {2011})%
  \bibAnnoteFile{NoStop}{Klunder2011}%
\bibitem{Kuleff2011}%
  \BibitemOpen
  \bibfield{author}{%
  \bibinfo {author} {\bibfnamefont{A.~I.}\ \bibnamefont{Kuleff}}\ and\ \bibinfo
  {author} {\bibfnamefont{L.~S.}\ \bibnamefont{Cederbaum}},\ }%
  \bibfield{journal}{%
  \Doi{10.1103/PhysRevLett.106.053001}{\bibinfo {journal} {Phys. Rev. Lett.}}\
  }%
  \textbf{\bibinfo {volume} {106}},\ \bibinfo {pages} {053001} (\bibinfo
  {month} {Jan.}\ \bibinfo {year} {2011}),\ ISSN \bibinfo {issn} {0031-9007},\
  \url{http://link.aps.org/doi/10.1103/PhysRevLett.106.053001}%
  \bibAnnoteFile{NoStop}{Kuleff2011}%
\bibitem{Remetter2006}%
  \BibitemOpen
  \bibfield{author}{%
  \bibinfo {author} {\bibfnamefont{T.}~\bibnamefont{Remetter}}, \bibinfo
  {author} {\bibfnamefont{P.}~\bibnamefont{Johnsson}}, \bibinfo {author}
  {\bibfnamefont{J.}~\bibnamefont{Mauritsson}}, \bibinfo {author}
  {\bibfnamefont{K.}~\bibnamefont{Varj\'{u}}}, \bibinfo {author}
  {\bibfnamefont{F.}~\bibnamefont{L\'{e}pine}}, \bibinfo {author}
  {\bibfnamefont{E.}~\bibnamefont{Gustafsson}}, \bibinfo {author}
  {\bibfnamefont{M.~F.}\ \bibnamefont{Kling}}, \bibinfo {author}
  {\bibfnamefont{J.~I.}\ \bibnamefont{Khan}}, \bibinfo {author}
  {\bibfnamefont{R.}~\bibnamefont{L\'{o}pez-Martens}}, \bibinfo {author}
  {\bibfnamefont{K.~J.}\ \bibnamefont{Schafer}}, \bibinfo {author}
  {\bibfnamefont{M.~J.~J.}\ \bibnamefont{Vrakking}},\ and\ \bibinfo {author}
  {\bibfnamefont{A.}~\bibnamefont{L'Huillier}},\ }%
  \bibfield{journal}{%
  \Doi{10.1038/nphys290}{\bibinfo {journal} {Nature Physics}}\ }%
  \textbf{\bibinfo {volume} {2}},\ \bibinfo {pages} {323} (\bibinfo {year}
  {2006})%
  \bibAnnoteFile{NoStop}{Remetter2006}%
\bibitem{Mauritsson2008}%
  \BibitemOpen
  \bibfield{author}{%
  \bibinfo {author} {\bibfnamefont{J.}~\bibnamefont{Mauritsson}}, \bibinfo
  {author} {\bibfnamefont{P.}~\bibnamefont{Johnsson}}, \bibinfo {author}
  {\bibfnamefont{E.}~\bibnamefont{Mansten}}, \bibinfo {author}
  {\bibfnamefont{M.}~\bibnamefont{Swoboda}}, \bibinfo {author}
  {\bibfnamefont{T.}~\bibnamefont{Ruchon}}, \bibinfo {author}
  {\bibfnamefont{A.}~\bibnamefont{L'Huillier}},\ and\ \bibinfo {author}
  {\bibfnamefont{K.~J.}\ \bibnamefont{Schafer}},\ }%
  \bibfield{journal}{%
  \Doi{10.1103/PhysRevLett.100.073003}{\bibinfo {journal} {Phys. Rev. Lett.}}\
  }%
  \textbf{\bibinfo {volume} {100}},\ \bibinfo {pages} {073003} (\bibinfo {year}
  {2008})%
  \bibAnnoteFile{NoStop}{Mauritsson2008}%
\bibitem{FeiNagPaz2009}%
  \BibitemOpen
  \bibfield{author}{%
  \bibinfo {author} {\bibfnamefont{J.}~\bibnamefont{Feist}}, \bibinfo {author}
  {\bibfnamefont{S.}~\bibnamefont{Nagele}}, \bibinfo {author}
  {\bibfnamefont{R.}~\bibnamefont{Pazourek}}, \bibinfo {author}
  {\bibfnamefont{E.}~\bibnamefont{Persson}}, \bibinfo {author}
  {\bibfnamefont{B.~I.}\ \bibnamefont{Schneider}}, \bibinfo {author}
  {\bibfnamefont{L.~A.}\ \bibnamefont{Collins}},\ and\ \bibinfo {author}
  {\bibfnamefont{J.}~\bibnamefont{Burgd\"{o}rfer}},\ }%
  \bibfield{journal}{%
  \Doi{10.1103/PhysRevLett.103.063002}{\bibinfo {journal} {Phys. Rev. Lett.}}\
  }%
  \textbf{\bibinfo {volume} {103}},\ \bibinfo {pages} {063002} (\bibinfo
  {month} {Aug.}\ \bibinfo {year} {2009})%
  \bibAnnoteFile{NoStop}{FeiNagPaz2009}%
\bibitem{Argenti2010}%
  \BibitemOpen
  \bibfield{author}{%
  \bibinfo {author} {\bibfnamefont{L.}~\bibnamefont{Argenti}}\ and\ \bibinfo
  {author} {\bibfnamefont{E.}~\bibnamefont{Lindroth}},\ }%
  \bibfield{journal}{%
  \Doi{10.1103/PhysRevLett.105.053002}{\bibinfo {journal} {Phys. Rev. Lett.}}\
  }%
  \textbf{\bibinfo {volume} {105}},\ \bibinfo {pages} {053002} (\bibinfo
  {month} {Jul.}\ \bibinfo {year} {2010}),\ ISSN \bibinfo {issn} {0031-9007},\
  \url{http://link.aps.org/doi/10.1103/PhysRevLett.105.053002}%
  \bibAnnoteFile{NoStop}{Argenti2010}%
\bibitem{Sansone2010}%
  \BibitemOpen
  \bibfield{author}{%
  \bibinfo {author} {\bibfnamefont{G.}~\bibnamefont{Sansone}}, \bibinfo
  {author} {\bibfnamefont{F.}~\bibnamefont{Kelkensberg}}, \bibinfo {author}
  {\bibfnamefont{J.~F.}\ \bibnamefont{P\'{e}rez-Torres}}, \bibinfo {author}
  {\bibfnamefont{F.}~\bibnamefont{Morales}}, \bibinfo {author}
  {\bibfnamefont{M.~F.}\ \bibnamefont{Kling}}, \bibinfo {author}
  {\bibfnamefont{W.}~\bibnamefont{Siu}}, \bibinfo {author}
  {\bibfnamefont{O.}~\bibnamefont{Ghafur}}, \bibinfo {author}
  {\bibfnamefont{P.}~\bibnamefont{Johnsson}}, \bibinfo {author}
  {\bibfnamefont{M.}~\bibnamefont{Swoboda}}, \bibinfo {author}
  {\bibfnamefont{E.}~\bibnamefont{Benedetti}}, \bibinfo {author}
  {\bibfnamefont{F.}~\bibnamefont{Ferrari}}, \bibinfo {author}
  {\bibfnamefont{F.}~\bibnamefont{L\'{e}pine}}, \bibinfo {author}
  {\bibfnamefont{J.~L.}\ \bibnamefont{Sanz-Vicario}}, \bibinfo {author}
  {\bibfnamefont{S.}~\bibnamefont{Zherebtsov}}, \bibinfo {author}
  {\bibfnamefont{I.}~\bibnamefont{Znakovskaya}}, \bibinfo {author}
  {\bibfnamefont{A.}~\bibnamefont{L'Huillier}}, \bibinfo {author}
  {\bibfnamefont{M.~Y.}\ \bibnamefont{Ivanov}}, \bibinfo {author}
  {\bibfnamefont{M.}~\bibnamefont{Nisoli}}, \bibinfo {author}
  {\bibfnamefont{F.}~\bibnamefont{Mart\'in}},\ and\ \bibinfo {author}
  {\bibfnamefont{M.~J.~J.}\ \bibnamefont{Vrakking}},\ }%
  \bibfield{journal}{%
  \Doi{10.1038/nature09084}{\bibinfo {journal} {Nature}}\ }%
  \textbf{\bibinfo {volume} {465}},\ \bibinfo {pages} {763} (\bibinfo {month}
  {Jun.}\ \bibinfo {year} {2010}),\ ISSN \bibinfo {issn} {0028-0836},\
  \url{http://www.nature.com/doifinder/10.1038/nature09084}%
  \bibAnnoteFile{NoStop}{Sansone2010}%
\bibitem{Kelkensberg2011}%
  \BibitemOpen
  \bibfield{author}{%
  \bibinfo {author} {\bibfnamefont{F.}~\bibnamefont{Kelkensberg}}, \bibinfo
  {author} {\bibfnamefont{W.}~\bibnamefont{Siu}}, \bibinfo {author}
  {\bibfnamefont{J.~F.}\ \bibnamefont{P\'{e}rez-Torres}}, \bibinfo {author}
  {\bibfnamefont{F.}~\bibnamefont{Morales}}, \bibinfo {author}
  {\bibfnamefont{G.}~\bibnamefont{Gademann}}, \bibinfo {author}
  {\bibfnamefont{A.}~\bibnamefont{Rouz\'{e}e}}, \bibinfo {author}
  {\bibfnamefont{P.}~\bibnamefont{Johnsson}}, \bibinfo {author}
  {\bibfnamefont{M.}~\bibnamefont{Lucchini}}, \bibinfo {author}
  {\bibfnamefont{F.}~\bibnamefont{Calegari}}, \bibinfo {author}
  {\bibfnamefont{J.~L.}\ \bibnamefont{Sanz-Vicario}}, \bibinfo {author}
  {\bibfnamefont{F.}~\bibnamefont{Mart\'in}},\ and\ \bibinfo {author}
  {\bibfnamefont{M.~J.~J.}\ \bibnamefont{Vrakking}},\ }%
  \bibfield{journal}{%
  \Doi{10.1103/PhysRevLett.107.043002}{\bibinfo {journal} {Phys. Rev. Lett.}}\
  }%
  \textbf{\bibinfo {volume} {107}},\ \bibinfo {pages} {043002} (\bibinfo {year}
  {2011})%
  \bibAnnoteFile{NoStop}{Kelkensberg2011}%
\bibitem{Fischer2010}%
  \BibitemOpen
  \bibfield{author}{%
  \bibinfo {author} {\bibfnamefont{B.}~\bibnamefont{Fischer}}, \bibinfo
  {author} {\bibfnamefont{M.}~\bibnamefont{Kremer}}, \bibinfo {author}
  {\bibfnamefont{T.}~\bibnamefont{Pfeifer}}, \bibinfo {author}
  {\bibfnamefont{B.}~\bibnamefont{Feuerstein}}, \bibinfo {author}
  {\bibfnamefont{V.}~\bibnamefont{Sharma}}, \bibinfo {author}
  {\bibfnamefont{U.}~\bibnamefont{Thumm}}, \bibinfo {author}
  {\bibfnamefont{C.-D.}\ \bibnamefont{Schr\"{o}ter}}, \bibinfo {author}
  {\bibfnamefont{R.}~\bibnamefont{Moshammer}},\ and\ \bibinfo {author}
  {\bibfnamefont{J.}~\bibnamefont{Ullrich}},\ }%
  \bibfield{journal}{%
  \Doi{10.1103/PhysRevLett.105.223001}{\bibinfo {journal} {Phys. Rev. Lett.}}\
  }%
  \textbf{\bibinfo {volume} {105}},\ \bibinfo {pages} {223001} (\bibinfo
  {month} {Nov.}\ \bibinfo {year} {2010}),\ ISSN \bibinfo {issn} {0031-9007},\
  \url{http://link.aps.org/doi/10.1103/PhysRevLett.105.223001}%
  \bibAnnoteFile{NoStop}{Fischer2010}%
\bibitem{Kruger2011}%
  \BibitemOpen
  \bibfield{author}{%
  \bibinfo {author} {\bibfnamefont{M.}~\bibnamefont{Kr\"{u}ger}}, \bibinfo
  {author} {\bibfnamefont{M.}~\bibnamefont{Schenk}},\ and\ \bibinfo {author}
  {\bibfnamefont{P.}~\bibnamefont{Hommelhoff}},\ }%
  \bibfield{journal}{%
  \Doi{10.1038/nature10196}{\bibinfo {journal} {Nature}}\ }%
  \textbf{\bibinfo {volume} {475}},\ \bibinfo {pages} {78} (\bibinfo {month}
  {Jul.}\ \bibinfo {year} {2011}),\ ISSN \bibinfo {issn} {0028-0836},\
  \url{http://www.nature.com/doifinder/10.1038/nature10196}%
  \bibAnnoteFile{NoStop}{Kruger2011}%
\bibitem{Jiang2010b}%
  \BibitemOpen
  \bibfield{author}{%
  \bibinfo {author} {\bibfnamefont{Y.~H.}\ \bibnamefont{Jiang}}, \bibinfo
  {author} {\bibfnamefont{A.}~\bibnamefont{Rudenko}}, \bibinfo {author}
  {\bibfnamefont{O.}~\bibnamefont{Herrwerth}}, \bibinfo {author}
  {\bibfnamefont{L.}~\bibnamefont{Foucar}}, \bibinfo {author}
  {\bibfnamefont{M.}~\bibnamefont{Kurka}}, \bibinfo {author}
  {\bibfnamefont{K.-U.}\ \bibnamefont{K\"{u}hnel}}, \bibinfo {author}
  {\bibfnamefont{M.}~\bibnamefont{Lezius}}, \bibinfo {author}
  {\bibfnamefont{M.~F.}\ \bibnamefont{Kling}}, \bibinfo {author}
  {\bibfnamefont{J.}~\bibnamefont{van Tilborg}}, \bibinfo {author}
  {\bibfnamefont{A.}~\bibnamefont{Belkacem}}, \bibinfo {author}
  {\bibfnamefont{K.}~\bibnamefont{Ueda}}, \bibinfo {author}
  {\bibfnamefont{S.}~\bibnamefont{D\"{u}sterer}}, \bibinfo {author}
  {\bibfnamefont{R.}~\bibnamefont{Treusch}}, \bibinfo {author}
  {\bibfnamefont{C.-D.}\ \bibnamefont{Schr\"{o}ter}}, \bibinfo {author}
  {\bibfnamefont{R.}~\bibnamefont{Moshammer}},\ and\ \bibinfo {author}
  {\bibfnamefont{J.}~\bibnamefont{Ullrich}},\ }%
  \bibfield{journal}{%
  \Doi{10.1103/PhysRevLett.105.263002}{\bibinfo {journal} {Phys. Rev. Lett.}}\
  }%
  \textbf{\bibinfo {volume} {105}},\ \bibinfo {pages} {263002} (\bibinfo
  {month} {Dec.}\ \bibinfo {year} {2010}),\ ISSN \bibinfo {issn} {0031-9007},\
  \url{http://link.aps.org/doi/10.1103/PhysRevLett.105.263002}%
  \bibAnnoteFile{NoStop}{Jiang2010b}%
\bibitem{Faisal}%
  \BibitemOpen
  \bibfield{author}{%
  \bibinfo {author} {\bibfnamefont{F.~H.~M.}\ \bibnamefont{Faisal}},\ }%
  \emph{\bibinfo {title} {Theory of Multiphoton Processes}}\ (\bibinfo
  {publisher} {Plenum Press},\ \bibinfo {address} {New York},\ \bibinfo {year}
  {1987})%
  \bibAnnoteFile{NoStop}{Faisal}%
\bibitem{Sansone2006}%
  \BibitemOpen
  \bibfield{author}{%
  \bibinfo {author} {\bibfnamefont{G.}~\bibnamefont{Sansone}}, \bibinfo
  {author} {\bibfnamefont{E.}~\bibnamefont{Benedetti}}, \bibinfo {author}
  {\bibfnamefont{F.}~\bibnamefont{Calegari}}, \bibinfo {author}
  {\bibfnamefont{C.}~\bibnamefont{Vozzi}}, \bibinfo {author}
  {\bibfnamefont{L.}~\bibnamefont{Avaldi}}, \bibinfo {author}
  {\bibfnamefont{R.}~\bibnamefont{Flammini}}, \bibinfo {author}
  {\bibfnamefont{L.}~\bibnamefont{Poletto}}, \bibinfo {author}
  {\bibfnamefont{P.}~\bibnamefont{Villoresi}}, \bibinfo {author}
  {\bibfnamefont{C.}~\bibnamefont{Altucci}}, \bibinfo {author}
  {\bibfnamefont{R.}~\bibnamefont{Velotta}}, \bibinfo {author}
  {\bibfnamefont{S.}~\bibnamefont{Stagira}}, \bibinfo {author}
  {\bibfnamefont{S.}~\bibnamefont{{De Silvestri}}},\ and\ \bibinfo {author}
  {\bibfnamefont{M.}~\bibnamefont{Nisoli}},\ }%
  \bibfield{journal}{%
  \Doi{10.1126/science.1132838}{\bibinfo {journal} {Science}}\ }%
  \textbf{\bibinfo {volume} {314}},\ \bibinfo {pages} {443} (\bibinfo {year}
  {2006})%
  \bibAnnoteFile{NoStop}{Sansone2006}%
\bibitem{Goulielmakis2008}%
  \BibitemOpen
  \bibfield{author}{%
  \bibinfo {author} {\bibfnamefont{E.}~\bibnamefont{Goulielmakis}}, \bibinfo
  {author} {\bibfnamefont{M.}~\bibnamefont{Schultze}}, \bibinfo {author}
  {\bibfnamefont{M.}~\bibnamefont{Hofstetter}}, \bibinfo {author}
  {\bibfnamefont{V.~S.}\ \bibnamefont{Yakovlev}}, \bibinfo {author}
  {\bibfnamefont{J.}~\bibnamefont{Gagnon}}, \bibinfo {author}
  {\bibfnamefont{M.}~\bibnamefont{Uiberacker}}, \bibinfo {author}
  {\bibfnamefont{A.}~\bibnamefont{Aquila}}, \bibinfo {author}
  {\bibfnamefont{E.~M.}\ \bibnamefont{Gullikson}}, \bibinfo {author}
  {\bibfnamefont{D.~T.}\ \bibnamefont{Attwood}}, \bibinfo {author}
  {\bibfnamefont{R.}~\bibnamefont{Kienberger}}, \bibinfo {author}
  {\bibfnamefont{F.}~\bibnamefont{Krausz}},\ and\ \bibinfo {author}
  {\bibfnamefont{U.}~\bibnamefont{Kleineberg}},\ }%
  \bibfield{journal}{%
  \Doi{10.1126/science.1157846}{\bibinfo {journal} {Science}}\ }%
  \textbf{\bibinfo {volume} {320}},\ \bibinfo {pages} {1614} (\bibinfo {year}
  {2008})%
  \bibAnnoteFile{NoStop}{Goulielmakis2008}%
\bibitem{Burke1991}%
  \BibitemOpen
  \bibfield{author}{%
  \bibinfo {author} {\bibfnamefont{P.~G.}\ \bibnamefont{Burke}}, \bibinfo
  {author} {\bibfnamefont{P.}~\bibnamefont{Francken}},\ and\ \bibinfo {author}
  {\bibfnamefont{C.~J.}\ \bibnamefont{Joachain}},\ }%
  \bibfield{journal}{%
  \bibinfo {journal} {J. Phys. B: At. Mol. Opt. Phys.}\ }%
  \textbf{\bibinfo {volume} {24}},\ \bibinfo {pages} {761} (\bibinfo {year}
  {1991})%
  \bibAnnoteFile{NoStop}{Burke1991}%
\bibitem{Smyth1998}%
  \BibitemOpen
  \bibfield{author}{%
  \bibinfo {author} {\bibfnamefont{E.~S.}\ \bibnamefont{Smyth}}, \bibinfo
  {author} {\bibfnamefont{J.~S.}\ \bibnamefont{Parker}},\ and\ \bibinfo
  {author} {\bibfnamefont{K.~T.}\ \bibnamefont{Taylor}},\ }%
  \bibfield{journal}{%
  \bibinfo {journal} {Comp. Phys. Commun.}\ }%
  \textbf{\bibinfo {volume} {114}},\ \bibinfo {pages} {1} (\bibinfo {year}
  {1998})%
  \bibAnnoteFile{NoStop}{Smyth1998}%
\bibitem{LagmagoKamta2002}%
  \BibitemOpen
  \bibfield{author}{%
  \bibinfo {author} {\bibfnamefont{G.~L.}\ \bibnamefont{Kamta}}\ and\ \bibinfo
  {author} {\bibfnamefont{A.~F.}\ \bibnamefont{Starace}},\ }%
  \bibfield{journal}{%
  \Doi{10.1103/PhysRevA.65.053418}{\bibinfo {journal} {Phys. Rev. A}}\ }%
  \textbf{\bibinfo {volume} {65}},\ \bibinfo {pages} {053418} (\bibinfo {year}
  {2002})%
  \bibAnnoteFile{NoStop}{LagmagoKamta2002}%
\bibitem{Laulan2004}%
  \BibitemOpen
  \bibfield{author}{%
  \bibinfo {author} {\bibfnamefont{S.}~\bibnamefont{Laulan}}\ and\ \bibinfo
  {author} {\bibfnamefont{H.}~\bibnamefont{Bachau}},\ }%
  \bibfield{journal}{%
  \Doi{10.1103/PhysRevA.69.033408}{\bibinfo {journal} {Physical Review A}}\ }%
  \textbf{\bibinfo {volume} {69}},\ \bibinfo {pages} {033408} (\bibinfo {month}
  {Mar.}\ \bibinfo {year} {2004}),\ ISSN \bibinfo {issn} {1050-2947},\
  \url{http://link.aps.org/doi/10.1103/PhysRevA.69.033408}%
  \bibAnnoteFile{NoStop}{Laulan2004}%
\bibitem{Sanz-Vicario2006}%
  \BibitemOpen
  \bibfield{author}{%
  \bibinfo {author} {\bibfnamefont{J.~L.}\ \bibnamefont{Sanz-Vicario}},
  \bibinfo {author} {\bibfnamefont{H.}~\bibnamefont{Bachau}},\ and\ \bibinfo
  {author} {\bibfnamefont{F.}~\bibnamefont{Mart\'in}},\ }%
  \bibfield{journal}{%
  \Doi{10.1103/PhysRevA.73.033410}{\bibinfo {journal} {Phys. Rev. A}}\ }%
  \textbf{\bibinfo {volume} {73}},\ \bibinfo {pages} {033410} (\bibinfo {year}
  {2006})%
  \bibAnnoteFile{NoStop}{Sanz-Vicario2006}%
\bibitem{Tong2006}%
  \BibitemOpen
  \bibfield{author}{%
  \bibinfo {author} {\bibfnamefont{X.-M.}\ \bibnamefont{Tong}}, \bibinfo
  {author} {\bibfnamefont{K.}~\bibnamefont{Hino}},\ and\ \bibinfo {author}
  {\bibfnamefont{N.}~\bibnamefont{Toshima}},\ }%
  \bibfield{journal}{%
  \Doi{10.1103/PhysRevA.74.031405}{\bibinfo {journal} {Phys. Rev. A}}\ }%
  \textbf{\bibinfo {volume} {74}},\ \bibinfo {pages} {031405(R)} (\bibinfo
  {month} {Sep.}\ \bibinfo {year} {2006}),\ ISSN \bibinfo {issn} {1050-2947},\
  \url{http://link.aps.org/doi/10.1103/PhysRevA.74.031405}%
  \bibAnnoteFile{NoStop}{Tong2006}%
\bibitem{Foumouo2006}%
  \BibitemOpen
  \bibfield{author}{%
  \bibinfo {author} {\bibfnamefont{E.}~\bibnamefont{Foumouo}}, \bibinfo
  {author} {\bibfnamefont{G.~L.}\ \bibnamefont{{Kamta}}}, \bibinfo {author}
  {\bibfnamefont{G.}~\bibnamefont{Edah}},\ and\ \bibinfo {author}
  {\bibfnamefont{B.}~\bibnamefont{Piraux}},\ }%
  \bibfield{journal}{%
  \Doi{10.1103/PhysRevA.74.063409}{\bibinfo {journal} {Phys. Rev. A}}\ }%
  \textbf{\bibinfo {volume} {74}},\ \bibinfo {pages} {063409} (\bibinfo {year}
  {2006})%
  \bibAnnoteFile{NoStop}{Foumouo2006}%
\bibitem{Lambropoulos2008a}%
  \BibitemOpen
  \bibfield{author}{%
  \bibinfo {author} {\bibfnamefont{P.}~\bibnamefont{Lambropoulos}}\ and\
  \bibinfo {author} {\bibfnamefont{L.~A.~A.}\ \bibnamefont{Nikolopoulos}},\ }%
  \bibfield{journal}{%
  \Doi{10.1088/1367-2630/10/2/025012}{\bibinfo {journal} {New J. Phys.}}\ }%
  \textbf{\bibinfo {volume} {10}},\ \bibinfo {pages} {025012} (\bibinfo {year}
  {2008})%
  \bibAnnoteFile{NoStop}{Lambropoulos2008a}%
\bibitem{Lysaght2008}%
  \BibitemOpen
  \bibfield{author}{%
  \bibinfo {author} {\bibfnamefont{M.~A.}\ \bibnamefont{Lysaght}}, \bibinfo
  {author} {\bibfnamefont{P.~G.}\ \bibnamefont{Burke}},\ and\ \bibinfo {author}
  {\bibfnamefont{H.~W.}\ \bibnamefont{van~der Hart}},\ }%
  \bibfield{journal}{%
  \Doi{10.1103/PhysRevLett.101.253001}{\bibinfo {journal} {Phys. Rev. Lett.}}\
  }%
  \textbf{\bibinfo {volume} {101}},\ \bibinfo {pages} {253001} (\bibinfo {year}
  {2008})%
  \bibAnnoteFile{NoStop}{Lysaght2008}%
\bibitem{Feist2008}%
  \BibitemOpen
  \bibfield{author}{%
  \bibinfo {author} {\bibfnamefont{J.}~\bibnamefont{Feist}}, \bibinfo {author}
  {\bibfnamefont{S.}~\bibnamefont{Nagele}}, \bibinfo {author}
  {\bibfnamefont{R.}~\bibnamefont{Pazourek}}, \bibinfo {author}
  {\bibfnamefont{E.}~\bibnamefont{Persson}}, \bibinfo {author}
  {\bibfnamefont{B.~I.}\ \bibnamefont{Schneider}}, \bibinfo {author}
  {\bibfnamefont{L.~A.}\ \bibnamefont{Collins}},\ and\ \bibinfo {author}
  {\bibfnamefont{J.}~\bibnamefont{Burgd\"{o}rfer}},\ }%
  \bibfield{journal}{%
  \Doi{10.1103/PhysRevA.77.043420}{\bibinfo {journal} {Phys. Rev. A}}\ }%
  \textbf{\bibinfo {volume} {77}},\ \bibinfo {pages} {043420} (\bibinfo {month}
  {Apr.}\ \bibinfo {year} {2008}),\ ISSN \bibinfo {issn} {1050-2947},\
  \url{http://link.aps.org/doi/10.1103/PhysRevA.77.043420}%
  \bibAnnoteFile{NoStop}{Feist2008}%
\bibitem{Palacios2008}%
  \BibitemOpen
  \bibfield{author}{%
  \bibinfo {author} {\bibfnamefont{A.}~\bibnamefont{Palacios}}, \bibinfo
  {author} {\bibfnamefont{T.~N.}\ \bibnamefont{Rescigno}},\ and\ \bibinfo
  {author} {\bibfnamefont{C.~W.}\ \bibnamefont{McCurdy}},\ }%
  \bibfield{journal}{%
  \Doi{10.1103/PhysRevA.77.032716}{\bibinfo {journal} {Phys. Rev. A}}\ }%
  \textbf{\bibinfo {volume} {77}},\ \bibinfo {pages} {032716} (\bibinfo {month}
  {Mar.}\ \bibinfo {year} {2008}),\ ISSN \bibinfo {issn} {1050-2947},\
  \url{http://link.aps.org/doi/10.1103/PhysRevA.77.032716}%
  \bibAnnoteFile{NoStop}{Palacios2008}%
\bibitem{Nepstad2010}%
  \BibitemOpen
  \bibfield{author}{%
  \bibinfo {author} {\bibfnamefont{R.}~\bibnamefont{Nepstad}}, \bibinfo
  {author} {\bibfnamefont{T.}~\bibnamefont{Birkeland}},\ and\ \bibinfo {author}
  {\bibfnamefont{M.}~\bibnamefont{F\o~rre}},\ }%
  \bibfield{journal}{%
  \Doi{10.1103/PhysRevA.81.063402}{\bibinfo {journal} {Phys. Rev. A}}\ }%
  \textbf{\bibinfo {volume} {81}},\ \bibinfo {pages} {063402} (\bibinfo {month}
  {Jun.}\ \bibinfo {year} {2010}),\ ISSN \bibinfo {issn} {1050-2947},\
  \url{http://link.aps.org/doi/10.1103/PhysRevA.81.063402}%
  \bibAnnoteFile{NoStop}{Nepstad2010}%
\bibitem{Eppink1997}%
  \BibitemOpen
  \bibfield{author}{%
  \bibinfo {author} {\bibfnamefont{A.~T. J.~B.}\ \bibnamefont{Eppink}}\ and\
  \bibinfo {author} {\bibfnamefont{D.~H.}\ \bibnamefont{Parker}},\ }%
  \bibfield{journal}{%
  \bibinfo {journal} {Rev. Sci. Instrum.}\ }%
  \textbf{\bibinfo {volume} {68}},\ \bibinfo {pages} {3477} (\bibinfo {year}
  {1997})%
  \bibAnnoteFile{NoStop}{Eppink1997}%
\bibitem{Ullrich2003}%
  \BibitemOpen
  \bibfield{author}{%
  \bibinfo {author} {\bibfnamefont{J.}~\bibnamefont{Ullrich}}, \bibinfo
  {author} {\bibfnamefont{R.}~\bibnamefont{Moshammer}}, \bibinfo {author}
  {\bibfnamefont{A.}~\bibnamefont{Dorn}}, \bibinfo {author}
  {\bibfnamefont{R.~D.}\ \bibnamefont{D\"{o}rner}}, \bibinfo {author}
  {\bibfnamefont{L.~P.~H.}\ \bibnamefont{Schmidt}},\ and\ \bibinfo {author}
  {\bibfnamefont{H.}~\bibnamefont{Schmidt-B\"{o}cking}},\ }%
  \bibfield{journal}{%
  \bibinfo {journal} {Rep. Prog. Phys.}\ }%
  \textbf{\bibinfo {volume} {66}},\ \bibinfo {pages} {1463} (\bibinfo {year}
  {2003})%
  \bibAnnoteFile{NoStop}{Ullrich2003}%
\bibitem{Goulielmakis2010}%
  \BibitemOpen
  \bibfield{author}{%
  \bibinfo {author} {\bibfnamefont{E.}~\bibnamefont{Goulielmakis}}, \bibinfo
  {author} {\bibfnamefont{Z.-H.}\ \bibnamefont{Loh}}, \bibinfo {author}
  {\bibfnamefont{A.}~\bibnamefont{Wirth}}, \bibinfo {author}
  {\bibfnamefont{R.}~\bibnamefont{Santra}}, \bibinfo {author}
  {\bibfnamefont{N.}~\bibnamefont{Rohringer}}, \bibinfo {author}
  {\bibfnamefont{V.~S.}\ \bibnamefont{Yakovlev}}, \bibinfo {author}
  {\bibfnamefont{S.}~\bibnamefont{Zherebtsov}}, \bibinfo {author}
  {\bibfnamefont{T.}~\bibnamefont{Pfeifer}}, \bibinfo {author}
  {\bibfnamefont{A.~M.}\ \bibnamefont{Azzeer}}, \bibinfo {author}
  {\bibfnamefont{M.~F.}\ \bibnamefont{Kling}}, \bibinfo {author}
  {\bibfnamefont{S.~R.}\ \bibnamefont{Leone}},\ and\ \bibinfo {author}
  {\bibfnamefont{F.}~\bibnamefont{Krausz}},\ }%
  \bibfield{journal}{%
  \Doi{10.1038/nature09212}{\bibinfo {journal} {Nature}}\ }%
  \textbf{\bibinfo {volume} {466}},\ \bibinfo {pages} {739} (\bibinfo {month}
  {Aug.}\ \bibinfo {year} {2010}),\ ISSN \bibinfo {issn} {0028-0836},\
  \url{http://www.nature.com/doifinder/10.1038/nature09212}%
  \bibAnnoteFile{NoStop}{Goulielmakis2010}%
\bibitem{Nikolopoulos2007}%
  \BibitemOpen
  \bibfield{author}{%
  \bibinfo {author} {\bibfnamefont{L.~A.~A.}\ \bibnamefont{Nikolopoulos}},
  \bibinfo {author} {\bibfnamefont{T.~K.}\ \bibnamefont{Kjeldsen}},\ and\
  \bibinfo {author} {\bibfnamefont{L.~B.}\ \bibnamefont{Madsen}},\ }%
  \bibfield{journal}{%
  \Doi{10.1103/PhysRevA.75.063426}{\bibinfo {journal} {Phys. Rev. A}}\ }%
  \textbf{\bibinfo {volume} {75}},\ \bibinfo {pages} {063426} (\bibinfo {month}
  {Jun.}\ \bibinfo {year} {2007}),\ ISSN \bibinfo {issn} {1050-2947},\
  \url{http://link.aps.org/doi/10.1103/PhysRevA.75.063426}%
  \bibAnnoteFile{NoStop}{Nikolopoulos2007}%
\bibitem{Kjeldsen2006}%
  \BibitemOpen
  \bibfield{author}{%
  \bibinfo {author} {\bibfnamefont{T.~K.}\ \bibnamefont{Kjeldsen}}, \bibinfo
  {author} {\bibfnamefont{L.~B.}\ \bibnamefont{Madsen}},\ and\ \bibinfo
  {author} {\bibfnamefont{J.~P.}\ \bibnamefont{Hansen}},\ }%
  \bibfield{journal}{%
  \Doi{10.1103/PhysRevA.74.035402}{\bibinfo {journal} {Phys. Rev. A}}\ }%
  \textbf{\bibinfo {volume} {74}},\ \bibinfo {pages} {035402} (\bibinfo {month}
  {Sep.}\ \bibinfo {year} {2006}),\ ISSN \bibinfo {issn} {1050-2947},\
  \url{http://link.aps.org/doi/10.1103/PhysRevA.74.035402}%
  \bibAnnoteFile{NoStop}{Kjeldsen2006}%
\bibitem{Madsen2007}%
  \BibitemOpen
  \bibfield{author}{%
  \bibinfo {author} {\bibfnamefont{L.~B.}\ \bibnamefont{Madsen}}, \bibinfo
  {author} {\bibfnamefont{L.~A.~A.}\ \bibnamefont{Nikolopoulos}}, \bibinfo
  {author} {\bibfnamefont{T.~K.}\ \bibnamefont{Kjeldsen}},\ and\ \bibinfo
  {author} {\bibfnamefont{J.}~\bibnamefont{Fern\'{a}ndez}},\ }%
  \bibfield{journal}{%
  \Doi{10.1103/PhysRevA.76.063407}{\bibinfo {journal} {Phys. Rev. A}}\ }%
  \textbf{\bibinfo {volume} {76}},\ \bibinfo {pages} {063407} (\bibinfo {year}
  {2007})%
  \bibAnnoteFile{NoStop}{Madsen2007}%
\bibitem{Catoire2012}%
  \BibitemOpen
  \bibfield{author}{%
  \bibinfo {author} {\bibfnamefont{F.}~\bibnamefont{Catoire}}\ and\ \bibinfo
  {author} {\bibfnamefont{H.}~\bibnamefont{Bachau}},\ }%
  \bibfield{journal}{%
  \Doi{10.1103/PhysRevA.85.023422}{\bibinfo {journal} {Phys. Rev. A}}\ }%
  \textbf{\bibinfo {volume} {85}},\ \bibinfo {pages} {023422} (\bibinfo {month}
  {Feb}\ \bibinfo {year} {2012}),\
  \url{http://link.aps.org/doi/10.1103/PhysRevA.85.023422}%
  \bibAnnoteFile{NoStop}{Catoire2012}%
\bibitem{McCurdy2004b}%
  \BibitemOpen
  \bibfield{author}{%
  \bibinfo {author} {\bibfnamefont{C.~W.}\ \bibnamefont{McCurdy}}, \bibinfo
  {author} {\bibfnamefont{M.}~\bibnamefont{Baertschy}},\ and\ \bibinfo {author}
  {\bibfnamefont{T.~N.}\ \bibnamefont{Rescigno}},\ }%
  \bibfield{journal}{%
  \Doi{10.1088/0953-4075/37/17/R01}{\bibinfo {journal} {J. Phys. B: At. Mol.
  Opt. Phys.}}\ }%
  \textbf{\bibinfo {volume} {37}},\ \bibinfo {pages} {R137} (\bibinfo {month}
  {Sep.}\ \bibinfo {year} {2004}),\ ISSN \bibinfo {issn} {0953-4075},\
  \url{http://stacks.iop.org/0953-4075/37/i=17/a=R01?key=crossref.2b47a7395a82%
01d53f3f3b2697689fbe}%
  \bibAnnoteFile{NoStop}{McCurdy2004b}%
\bibitem{Palacios2007a}%
  \BibitemOpen
  \bibfield{author}{%
  \bibinfo {author} {\bibfnamefont{A.}~\bibnamefont{Palacios}}, \bibinfo
  {author} {\bibfnamefont{C.~W.}\ \bibnamefont{McCurdy}},\ and\ \bibinfo
  {author} {\bibfnamefont{T.~N.}\ \bibnamefont{Rescigno}},\ }%
  \bibfield{journal}{%
  \Doi{10.1103/PhysRevA.76.043420}{\bibinfo {journal} {Phys. Rev. A}}\ }%
  \textbf{\bibinfo {volume} {76}},\ \bibinfo {pages} {043420} (\bibinfo {month}
  {Oct.}\ \bibinfo {year} {2007}),\ ISSN \bibinfo {issn} {1050-2947},\
  \url{http://link.aps.org/doi/10.1103/PhysRevA.76.043420}%
  \bibAnnoteFile{NoStop}{Palacios2007a}%
\bibitem{Palacios2009}%
  \BibitemOpen
  \bibfield{author}{%
  \bibinfo {author} {\bibfnamefont{A.}~\bibnamefont{Palacios}}, \bibinfo
  {author} {\bibfnamefont{T.~N.}\ \bibnamefont{Rescigno}},\ and\ \bibinfo
  {author} {\bibfnamefont{C.~W.}\ \bibnamefont{McCurdy}},\ }%
  \bibfield{journal}{%
  \Doi{10.1103/PhysRevA.79.033402}{\bibinfo {journal} {Phys. Rev. A}}\ }%
  \textbf{\bibinfo {volume} {79}},\ \bibinfo {pages} {033402} (\bibinfo {month}
  {Mar.}\ \bibinfo {year} {2009}),\ ISSN \bibinfo {issn} {1050-2947},\
  \url{http://link.aps.org/doi/10.1103/PhysRevA.79.033402}%
  \bibAnnoteFile{NoStop}{Palacios2009}%
\bibitem{Kazansky2003}%
  \BibitemOpen
  \bibfield{author}{%
  \bibinfo {author} {\bibfnamefont{A.~K.}\ \bibnamefont{Kazansky}}, \bibinfo
  {author} {\bibfnamefont{P.}~\bibnamefont{Selles}},\ and\ \bibinfo {author}
  {\bibfnamefont{L.}~\bibnamefont{Malegat}},\ }%
  \bibfield{journal}{%
  \Doi{10.1103/PhysRevA.68.052701}{\bibinfo {journal} {Phys. Rev. A}}\ }%
  \textbf{\bibinfo {volume} {68}},\ \bibinfo {pages} {052701} (\bibinfo {month}
  {Nov.}\ \bibinfo {year} {2003}),\ ISSN \bibinfo {issn} {1050-2947},\
  \url{http://link.aps.org/doi/10.1103/PhysRevA.68.052701}%
  \bibAnnoteFile{NoStop}{Kazansky2003}%
\bibitem{Fernandez2009a}%
  \BibitemOpen
  \bibfield{author}{%
  \bibinfo {author} {\bibfnamefont{J.}~\bibnamefont{Fern\'{a}ndez}}\ and\
  \bibinfo {author} {\bibfnamefont{L.~B.}\ \bibnamefont{Madsen}},\ }%
  \bibfield{journal}{%
  \Doi{10.1088/0953-4075/42/8/085602}{\bibinfo {journal} {J. Phys. B: At. Mol.
  Opt. Phys.}}\ }%
  \textbf{\bibinfo {volume} {42}},\ \bibinfo {pages} {085602} (\bibinfo {month}
  {Apr.}\ \bibinfo {year} {2009}),\ ISSN \bibinfo {issn} {0953-4075},\
  \url{http://stacks.iop.org/0953-4075/42/i=8/a=085602?key=crossref.dd08a12f4c%
22eab66e8a422af0f4b69d}%
  \bibAnnoteFile{NoStop}{Fernandez2009a}%
\bibitem{Tao2012}%
  \BibitemOpen
  \bibfield{author}{%
  \bibinfo {author} {\bibfnamefont{L.}~\bibnamefont{Tao}}\ and\ \bibinfo
  {author} {\bibfnamefont{A.}~\bibnamefont{Scrinzi}},\ }%
  \bibfield{journal}{%
  \bibinfo {journal} {New Journal of Physics}\ }%
  \textbf{\bibinfo {volume} {14}},\ \bibinfo {pages} {013021} (\bibinfo {year}
  {2012}),\ \url{http://stacks.iop.org/1367-2630/14/i=1/a=013021}%
  \bibAnnoteFile{NoStop}{Tao2012}%
\bibitem{Scrinzi2012}%
  \BibitemOpen
  \bibfield{author}{%
  \bibinfo {author} {\bibfnamefont{A.}~\bibnamefont{Scrinzi}},\ }%
  \bibfield{journal}{%
  \bibinfo {journal} {New Journal of Physics}\ }%
  \textbf{\bibinfo {volume} {14}},\ \bibinfo {pages} {085008} (\bibinfo {year}
  {2012}),\ \url{http://stacks.iop.org/1367-2630/14/i=8/a=085008}%
  \bibAnnoteFile{NoStop}{Scrinzi2012}%
\bibitem{Fei2009}%
  \BibitemOpen
  \bibfield{author}{%
  \bibinfo {author} {\bibfnamefont{J.}~\bibnamefont{Feist}},\ }%
  \emph{\bibinfo {title} {Two-photon double ionization of helium}},\ Ph.D.
  thesis,\ \bibinfo {school} {Vienna Univ. of Technology} (\bibinfo {year}
  {2009})%
  \bibAnnoteFile{NoStop}{Fei2009}%
\bibitem{SchFeiNag2011}%
  \BibitemOpen
  \bibfield{author}{%
  \bibinfo {author} {\bibfnamefont{B.~I.}\ \bibnamefont{Schneider}}, \bibinfo
  {author} {\bibfnamefont{J.}~\bibnamefont{Feist}}, \bibinfo {author}
  {\bibfnamefont{S.}~\bibnamefont{Nagele}}, \bibinfo {author}
  {\bibfnamefont{R.}~\bibnamefont{Pazourek}}, \bibinfo {author}
  {\bibfnamefont{S.~X.}\ \bibnamefont{Hu}}, \bibinfo {author}
  {\bibfnamefont{L.~A.}\ \bibnamefont{Collins}},\ and\ \bibinfo {author}
  {\bibfnamefont{J.}~\bibnamefont{Burgd\"{o}rfer}},\ }%
  in\ \Doi{10.1007/978-1-4419-9491-2\_10}{\emph{\bibinfo {booktitle} {Quantum
  Dynamic Imaging}}},\ \bibinfo {series and number} {CRM Series in Mathematical
  Physics},\ \bibinfo {editor} {edited by\ \bibinfo {editor}
  {\bibfnamefont{A.~D.}\ \bibnamefont{Bandrauk}}\ and\ \bibinfo {editor}
  {\bibfnamefont{M.}~\bibnamefont{Ivanov}}}\ (\bibinfo {publisher} {Springer},\
  \bibinfo {year} {2011})\ Chap.~\bibinfo {chapter} {10}%
  \bibAnnoteFile{NoStop}{SchFeiNag2011}%
\bibitem{ColPin2002}%
  \BibitemOpen
  \bibfield{author}{%
  \bibinfo {author} {\bibfnamefont{J.}~\bibnamefont{Colgan}}\ and\ \bibinfo
  {author} {\bibfnamefont{M.~S.}\ \bibnamefont{Pindzola}},\ }%
  \bibfield{journal}{%
  \Doi{10.1103/PhysRevLett.88.173002}{\bibinfo {journal} {Phys. Rev. Lett.}}\
  }%
  \textbf{\bibinfo {volume} {88}},\ \bibinfo {pages} {173002} (\bibinfo {year}
  {2002})%
  \bibAnnoteFile{NoStop}{ColPin2002}%
\bibitem{LauBac2003}%
  \BibitemOpen
  \bibfield{author}{%
  \bibinfo {author} {\bibfnamefont{S.}~\bibnamefont{Laulan}}\ and\ \bibinfo
  {author} {\bibfnamefont{H.}~\bibnamefont{Bachau}},\ }%
  \bibfield{journal}{%
  \Doi{10.1103/PhysRevA.68.013409}{\bibinfo {journal} {Phys. Rev. A}}\ }%
  \textbf{\bibinfo {volume} {68}},\ \bibinfo {pages} {013409} (\bibinfo {month}
  {Jul.}\ \bibinfo {year} {2003})%
  \bibAnnoteFile{NoStop}{LauBac2003}%
\bibitem{HuColCol2005}%
  \BibitemOpen
  \bibfield{author}{%
  \bibinfo {author} {\bibfnamefont{S.~X.}\ \bibnamefont{Hu}}, \bibinfo {author}
  {\bibfnamefont{J.}~\bibnamefont{Colgan}},\ and\ \bibinfo {author}
  {\bibfnamefont{L.~A.}\ \bibnamefont{Collins}},\ }%
  \bibfield{journal}{%
  \Doi{10.1088/0953-4075/38/1/L05}{\bibinfo {journal} {J. Phys. B}}\ }%
  \textbf{\bibinfo {volume} {38}},\ \bibinfo {pages} {L35} (\bibinfo {year}
  {2005})%
  \bibAnnoteFile{NoStop}{HuColCol2005}%
\bibitem{PinRobLoc2007}%
  \BibitemOpen
  \bibfield{author}{%
  \bibinfo {author} {\bibfnamefont{M.~S.}\ \bibnamefont{Pindzola}}, \bibinfo
  {author} {\bibfnamefont{F.}~\bibnamefont{Robicheaux}}, \bibinfo {author}
  {\bibfnamefont{S.~D.}\ \bibnamefont{Loch}}, \bibinfo {author}
  {\bibfnamefont{J.~C.}\ \bibnamefont{Berengut}}, \bibinfo {author}
  {\bibfnamefont{T.}~\bibnamefont{Topcu}}, \bibinfo {author}
  {\bibfnamefont{J.}~\bibnamefont{Colgan}}, \bibinfo {author}
  {\bibfnamefont{M.}~\bibnamefont{Foster}}, \bibinfo {author}
  {\bibfnamefont{D.~C.}\ \bibnamefont{Griffin}}, \bibinfo {author}
  {\bibfnamefont{C.~P.}\ \bibnamefont{Ballance}}, \bibinfo {author}
  {\bibfnamefont{D.~R.}\ \bibnamefont{Schultz}}, \bibinfo {author}
  {\bibfnamefont{T.}~\bibnamefont{Minami}}, \bibinfo {author}
  {\bibfnamefont{N.~R.}\ \bibnamefont{Badnell}}, \bibinfo {author}
  {\bibfnamefont{M.~C.}\ \bibnamefont{Witthoeft}}, \bibinfo {author}
  {\bibfnamefont{D.~R.}\ \bibnamefont{Plante}}, \bibinfo {author}
  {\bibfnamefont{D.~M.}\ \bibnamefont{Mitnik}}, \bibinfo {author}
  {\bibfnamefont{J.~A.}\ \bibnamefont{Ludlow}},\ and\ \bibinfo {author}
  {\bibfnamefont{U.}~\bibnamefont{Kleiman}},\ }%
  \bibfield{journal}{%
  \Doi{10.1088/0953-4075/40/7/R01}{\bibinfo {journal} {J. Phys. B}}\ }%
  \textbf{\bibinfo {volume} {40}},\ \bibinfo {pages} {R39} (\bibinfo {year}
  {2007})%
  \bibAnnoteFile{NoStop}{PinRobLoc2007}%
\bibitem{ResMcc2000}%
  \BibitemOpen
  \bibfield{author}{%
  \bibinfo {author} {\bibfnamefont{T.~N.}\ \bibnamefont{Rescigno}}\ and\
  \bibinfo {author} {\bibfnamefont{C.~W.}\ \bibnamefont{McCurdy}},\ }%
  \bibfield{journal}{%
  \Doi{10.1103/PhysRevA.62.032706}{\bibinfo {journal} {Phys. Rev. A}}\ }%
  \textbf{\bibinfo {volume} {62}},\ \bibinfo {pages} {032706} (\bibinfo {year}
  {2000})%
  \bibAnnoteFile{NoStop}{ResMcc2000}%
\bibitem{MccHorRes2001}%
  \BibitemOpen
  \bibfield{author}{%
  \bibinfo {author} {\bibfnamefont{C.~W.}\ \bibnamefont{McCurdy}}, \bibinfo
  {author} {\bibfnamefont{D.~A.}\ \bibnamefont{Horner}},\ and\ \bibinfo
  {author} {\bibfnamefont{T.~N.}\ \bibnamefont{Rescigno}},\ }%
  \bibfield{journal}{%
  \Doi{10.1103/PhysRevA.63.022711}{\bibinfo {journal} {Phys. Rev. A}}\ }%
  \textbf{\bibinfo {volume} {63}},\ \bibinfo {pages} {022711} (\bibinfo {year}
  {2001})%
  \bibAnnoteFile{NoStop}{MccHorRes2001}%
\bibitem{SchColHu2006}%
  \BibitemOpen
  \bibfield{author}{%
  \bibinfo {author} {\bibfnamefont{B.~I.}\ \bibnamefont{Schneider}}, \bibinfo
  {author} {\bibfnamefont{L.~A.}\ \bibnamefont{Collins}},\ and\ \bibinfo
  {author} {\bibfnamefont{S.~X.}\ \bibnamefont{Hu}},\ }%
  \bibfield{journal}{%
  \Doi{10.1103/PhysRevE.73.036708}{\bibinfo {journal} {Phys. Rev. E}}\ }%
  \textbf{\bibinfo {volume} {73}},\ \bibinfo {pages} {036708} (\bibinfo {year}
  {2006})%
  \bibAnnoteFile{NoStop}{SchColHu2006}%
\bibitem{ParkLight86}%
  \BibitemOpen
  \bibfield{author}{%
  \bibinfo {author} {\bibfnamefont{T.~J.}\ \bibnamefont{Park}}\ and\ \bibinfo
  {author} {\bibfnamefont{J.~C.}\ \bibnamefont{Light}},\ }%
  \bibfield{journal}{%
  \Doi{10.1063/1.451548}{\bibinfo {journal} {J. Chem. Phys.}}\ }%
  \textbf{\bibinfo {volume} {85}},\ \bibinfo {pages} {5870} (\bibinfo {year}
  {1986})%
  \bibAnnoteFile{NoStop}{ParkLight86}%
\bibitem{SmyParTay1998}%
  \BibitemOpen
  \bibfield{author}{%
  \bibinfo {author} {\bibfnamefont{E.~S.}\ \bibnamefont{Smyth}}, \bibinfo
  {author} {\bibfnamefont{J.~S.}\ \bibnamefont{Parker}},\ and\ \bibinfo
  {author} {\bibfnamefont{K.~T.}\ \bibnamefont{Taylor}},\ }%
  \bibfield{journal}{%
  \bibinfo {journal} {Comput. Phys. Commun.}\ }%
  \textbf{\bibinfo {volume} {114}},\ \bibinfo {pages} {1} (\bibinfo {month}
  {Nov.}\ \bibinfo {year} {1998})%
  \bibAnnoteFile{NoStop}{SmyParTay1998}%
\bibitem{Lefo90}%
  \BibitemOpen
  \bibfield{author}{%
  \bibinfo {author} {\bibfnamefont{C.}~\bibnamefont{Leforestier}}, \bibinfo
  {author} {\bibfnamefont{R.~H.}\ \bibnamefont{Bisseling}}, \bibinfo {author}
  {\bibfnamefont{C.}~\bibnamefont{Cerjan}}, \bibinfo {author}
  {\bibfnamefont{M.~D.}\ \bibnamefont{Feit}}, \bibinfo {author}
  {\bibfnamefont{R.}~\bibnamefont{Friesner}}, \bibinfo {author}
  {\bibfnamefont{A.}~\bibnamefont{Guldberg}}, \bibinfo {author}
  {\bibfnamefont{A.}~\bibnamefont{Hammerich}}, \bibinfo {author}
  {\bibfnamefont{G.}~\bibnamefont{Jolicard}}, \bibinfo {author}
  {\bibfnamefont{W.}~\bibnamefont{Karrlein}}, \bibinfo {author}
  {\bibfnamefont{H.-D.}\ \bibnamefont{Meyer}}, \bibinfo {author}
  {\bibfnamefont{N.}~\bibnamefont{Lipkin}}, \bibinfo {author}
  {\bibfnamefont{O.}~\bibnamefont{Roncero}},\ and\ \bibinfo {author}
  {\bibfnamefont{R.}~\bibnamefont{Kosloff}},\ }%
  \bibfield{journal}{%
  \Doi{10.1016/0021-9991(91)90137-A}{\bibinfo {journal} {J. Comp. Phys.}}\ }%
  \textbf{\bibinfo {volume} {94}},\ \bibinfo {pages} {59} (\bibinfo {year}
  {1991})%
  \bibAnnoteFile{NoStop}{Lefo90}%
\bibitem{HerRomVid2005}%
  \BibitemOpen
  \bibfield{author}{%
  \bibinfo {author} {\bibfnamefont{V.}~\bibnamefont{Hernandez}}, \bibinfo
  {author} {\bibfnamefont{J.~E.}\ \bibnamefont{Roman}},\ and\ \bibinfo {author}
  {\bibfnamefont{V.}~\bibnamefont{Vidal}},\ }%
  \bibfield{journal}{%
  \Doi{10.1145/1089014.1089019}{\bibinfo {journal} {ACM Trans. Math. Softw.}}\
  }%
  \textbf{\bibinfo {volume} {31}},\ \bibinfo {pages} {351} (\bibinfo {month}
  {Sep.}\ \bibinfo {year} {2005})%
  \bibAnnoteFile{NoStop}{HerRomVid2005}%
\bibitem{Newton}%
  \BibitemOpen
  \bibfield{author}{%
  \bibinfo {author} {\bibfnamefont{R.~G.}\ \bibnamefont{Newton}},\ }%
  \emph{\bibinfo {title} {Scattering Theory of Waves and Particles}}\ (\bibinfo
  {publisher} {McGraw-Hill},\ \bibinfo {address} {New York},\ \bibinfo {year}
  {1966})%
  \bibAnnoteFile{NoStop}{Newton}%
\bibitem{Breit1954}%
  \BibitemOpen
  \bibfield{author}{%
  \bibinfo {author} {\bibfnamefont{G.}~\bibnamefont{Breit}}\ and\ \bibinfo
  {author} {\bibfnamefont{H.~A.}\ \bibnamefont{Bethe}},\ }%
  \bibfield{journal}{%
  \Doi{10.1103/PhysRev.93.888}{\bibinfo {journal} {Phys. Rev.}}\ }%
  \textbf{\bibinfo {volume} {93}},\ \bibinfo {pages} {888} (\bibinfo {month}
  {Feb}\ \bibinfo {year} {1954}),\
  \url{http://link.aps.org/doi/10.1103/PhysRev.93.888}%
  \bibAnnoteFile{NoStop}{Breit1954}%
\bibitem{Altshuler1956}%
  \BibitemOpen
  \bibfield{author}{%
  \bibinfo {author} {\bibfnamefont{S.}~\bibnamefont{Altshuler}},\ }%
  \bibfield{journal}{%
  \bibinfo {journal} {Il Nuovo Cimento (1955-1965)}\ }%
  \textbf{\bibinfo {volume} {3}},\ \bibinfo {pages} {246} (\bibinfo {year}
  {1956})%
  \bibAnnoteFile{NoStop}{Altshuler1956}%
\bibitem{pra.14.2159}%
  \BibitemOpen
  \bibfield{author}{%
  \bibinfo {author} {\bibfnamefont{J.~T.}\ \bibnamefont{Broad}}\ and\ \bibinfo
  {author} {\bibfnamefont{W.~P.}\ \bibnamefont{Reinhardt}},\ }%
  \bibfield{journal}{%
  \bibinfo {journal} {Phys. Rev. A}\ }%
  \textbf{\bibinfo {volume} {14}},\ \bibinfo {pages} {2159} (\bibinfo {year}
  {1976})%
  \bibAnnoteFile{NoStop}{pra.14.2159}%
\bibitem{pra.43.3474}%
  \BibitemOpen
  \bibfield{author}{%
  \bibinfo {author} {\bibfnamefont{M.~L.}\ \bibnamefont{Du}}\ and\ \bibinfo
  {author} {\bibfnamefont{A.}~\bibnamefont{Dalgarno}},\ }%
  \bibfield{journal}{%
  \bibinfo {journal} {Phys. Rev. A}\ }%
  \textbf{\bibinfo {volume} {43}},\ \bibinfo {pages} {3474} (\bibinfo {year}
  {1991})%
  \bibAnnoteFile{NoStop}{pra.43.3474}%
\bibitem{pra.43.1301}%
  \BibitemOpen
  \bibfield{author}{%
  \bibinfo {author} {\bibfnamefont{I.}~\bibnamefont{Bray}}, \bibinfo {author}
  {\bibfnamefont{D.~A.}\ \bibnamefont{Konovalov}},\ and\ \bibinfo {author}
  {\bibfnamefont{I.~E.}\ \bibnamefont{McCarthy}},\ }%
  \bibfield{journal}{%
  \bibinfo {journal} {Phys. Rev. A}\ }%
  \textbf{\bibinfo {volume} {43}},\ \bibinfo {pages} {1301} (\bibinfo {year}
  {1991})%
  \bibAnnoteFile{NoStop}{pra.43.1301}%
\bibitem{Moccia1991}%
  \BibitemOpen
  \bibfield{author}{%
  \bibinfo {author} {\bibfnamefont{R.}~\bibnamefont{Moccia}}\ and\ \bibinfo
  {author} {\bibfnamefont{P.}~\bibnamefont{Spizzo}},\ }%
  \bibfield{journal}{%
  \Doi{10.1103/PhysRevA.43.2199}{\bibinfo {journal} {Phys. Rev. A}}\ }%
  \textbf{\bibinfo {volume} {43}},\ \bibinfo {pages} {2199} (\bibinfo {month}
  {Mar}\ \bibinfo {year} {1991}),\
  \url{http://link.aps.org/doi/10.1103/PhysRevA.43.2199}%
  \bibAnnoteFile{NoStop}{Moccia1991}%
\bibitem{pra.44.R13}%
  \BibitemOpen
  \bibfield{author}{%
  \bibinfo {author} {\bibfnamefont{I.}~\bibnamefont{S{\'a}nchez}}\ and\
  \bibinfo {author} {\bibfnamefont{F.}~\bibnamefont{Mart\'in}},\ }%
  \bibfield{journal}{%
  \bibinfo {journal} {Phys. Rev. A}\ }%
  \textbf{\bibinfo {volume} {44}},\ \bibinfo {pages} {13(R)} (\bibinfo {year}
  {1991})%
  \bibAnnoteFile{NoStop}{pra.44.R13}%
\bibitem{Burke_book}%
  \BibitemOpen
  \bibfield{author}{%
  \bibinfo {author} {\bibfnamefont{P.~G.}\ \bibnamefont{Burke}},\ }%
  \emph{\bibinfo {title} {R-Matrix Theory of Atomic Collisions}}\ (\bibinfo
  {publisher} {Springer-Verlag},\ \bibinfo {address} {Heidelberg},\ \bibinfo
  {year} {2011})%
  \bibAnnoteFile{NoStop}{Burke_book}%
\bibitem{Vanroose2002}%
  \BibitemOpen
  \bibfield{author}{%
  \bibinfo {author} {\bibfnamefont{W.}~\bibnamefont{Vanroose}}, \bibinfo
  {author} {\bibfnamefont{J.}~\bibnamefont{Broeckhove}},\ and\ \bibinfo
  {author} {\bibfnamefont{F.}~\bibnamefont{Arickx}},\ }%
  \bibfield{journal}{%
  \Doi{10.1103/PhysRevLett.88.010404}{\bibinfo {journal} {Phys. Rev. Lett.}}\
  }%
  \textbf{\bibinfo {volume} {88}},\ \bibinfo {pages} {010404} (\bibinfo {month}
  {Dec.}\ \bibinfo {year} {2001}),\ ISSN \bibinfo {issn} {0031-9007},\
  \url{http://link.aps.org/doi/10.1103/PhysRevLett.88.010404}%
  \bibAnnoteFile{NoStop}{Vanroose2002}%
\bibitem{Cacelli1991}%
  \BibitemOpen
  \bibfield{author}{%
  \bibinfo {author} {\bibfnamefont{I.}~\bibnamefont{Cacelli}}, \bibinfo
  {author} {\bibfnamefont{V.}~\bibnamefont{Carravetta}}, \bibinfo {author}
  {\bibfnamefont{A.}~\bibnamefont{Rizzo}},\ and\ \bibinfo {author}
  {\bibfnamefont{R.}~\bibnamefont{Moccia}},\ }%
  \bibfield{journal}{%
  \bibinfo {journal} {Phys. Rep.}\ }%
  \textbf{\bibinfo {volume} {205}},\ \bibinfo {pages} {283} (\bibinfo {year}
  {1991})%
  \bibAnnoteFile{NoStop}{Cacelli1991}%
\bibitem{Argenti2006}%
  \BibitemOpen
  \bibfield{author}{%
  \bibinfo {author} {\bibfnamefont{L.}~\bibnamefont{Argenti}}\ and\ \bibinfo
  {author} {\bibfnamefont{R.}~\bibnamefont{Moccia}},\ }%
  \bibfield{journal}{%
  \Doi{10.1088/0953-4075/39/12/012}{\bibinfo {journal} {J. Phys. B: At. Mol.
  Opt. Phys.}}\ }%
  \textbf{\bibinfo {volume} {39}},\ \bibinfo {pages} {2773} (\bibinfo {year}
  {2006})%
  \bibAnnoteFile{NoStop}{Argenti2006}%
\bibitem{Martin1993}%
  \BibitemOpen
  \bibfield{author}{%
  \bibinfo {author} {\bibfnamefont{F.}~\bibnamefont{Mart\'in}},\ }%
  \bibfield{journal}{%
  \bibinfo {journal} {Phys. Rev. A}\ }%
  \textbf{\bibinfo {volume} {48}},\ \bibinfo {pages} {331} (\bibinfo {year}
  {1993})%
  \bibAnnoteFile{NoStop}{Martin1993}%
\bibitem{Venuti1996}%
  \BibitemOpen
  \bibfield{author}{%
  \bibinfo {author} {\bibfnamefont{M.}~\bibnamefont{Venuti}}, \bibinfo {author}
  {\bibfnamefont{P.}~\bibnamefont{Decleva}},\ and\ \bibinfo {author}
  {\bibfnamefont{A.}~\bibnamefont{Lisini}},\ }%
  \bibfield{journal}{%
  \bibinfo {journal} {J. Phys. B: At. Mol. Opt. Phys.}\ }%
  \textbf{\bibinfo {volume} {29}},\ \bibinfo {pages} {5315} (\bibinfo {year}
  {1996})%
  \bibAnnoteFile{NoStop}{Venuti1996}%
\bibitem{petsc-web-page}%
  \BibitemOpen
  \bibfield{author}{%
  \bibinfo {author} {\bibfnamefont{S.}~\bibnamefont{Balay}}, \bibinfo {author}
  {\bibfnamefont{J.}~\bibnamefont{Brown}}, \bibinfo {author}
  {\bibfnamefont{K.}~\bibnamefont{Buschelman}}, \bibinfo {author}
  {\bibfnamefont{W.~D.}\ \bibnamefont{Gropp}}, \bibinfo {author}
  {\bibfnamefont{D.}~\bibnamefont{Kaushik}}, \bibinfo {author}
  {\bibfnamefont{M.~G.}\ \bibnamefont{Knepley}}, \bibinfo {author}
  {\bibfnamefont{L.~C.}\ \bibnamefont{McInnes}}, \bibinfo {author}
  {\bibfnamefont{B.~F.}\ \bibnamefont{Smith}},\ and\ \bibinfo {author}
  {\bibfnamefont{H.}~\bibnamefont{Zhang}},\ }%
  \enquote{\bibinfo {title} {{PETSc} {W}eb page},}\  (\bibinfo {year} {2011})%
  \bibAnnoteFile{NoStop}{petsc-web-page}%
\bibitem{Brauner1989}%
  \BibitemOpen
  \bibfield{author}{%
  \bibinfo {author} {\bibfnamefont{M.}~\bibnamefont{Brauner}}, \bibinfo
  {author} {\bibfnamefont{J.~S.}\ \bibnamefont{Briggs}},\ and\ \bibinfo
  {author} {\bibfnamefont{H.}~\bibnamefont{Klar}},\ }%
  \bibfield{journal}{%
  \bibinfo {journal} {J. Phys. B: At. Mol. Opt. Phys.}\ }%
  \textbf{\bibinfo {volume} {22}},\ \bibinfo {pages} {2265} (\bibinfo {year}
  {1989})%
  \bibAnnoteFile{NoStop}{Brauner1989}%
\bibitem{Drake1969}%
  \BibitemOpen
  \bibfield{author}{%
  \bibinfo {author} {\bibfnamefont{G.~W.~F.}\ \bibnamefont{Drake}}, \bibinfo
  {author} {\bibfnamefont{G.~A.}\ \bibnamefont{Victor}},\ and\ \bibinfo
  {author} {\bibfnamefont{A.}~\bibnamefont{Dalgarno}},\ }%
  \bibfield{journal}{%
  \bibinfo {journal} {Phys. Rev.}\ }%
  \textbf{\bibinfo {volume} {180}},\ \bibinfo {pages} {25} (\bibinfo {year}
  {1969})%
  \bibAnnoteFile{NoStop}{Drake1969}%
\bibitem{Rost1997}%
  \BibitemOpen
  \bibfield{author}{%
  \bibinfo {author} {\bibfnamefont{J.~M.}\ \bibnamefont{Rost}}, \bibinfo
  {author} {\bibfnamefont{K.}~\bibnamefont{Schulz}}, \bibinfo {author}
  {\bibfnamefont{M.}~\bibnamefont{Domke}},\ and\ \bibinfo {author}
  {\bibfnamefont{G.}~\bibnamefont{Kaindl}},\ }%
  \bibfield{journal}{%
  \bibinfo {journal} {Journal of Physics B: Atomic, Molecular and Optical
  Physics}\ }%
  \textbf{\bibinfo {volume} {30}},\ \bibinfo {pages} {4663} (\bibinfo {year}
  {1997}),\ \url{http://stacks.iop.org/0953-4075/30/i=21/a=010}%
  \bibAnnoteFile{NoStop}{Rost1997}%
\bibitem{TanRicRos2000}%
  \BibitemOpen
  \bibfield{author}{%
  \bibinfo {author} {\bibfnamefont{G.}~\bibnamefont{Tanner}}, \bibinfo {author}
  {\bibfnamefont{K.}~\bibnamefont{Richter}},\ and\ \bibinfo {author}
  {\bibfnamefont{J.~M.}\ \bibnamefont{Rost}},\ }%
  \bibfield{journal}{%
  \Doi{10.1103/RevModPhys.72.497}{\bibinfo {journal} {Rev. Mod. Phys}}\ }%
  \textbf{\bibinfo {volume} {72}},\ \bibinfo {pages} {497} (\bibinfo {year}
  {2000})%
  \bibAnnoteFile{NoStop}{TanRicRos2000}%
\bibitem{DomXuePus1991}%
  \BibitemOpen
  \bibfield{author}{%
  \bibinfo {author} {\bibfnamefont{M.}~\bibnamefont{Domke}}, \bibinfo {author}
  {\bibfnamefont{C.}~\bibnamefont{Xue}}, \bibinfo {author}
  {\bibfnamefont{A.}~\bibnamefont{Puschmann}}, \bibinfo {author}
  {\bibfnamefont{T.}~\bibnamefont{Mandel}}, \bibinfo {author}
  {\bibfnamefont{E.}~\bibnamefont{Hudson}}, \bibinfo {author}
  {\bibfnamefont{D.~A.}\ \bibnamefont{Shirley}}, \bibinfo {author}
  {\bibfnamefont{G.}~\bibnamefont{Kaindl}}, \bibinfo {author}
  {\bibfnamefont{C.~H.}\ \bibnamefont{Greene}}, \bibinfo {author}
  {\bibfnamefont{H.~R.}\ \bibnamefont{Sadeghpour}},\ and\ \bibinfo {author}
  {\bibfnamefont{H.}~\bibnamefont{Petersen}},\ }%
  \bibfield{journal}{%
  \Doi{10.1103/PhysRevLett.66.1306}{\bibinfo {journal} {Phys. Rev. Lett.}}\ }%
  \textbf{\bibinfo {volume} {66}},\ \bibinfo {pages} {1306} (\bibinfo {month}
  {Mar.}\ \bibinfo {year} {1991})%
  \bibAnnoteFile{NoStop}{DomXuePus1991}%
\bibitem{BizWuiDhe1982}%
  \BibitemOpen
  \bibfield{author}{%
  \bibinfo {author} {\bibfnamefont{J.~M.}\ \bibnamefont{Bizau}}, \bibinfo
  {author} {\bibfnamefont{F.}~\bibnamefont{Wuilleumier}}, \bibinfo {author}
  {\bibfnamefont{P.}~\bibnamefont{Dhez}}, \bibinfo {author}
  {\bibfnamefont{D.~L.}\ \bibnamefont{Ederer}}, \bibinfo {author}
  {\bibfnamefont{T.~N.}\ \bibnamefont{Chang}}, \bibinfo {author}
  {\bibfnamefont{S.}~\bibnamefont{Krummacher}},\ and\ \bibinfo {author}
  {\bibfnamefont{V.}~\bibnamefont{Schmidt}},\ }%
  \bibfield{journal}{%
  \Doi{10.1103/PhysRevLett.48.588}{\bibinfo {journal} {Phys. Rev. Lett.}}\ }%
  \textbf{\bibinfo {volume} {48}},\ \bibinfo {pages} {588} (\bibinfo {month}
  {Mar.}\ \bibinfo {year} {1982})%
  \bibAnnoteFile{NoStop}{BizWuiDhe1982}%
\bibitem{SanMar1991}%
  \BibitemOpen
  \bibfield{author}{%
  \bibinfo {author} {\bibfnamefont{I.}~\bibnamefont{S\'{a}nchez}}\ and\
  \bibinfo {author} {\bibfnamefont{F.}~\bibnamefont{Mart\'in}},\ }%
  \bibfield{journal}{%
  \Doi{10.1103/PhysRevA.44.7318}{\bibinfo {journal} {Phys. Rev. A}}\ }%
  \textbf{\bibinfo {volume} {44}},\ \bibinfo {pages} {7318} (\bibinfo {month}
  {Dec.}\ \bibinfo {year} {1991})%
  \bibAnnoteFile{NoStop}{SanMar1991}%
\bibitem{SanMar1992}%
  \BibitemOpen
  \bibfield{author}{%
  \bibinfo {author} {\bibfnamefont{I.}~\bibnamefont{S\'{a}nchez}}\ and\
  \bibinfo {author} {\bibfnamefont{F.}~\bibnamefont{Mart\'in}},\ }%
  \bibfield{journal}{%
  \Doi{10.1103/PhysRevA.45.4468}{\bibinfo {journal} {Phys. Rev. A}}\ }%
  \textbf{\bibinfo {volume} {45}},\ \bibinfo {pages} {4468} (\bibinfo {month}
  {Apr.}\ \bibinfo {year} {1992})%
  \bibAnnoteFile{NoStop}{SanMar1992}%
\bibitem{MenFriWhi1996}%
  \BibitemOpen
  \bibfield{author}{%
  \bibinfo {author} {\bibfnamefont{A.}~\bibnamefont{Menzel}}, \bibinfo {author}
  {\bibfnamefont{S.~P.}\ \bibnamefont{Frigo}}, \bibinfo {author}
  {\bibfnamefont{S.~B.}\ \bibnamefont{Whitfield}}, \bibinfo {author}
  {\bibfnamefont{C.~D.}\ \bibnamefont{Caldwell}},\ and\ \bibinfo {author}
  {\bibfnamefont{M.~O.}\ \bibnamefont{Krause}},\ }%
  \bibfield{journal}{%
  \Doi{10.1103/PhysRevA.54.2080}{\bibinfo {journal} {Phys. Rev. A}}\ }%
  \textbf{\bibinfo {volume} {54}},\ \bibinfo {pages} {2080} (\bibinfo {month}
  {Sep.}\ \bibinfo {year} {1996})%
  \bibAnnoteFile{NoStop}{MenFriWhi1996}%
\bibitem{ZubDawHal1991}%
  \BibitemOpen
  \bibfield{author}{%
  \bibinfo {author} {\bibfnamefont{M.}~\bibnamefont{Zubek}}, \bibinfo {author}
  {\bibfnamefont{G.}~\bibnamefont{Dawber}}, \bibinfo {author}
  {\bibfnamefont{R.~I.}\ \bibnamefont{Hall}}, \bibinfo {author}
  {\bibfnamefont{L.}~\bibnamefont{Avaldi}}, \bibinfo {author}
  {\bibfnamefont{K.}~\bibnamefont{Ellis}},\ and\ \bibinfo {author}
  {\bibfnamefont{G.~C.}\ \bibnamefont{King}},\ }%
  \bibfield{journal}{%
  \Doi{10.1088/0953-4075/24/14/002}{\bibinfo {journal} {J. Phys. B}}\ }%
  \textbf{\bibinfo {volume} {24}},\ \bibinfo {pages} {L337} (\bibinfo {month}
  {Jan.}\ \bibinfo {year} {1999})%
  \bibAnnoteFile{NoStop}{ZubDawHal1991}%
\bibitem{LinFerBec1985}%
  \BibitemOpen
  \bibfield{author}{%
  \bibinfo {author} {\bibfnamefont{D.~W.}\ \bibnamefont{Lindle}}, \bibinfo
  {author} {\bibfnamefont{T.~A.}\ \bibnamefont{Ferrett}}, \bibinfo {author}
  {\bibfnamefont{U.}~\bibnamefont{Becker}}, \bibinfo {author}
  {\bibfnamefont{P.~H.}\ \bibnamefont{Kobrin}}, \bibinfo {author}
  {\bibfnamefont{C.~M.}\ \bibnamefont{Truesdale}}, \bibinfo {author}
  {\bibfnamefont{H.~G.}\ \bibnamefont{Kerkhoff}},\ and\ \bibinfo {author}
  {\bibfnamefont{D.~A.}\ \bibnamefont{Shirley}},\ }%
  \bibfield{journal}{%
  \Doi{10.1103/PhysRevA.31.714}{\bibinfo {journal} {Phys. Rev. A}}\ }%
  \textbf{\bibinfo {volume} {31}},\ \bibinfo {pages} {714} (\bibinfo {month}
  {Feb.}\ \bibinfo {year} {1985})%
  \bibAnnoteFile{NoStop}{LinFerBec1985}%
\bibitem{VenDecLis1996}%
  \BibitemOpen
  \bibfield{author}{%
  \bibinfo {author} {\bibfnamefont{M.}~\bibnamefont{Venuti}}, \bibinfo {author}
  {\bibfnamefont{P.}~\bibnamefont{Decleva}},\ and\ \bibinfo {author}
  {\bibfnamefont{A.}~\bibnamefont{Lisini}},\ }%
  \bibfield{journal}{%
  \Doi{10.1088/0953-4075/29/22/011}{\bibinfo {journal} {J. Phys. B}}\ }%
  \textbf{\bibinfo {volume} {29}},\ \bibinfo {pages} {5315} (\bibinfo {month}
  {Jan.}\ \bibinfo {year} {1999})%
  \bibAnnoteFile{NoStop}{VenDecLis1996}%
\bibitem{Sanchez1993}%
  \BibitemOpen
  \bibfield{author}{%
  \bibinfo {author} {\bibfnamefont{I.}~\bibnamefont{S\'anchez}}\ and\ \bibinfo
  {author} {\bibfnamefont{F.}~\bibnamefont{Mart\'in}},\ }%
  \bibfield{journal}{%
  \Doi{10.1103/PhysRevA.47.1520}{\bibinfo {journal} {Phys. Rev. A}}\ }%
  \textbf{\bibinfo {volume} {47}},\ \bibinfo {pages} {1520} (\bibinfo {month}
  {Feb}\ \bibinfo {year} {1993}),\
  \url{http://link.aps.org/doi/10.1103/PhysRevA.47.1520}%
  \bibAnnoteFile{NoStop}{Sanchez1993}%
\bibitem{FanCha2000}%
  \BibitemOpen
  \bibfield{author}{%
  \bibinfo {author} {\bibfnamefont{T.~K.}\ \bibnamefont{Fang}}\ and\ \bibinfo
  {author} {\bibfnamefont{T.~N.}\ \bibnamefont{Chang}},\ }%
  \bibfield{journal}{%
  \Doi{10.1103/PhysRevA.61.062704}{\bibinfo {journal} {Phys. Rev. A}}\ }%
  \textbf{\bibinfo {volume} {61}},\ \bibinfo {pages} {062704} (\bibinfo {month}
  {May}\ \bibinfo {year} {2000})%
  \bibAnnoteFile{NoStop}{FanCha2000}%
\bibitem{ZhoLin1994}%
  \BibitemOpen
  \bibfield{author}{%
  \bibinfo {author} {\bibfnamefont{B.}~\bibnamefont{Zhou}}\ and\ \bibinfo
  {author} {\bibfnamefont{C.~D.}\ \bibnamefont{Lin}},\ }%
  \bibfield{journal}{%
  \Doi{10.1103/PhysRevA.49.1057}{\bibinfo {journal} {Phys. Rev. A}}\ }%
  \textbf{\bibinfo {volume} {49}},\ \bibinfo {pages} {1057} (\bibinfo {month}
  {Feb.}\ \bibinfo {year} {1994})%
  \bibAnnoteFile{NoStop}{ZhoLin1994}%
\bibitem{Alagia2009}%
  \BibitemOpen
  \bibfield{author}{%
  \bibinfo {author} {\bibfnamefont{M.}~\bibnamefont{Alagia}}, \bibinfo {author}
  {\bibfnamefont{M.}~\bibnamefont{Coreno}}, \bibinfo {author}
  {\bibfnamefont{H.}~\bibnamefont{Farrokhpour}}, \bibinfo {author}
  {\bibfnamefont{P.}~\bibnamefont{Franceschi}}, \bibinfo {author}
  {\bibfnamefont{A.}~\bibnamefont{Miheli\v{c}}}, \bibinfo {author}
  {\bibfnamefont{A.}~\bibnamefont{Moise}}, \bibinfo {author}
  {\bibfnamefont{R.}~\bibnamefont{Omidyan}}, \bibinfo {author}
  {\bibfnamefont{K.~C.}\ \bibnamefont{Prince}}, \bibinfo {author}
  {\bibfnamefont{R.}~\bibnamefont{Richter}}, \bibinfo {author}
  {\bibfnamefont{J.}~\bibnamefont{S\"{o}derstr\"{o}m}}, \bibinfo {author}
  {\bibfnamefont{S.}~\bibnamefont{Stranges}}, \bibinfo {author}
  {\bibfnamefont{M.}~\bibnamefont{Tabrizchi}},\ and\ \bibinfo {author}
  {\bibfnamefont{M.}~\bibnamefont{\v{Z}itnik}},\ }%
  \bibfield{journal}{%
  \Doi{10.1103/PhysRevLett.102.153001}{\bibinfo {journal} {Phys. Rev. Lett.}}\
  }%
  \textbf{\bibinfo {volume} {102}},\ \bibinfo {pages} {153001} (\bibinfo
  {month} {Apr.}\ \bibinfo {year} {2009}),\ ISSN \bibinfo {issn} {0031-9007},\
  \url{http://link.aps.org/doi/10.1103/PhysRevLett.102.153001}%
  \bibAnnoteFile{NoStop}{Alagia2009}%
\bibitem{Lin1986}%
  \BibitemOpen
  \bibfield{author}{%
  \bibinfo {author} {\bibfnamefont{C.~D.}\ \bibnamefont{Lin}},\ }%
  \bibfield{journal}{%
  \Doi{10.1016/S0065-2199(08)60335-8}{\bibinfo {journal} {Adv. At. Mol.
  Phys.}}\ }%
  \textbf{\bibinfo {volume} {22}},\ \bibinfo {pages} {77} (\bibinfo {year}
  {1986})%
  \bibAnnoteFile{NoStop}{Lin1986}%
\bibitem{Lin84}%
  \BibitemOpen
  \bibfield{author}{%
  \bibinfo {author} {\bibfnamefont{C.~D.}\ \bibnamefont{Lin}},\ }%
  \bibfield{journal}{%
  \Doi{10.1103/PhysRevA.29.1019}{\bibinfo {journal} {Phys. Rev. A}}\ }%
  \textbf{\bibinfo {volume} {29}},\ \bibinfo {pages} {1019} (\bibinfo {month}
  {Mar.}\ \bibinfo {year} {1984})%
  \bibAnnoteFile{NoStop}{Lin84}%
\bibitem{HerKelPol1980}%
  \BibitemOpen
  \bibfield{author}{%
  \bibinfo {author} {\bibfnamefont{D.~R.}\ \bibnamefont{Herrick}}, \bibinfo
  {author} {\bibfnamefont{M.~E.}\ \bibnamefont{Kellman}},\ and\ \bibinfo
  {author} {\bibfnamefont{R.~D.}\ \bibnamefont{Poliak}},\ }%
  \bibfield{journal}{%
  \Doi{10.1103/PhysRevA.22.1517}{\bibinfo {journal} {Phys. Rev. A}}\ }%
  \textbf{\bibinfo {volume} {22}},\ \bibinfo {pages} {1517} (\bibinfo {month}
  {Oct.}\ \bibinfo {year} {1980})%
  \bibAnnoteFile{NoStop}{HerKelPol1980}%
\bibitem{FeaBri1986}%
  \BibitemOpen
  \bibfield{author}{%
  \bibinfo {author} {\bibfnamefont{J.~M.}\ \bibnamefont{Feagin}}\ and\ \bibinfo
  {author} {\bibfnamefont{J.~S.}\ \bibnamefont{Briggs}},\ }%
  \bibfield{journal}{%
  \Doi{10.1103/PhysRevLett.57.984}{\bibinfo {journal} {Phys. Rev. Lett.}}\ }%
  \textbf{\bibinfo {volume} {57}},\ \bibinfo {pages} {984} (\bibinfo {month}
  {Aug.}\ \bibinfo {year} {1986})%
  \bibAnnoteFile{NoStop}{FeaBri1986}%
\bibitem{HerSin1975}%
  \BibitemOpen
  \bibfield{author}{%
  \bibinfo {author} {\bibfnamefont{D.~R.}\ \bibnamefont{Herrick}}\ and\
  \bibinfo {author} {\bibfnamefont{O.}~\bibnamefont{Sinano\v{g}lu}},\ }%
  \bibfield{journal}{%
  \Doi{10.1103/PhysRevA.11.97}{\bibinfo {journal} {Phys. Rev. A}}\ }%
  \textbf{\bibinfo {volume} {11}},\ \bibinfo {pages} {97} (\bibinfo {month}
  {Jan.}\ \bibinfo {year} {1975})%
  \bibAnnoteFile{NoStop}{HerSin1975}%
\bibitem{Fano1961}%
  \BibitemOpen
  \bibfield{author}{%
  \bibinfo {author} {\bibfnamefont{U.}~\bibnamefont{Fano}},\ }%
  \bibfield{journal}{%
  \Doi{10.1103/PhysRev.124.1866}{\bibinfo {journal} {Phys. Rev.}}\ }%
  \textbf{\bibinfo {volume} {124}},\ \bibinfo {pages} {1866} (\bibinfo {month}
  {Dec}\ \bibinfo {year} {1961}),\
  \url{http://link.aps.org/doi/10.1103/PhysRev.124.1866}%
  \bibAnnoteFile{NoStop}{Fano1961}%
\bibitem{Wickenhauser2005}%
  \BibitemOpen
  \bibfield{author}{%
  \bibinfo {author} {\bibfnamefont{M.}~\bibnamefont{Wickenhauser}}, \bibinfo
  {author} {\bibfnamefont{J.}~\bibnamefont{Burgd\"{o}rfer}}, \bibinfo {author}
  {\bibfnamefont{F.}~\bibnamefont{Krausz}},\ and\ \bibinfo {author}
  {\bibfnamefont{M.}~\bibnamefont{Drescher}},\ }%
  \bibfield{journal}{%
  \Doi{10.1103/PhysRevLett.94.023002}{\bibinfo {journal} {Phys. Rev. Lett.}}\
  }%
  \textbf{\bibinfo {volume} {94}},\ \bibinfo {pages} {023002} (\bibinfo {year}
  {2005})%
  \bibAnnoteFile{NoStop}{Wickenhauser2005}%
\bibitem{WicBurKra2005}%
  \BibitemOpen
  \bibfield{author}{%
  \bibinfo {author} {\bibfnamefont{M.}~\bibnamefont{Wickenhauser}}, \bibinfo
  {author} {\bibnamefont{Burgd\"{o}rfer}}, \bibinfo {author}
  {\bibfnamefont{F.}~\bibnamefont{Krausz}},\ and\ \bibinfo {author}
  {\bibfnamefont{M.}~\bibnamefont{Drescher}},\ }%
  \bibfield{journal}{%
  \Doi{10.1080/09500340500259870}{\bibinfo {journal} {J. Mod. Opt.}}\ }%
  \textbf{\bibinfo {volume} {53}},\ \bibinfo {pages} {247} (\bibinfo {month}
  {Jan.}\ \bibinfo {year} {2006})%
  \bibAnnoteFile{NoStop}{WicBurKra2005}%
\bibitem{Wic2006}%
  \BibitemOpen
  \bibfield{author}{%
  \bibinfo {author} {\bibfnamefont{M.}~\bibnamefont{Wickenhauser}},\ }%
  \emph{\bibinfo {title} {Ionization dynamics of atoms in femto- and attosecond
  pulses}},\ Ph.D. thesis,\ \bibinfo {school} {Vienna Univ. of Technology}
  (\bibinfo {year} {2006})%
  \bibAnnoteFile{NoStop}{Wic2006}%
\bibitem{MauRemSwo2010}%
  \BibitemOpen
  \bibfield{author}{%
  \bibinfo {author} {\bibfnamefont{J.}~\bibnamefont{Mauritsson}}, \bibinfo
  {author} {\bibfnamefont{T.}~\bibnamefont{Remetter}}, \bibinfo {author}
  {\bibfnamefont{M.}~\bibnamefont{Swoboda}}, \bibinfo {author}
  {\bibfnamefont{K.}~\bibnamefont{Kl\"{u}nder}}, \bibinfo {author}
  {\bibfnamefont{A.}~\bibnamefont{L'Huillier}}, \bibinfo {author}
  {\bibfnamefont{K.~J.}\ \bibnamefont{Schafer}}, \bibinfo {author}
  {\bibfnamefont{O.}~\bibnamefont{Ghafur}}, \bibinfo {author}
  {\bibfnamefont{F.}~\bibnamefont{Kelkensberg}}, \bibinfo {author}
  {\bibfnamefont{W.}~\bibnamefont{Siu}}, \bibinfo {author}
  {\bibfnamefont{P.}~\bibnamefont{Johnsson}}, \bibinfo {author}
  {\bibfnamefont{M.~J.~J.}\ \bibnamefont{Vrakking}}, \bibinfo {author}
  {\bibfnamefont{I.}~\bibnamefont{Znakovskaya}}, \bibinfo {author}
  {\bibfnamefont{T.}~\bibnamefont{Uphues}}, \bibinfo {author}
  {\bibfnamefont{S.}~\bibnamefont{Zherebtsov}}, \bibinfo {author}
  {\bibfnamefont{M.~F.}\ \bibnamefont{Kling}}, \bibinfo {author}
  {\bibfnamefont{F.}~\bibnamefont{L\'{e}pine}}, \bibinfo {author}
  {\bibfnamefont{E.}~\bibnamefont{Benedetti}}, \bibinfo {author}
  {\bibfnamefont{F.}~\bibnamefont{Ferrari}}, \bibinfo {author}
  {\bibfnamefont{G.}~\bibnamefont{Sansone}},\ and\ \bibinfo {author}
  {\bibfnamefont{M.}~\bibnamefont{Nisoli}},\ }%
  \bibfield{journal}{%
  \Doi{10.1103/PhysRevLett.105.053001}{\bibinfo {journal} {Phys. Rev. Lett.}}\
  }%
  \textbf{\bibinfo {volume} {105}},\ \bibinfo {pages} {053001} (\bibinfo
  {month} {Jul.}\ \bibinfo {year} {2010})%
  \bibAnnoteFile{NoStop}{MauRemSwo2010}%
\bibitem{Ott2012}%
  \BibitemOpen
  \bibfield{author}{%
  \bibinfo {author} {\bibfnamefont{C.}~\bibnamefont{Ott}}, \bibinfo {author}
  {\bibfnamefont{A.}~\bibnamefont{Kaldun}}, \bibinfo {author}
  {\bibfnamefont{P.}~\bibnamefont{Raith}}, \bibinfo {author}
  {\bibfnamefont{K.}~\bibnamefont{Meyer}}, \bibinfo {author}
  {\bibfnamefont{M.}~\bibnamefont{Laux}}, \bibinfo {author}
  {\bibfnamefont{Y.}~\bibnamefont{Zhang}}, \bibinfo {author}
  {\bibfnamefont{S.}~\bibnamefont{Hagstotz}}, \bibinfo {author}
  {\bibfnamefont{T.}~\bibnamefont{Ding}}, \bibinfo {author}
  {\bibfnamefont{R.}~\bibnamefont{Heck}},\ and\ \bibinfo {author}
  {\bibfnamefont{T.}~\bibnamefont{Pfeifer}},\ }%
  \bibfield{journal}{%
  \bibinfo {journal} {arXiv [physics.atom-ph]}\ }%
  \textbf{\bibinfo {volume} {05}},\ \bibinfo {pages} {0519v1} (\bibinfo {year}
  {2012})%
  \bibAnnoteFile{NoStop}{Ott2012}%
\bibitem{Argenti2012}%
  \BibitemOpen
  \bibfield{author}{%
  \bibinfo {author} {\bibfnamefont{L.}~\bibnamefont{Argenti}}, \bibinfo
  {author} {\bibfnamefont{C.}~\bibnamefont{Ott}}, \bibinfo {author}
  {\bibfnamefont{T.}~\bibnamefont{Pfeifer}},\ and\ \bibinfo {author}
  {\bibfnamefont{F.}~\bibnamefont{Mart\'in}},\ }%
  \bibfield{journal}{%
  \bibinfo {journal} {arXiv [physics.atom-ph]}\ }%
  \textbf{\bibinfo {volume} {11}},\ \bibinfo {pages} {1211.2566v1} (\bibinfo
  {year} {2012})%
  \bibAnnoteFile{NoStop}{Argenti2012}%
\bibitem{IshMid2005}%
  \BibitemOpen
  \bibfield{author}{%
  \bibinfo {author} {\bibfnamefont{K.~L.}\ \bibnamefont{Ishikawa}}\ and\
  \bibinfo {author} {\bibfnamefont{K.}~\bibnamefont{Midorikawa}},\ }%
  \bibfield{journal}{%
  \Doi{10.1103/PhysRevA.72.013407}{\bibinfo {journal} {Phys. Rev. A}}\ }%
  \textbf{\bibinfo {volume} {72}},\ \bibinfo {pages} {013407} (\bibinfo {year}
  {2005})%
  \bibAnnoteFile{NoStop}{IshMid2005}%
\bibitem{PalHorRes2010}%
  \BibitemOpen
  \bibfield{author}{%
  \bibinfo {author} {\bibfnamefont{A.}~\bibnamefont{Palacios}}, \bibinfo
  {author} {\bibfnamefont{D.~A.}\ \bibnamefont{Horner}}, \bibinfo {author}
  {\bibfnamefont{T.~N.}\ \bibnamefont{Rescigno}},\ and\ \bibinfo {author}
  {\bibfnamefont{C.~W.}\ \bibnamefont{McCurdy}},\ }%
  \bibfield{journal}{%
  \Doi{10.1088/0953-4075/43/19/194003}{\bibinfo {journal} {J. Phys. B}}\ }%
  \textbf{\bibinfo {volume} {43}},\ \bibinfo {pages} {194003} (\bibinfo {month}
  {Sep.}\ \bibinfo {year} {2010})%
  \bibAnnoteFile{NoStop}{PalHorRes2010}%
\bibitem{PazFeiNag2011}%
  \BibitemOpen
  \bibfield{author}{%
  \bibinfo {author} {\bibfnamefont{R.}~\bibnamefont{Pazourek}}, \bibinfo
  {author} {\bibfnamefont{J.}~\bibnamefont{Feist}}, \bibinfo {author}
  {\bibfnamefont{S.}~\bibnamefont{Nagele}}, \bibinfo {author}
  {\bibfnamefont{E.}~\bibnamefont{Persson}}, \bibinfo {author}
  {\bibfnamefont{B.~I.}\ \bibnamefont{Schneider}}, \bibinfo {author}
  {\bibfnamefont{L.~A.}\ \bibnamefont{Collins}},\ and\ \bibinfo {author}
  {\bibfnamefont{J.}~\bibnamefont{Burgd\"{o}rfer}},\ }%
  \bibfield{journal}{%
  \Doi{10.1103/PhysRevA.83.053418}{\bibinfo {journal} {Phys. Rev. A}}\ }%
  \textbf{\bibinfo {volume} {83}},\ \bibinfo {pages} {053418} (\bibinfo {month}
  {May}\ \bibinfo {year} {2011})%
  \bibAnnoteFile{NoStop}{PazFeiNag2011}%
\bibitem{Bachau2001}%
  \BibitemOpen
  \bibfield{author}{%
  \bibinfo {author} {\bibfnamefont{H.}~\bibnamefont{Bachau}}, \bibinfo {author}
  {\bibfnamefont{E.}~\bibnamefont{Cormier}}, \bibinfo {author}
  {\bibfnamefont{P.}~\bibnamefont{Decleva}}, \bibinfo {author}
  {\bibfnamefont{J.~E.}\ \bibnamefont{Hansen}},\ and\ \bibinfo {author}
  {\bibfnamefont{F.}~\bibnamefont{Mart\'in}},\ }%
  \bibfield{journal}{%
  \bibinfo {journal} {Rep. Prog. Phys.}\ }%
  \textbf{\bibinfo {volume} {64}},\ \bibinfo {pages} {1815} (\bibinfo {year}
  {2001})%
  \bibAnnoteFile{NoStop}{Bachau2001}%
\bibitem{Argenti2009}%
  \BibitemOpen
  \bibfield{author}{%
  \bibinfo {author} {\bibfnamefont{L.}~\bibnamefont{Argenti}}\ and\ \bibinfo
  {author} {\bibfnamefont{R.}~\bibnamefont{Colle}},\ }%
  \bibfield{journal}{%
  \Doi{10.1016/j.cpc.2009.03.002}{\bibinfo {journal} {Comp. Phys. Commun.}}\ }%
  \textbf{\bibinfo {volume} {180}},\ \bibinfo {pages} {1442} (\bibinfo {month}
  {Sep.}\ \bibinfo {year} {2009}),\ ISSN \bibinfo {issn} {00104655},\
  \url{http://linkinghub.elsevier.com/retrieve/pii/S0010465509000848}%
  \bibAnnoteFile{NoStop}{Argenti2009}%
\bibitem{Argenti2008b}%
  \BibitemOpen
  \bibfield{author}{%
  \bibinfo {author} {\bibfnamefont{L.}~\bibnamefont{Argenti}}\ and\ \bibinfo
  {author} {\bibfnamefont{R.}~\bibnamefont{Moccia}},\ }%
  \bibfield{journal}{%
  \Doi{10.1088/0953-4075/41/3/035002}{\bibinfo {journal} {J. Phys. B: At. Mol.
  Opt. Phys.}}\ }%
  \textbf{\bibinfo {volume} {41}},\ \bibinfo {pages} {035002} (\bibinfo {month}
  {Feb.}\ \bibinfo {year} {2008}),\ ISSN \bibinfo {issn} {0953-4075},\
  \url{http://stacks.iop.org/0953-4075/41/i=3/a=035002?key=crossref.5ab2921b96%
d1d4c22b002a1e537e96d7}%
  \bibAnnoteFile{NoStop}{Argenti2008b}%
\bibitem{Argenti2010b}%
  \BibitemOpen
  \bibfield{author}{%
  \bibinfo {author} {\bibfnamefont{L.}~\bibnamefont{Argenti}}\ and\ \bibinfo
  {author} {\bibfnamefont{R.}~\bibnamefont{Moccia}},\ }%
  \bibfield{journal}{%
  \Doi{10.1088/0953-4075/43/23/235006}{\bibinfo {journal} {J. Phys. B: At. Mol.
  Opt. Phys.}}\ }%
  \textbf{\bibinfo {volume} {43}},\ \bibinfo {pages} {235006} (\bibinfo {year}
  {2010})%
  \bibAnnoteFile{NoStop}{Argenti2010b}%
\bibitem{Lindroth2012}%
  \BibitemOpen
  \bibfield{author}{%
  \bibinfo {author} {\bibfnamefont{E.}~\bibnamefont{Lindroth}}\ and\ \bibinfo
  {author} {\bibfnamefont{L.}~\bibnamefont{Argenti}},\ }%
  \bibfield{journal}{%
  \bibinfo {journal} {Adv. Q. Chem.}\ }%
  \textbf{\bibinfo {volume} {63}},\ \bibinfo {pages} {247} (\bibinfo {year}
  {2012})%
  \bibAnnoteFile{NoStop}{Lindroth2012}%
\end{thebibliography}%

\end{document}